\documentclass[preprint,showpacs,preprintnumbers,amsmath,amssymb]{revtex4}
\usepackage{graphicx}
\usepackage{dcolumn}
\usepackage{bm}

\begin{document}

\title{\lq\lq Tomography\rq\rq~of the cluster structure of light nuclei via relativistic dissociation}

\author{P. I. Zarubin}
   \email{zarubin@lhe.jinr.ru}
   \affiliation{V. I. Veksler and A. M. Baldin Laboratory of High Energy Physics\\
Joint Institute for Nuclear Research, Dubna, Russia}

\begin{abstract}
\indent  These lecture notes present the capabilities of relativistic nuclear physics for the development of the physics of nuclear clusters. Nuclear track emulsion continues to be an effective technique for pilot studies that allows one, in particular, to study the cluster dissociation of a wide variety of light relativistic nuclei within a common approach. Despite the fact that the capabilities of the relativistic fragmentation for the study of nuclear clustering were recognized quite a long time ago, electronic experiments have not been able to come closer to an integrated analysis of ensembles of relativistic fragments. The continued pause in the investigation of the \lq\lq fine\rq\rq~structure of relativistic fragmentation has led to resumption of regular exposures of nuclear emulsions in beams of light nuclei produced for the first time at the  Nuclotron of the Joint Institute for Nuclear Research (JINR, Dubna). To date, an analysis of the peripheral interactions of relativistic isotopes of beryllium, boron, carbon and nitrogen, including radioactive ones, with nuclei of the emulsion composition, has been performed, which allows the clustering pattern to be presented for a whole family of light nuclei.\par
\indent \\
\end{abstract}

\maketitle
\section*{INTRODUCTION}

\begin{figure}
    \includegraphics[width=4in]{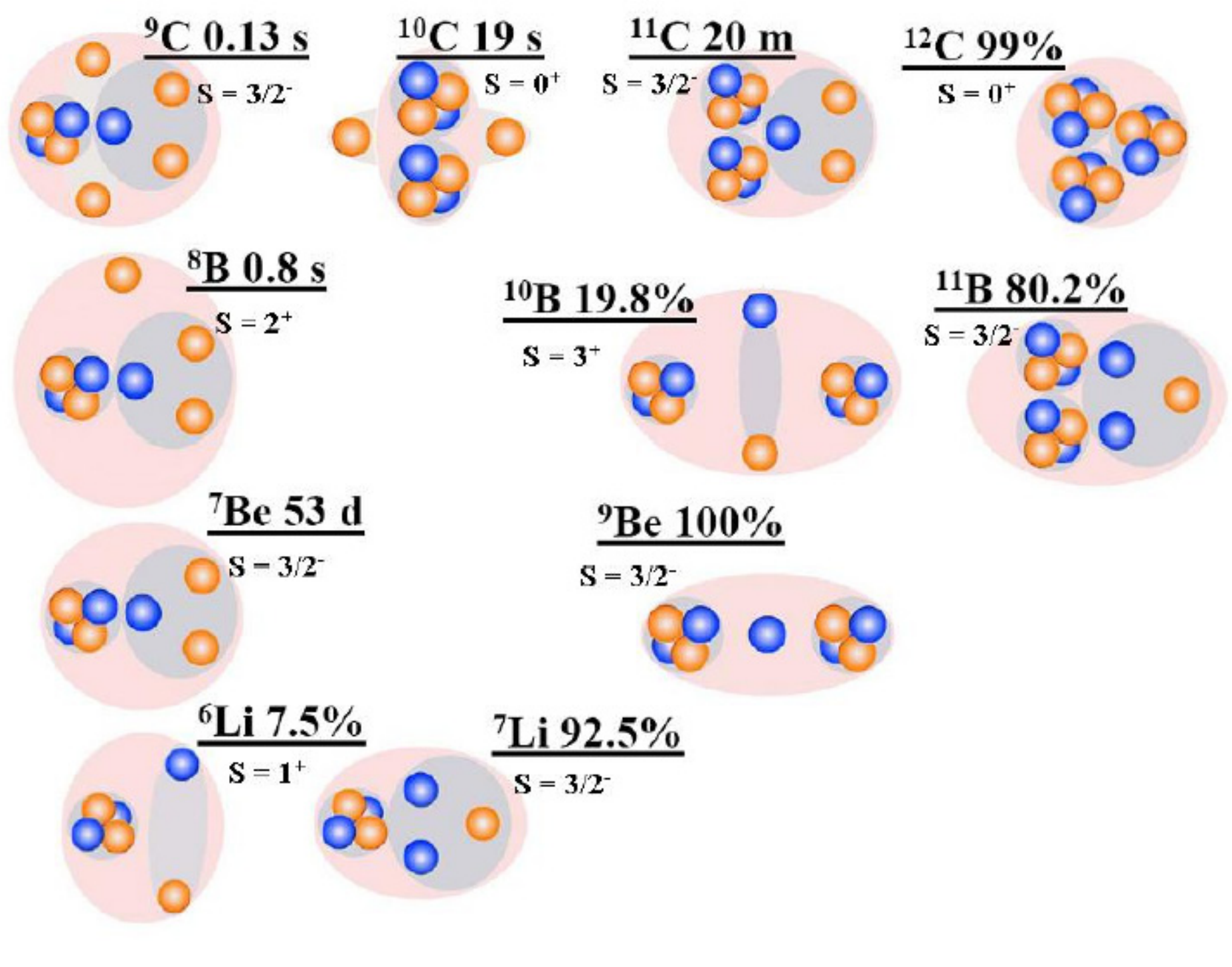}
    \caption{\label{Fig:1} (Color online) Diagram of cluster degrees of freedom in stable and neutron-deficient nuclei; abundances or lifetimes of isotopes, their spins and parities are indicated; orange circles correspond to protons and blue ones$~-~$ to neutrons; clusters are marked as dark background.}
    \end{figure}

\indent Collective degrees of freedom, in which groups of few nucleons behave as composing clusters, are a key aspect of nuclear structure. The fundamental \lq\lq building blocks\rq\rq~elements of clustering are the lightest nuclei having no excited states$~-~$first of all, the  $^4$He nucleus ($\alpha$ particles) as well as the deuteron ($d$), the triton ($t$) and the $^3$He nucleus  ($h$, helion). This feature is clearly seen in light nuclei, where the number of possible cluster configurations is small (Fig.~\ref{Fig:1}). In particular, the cluster separation thresholds in the nuclei of $^7$Be, $^{6,7}$Li, $^{11,10}$B, $^{11,12}$C and $^{16}$O are below the nucleon separation thresholds.  The stable $^9$Be, and unbound $^8$Be and $^9$B nuclei have a clearly pronounced cluster nature. In turn, the cluster nuclei $^7$Be, $^7$Li, and $^8$Be serve as cores in the isotopes $^8$B and $^{9-12}$C. Descriptions of the ground states of light nuclei in the shell and cluster models are complementary. In the cluster pattern the light nuclei are represented as superpositions of different cluster and nucleon configurations. The interest in such states is associated with the prediction of their  molecular-like properties \cite{Freer1,Freer2}. Nuclear clustering is traditionally regarded as the prerogative of the physics of nuclear reactions at low energies \cite{Beck}. The purpose of these lecture notes is to present the potential of one of the sections of high-energy physics$~-~$relativistic nuclear physics$~-~$for the development of the concepts of nuclear clustering.\par

\indent In the last decade, the concepts of ultracold dilute nuclear matter based on the condensation of nucleons in the lightest nuclei have been developed \cite{Horowitz,Shlomo,Yamada,Oertzen}. An $\alpha$-particle Bose-Einstein condensate ($\alpha$BEC) is considered as an analogue of atomic quantum gases \cite{Shlomo,Oertzen}. These developments put forward the problem of studying a variety of cluster ensembles and unbound nuclei as fundamental components of novel quantum matter. In a macroscopic scale coherent ensembles of clusters may play an intermediate role in nucleosynthesis, which makes the study of nuclear clustering more important and going beyond the scope of the problems of nuclear structure. At first glance, the studies of nuclear many-body systems seem to be impossible in laboratory conditions. Nevertheless, they can be studied indirectly in nuclear disintegration processes when the excitation is slightly above the appropriate thresholds. The configuration overlap of the ground state of a fragmenting nucleus with the final cluster states is fully manifested in interactions at the periphery of the target nucleus when the introduced perturbation is minimal. It appears that the phenomenon of peripheral dissociation of relativistic nuclei can serve as an alternative \lq\lq laboratory\rq\rq~for studying an unprecedented diversity of cluster ensembles.\par

\indent This idea is based on the following facts. At collisions of nuclei of the energy above  1$~A~$GeV, the kinematical regions of fragmentation of the projectile and target nuclei are clearly separated, and the momentum spectra of fragments come to asymptotic behavior. Thus, the regime of the limiting fragmentation of nuclei is reached, which also means that the isotopic composition of the fragments remains constant with increasing collision energy. Of particular value for the cluster physics are the events of peripheral dissociation of the incident nucleus with preservation of the number of nucleons in the region of its fragmentation. At a projectile energy above 1$~A~$GeV the probability of such dissociation reaches a few percent. Definition of interactions as peripheral ones is facilitated by increasing collimation of fragments. Thresholds of detection of relativistic fragments are absent, and their energy losses in the detectors are minimal. All these factors are essential for experimental studies.\par 	

\begin{figure}
    \includegraphics[width=5in]{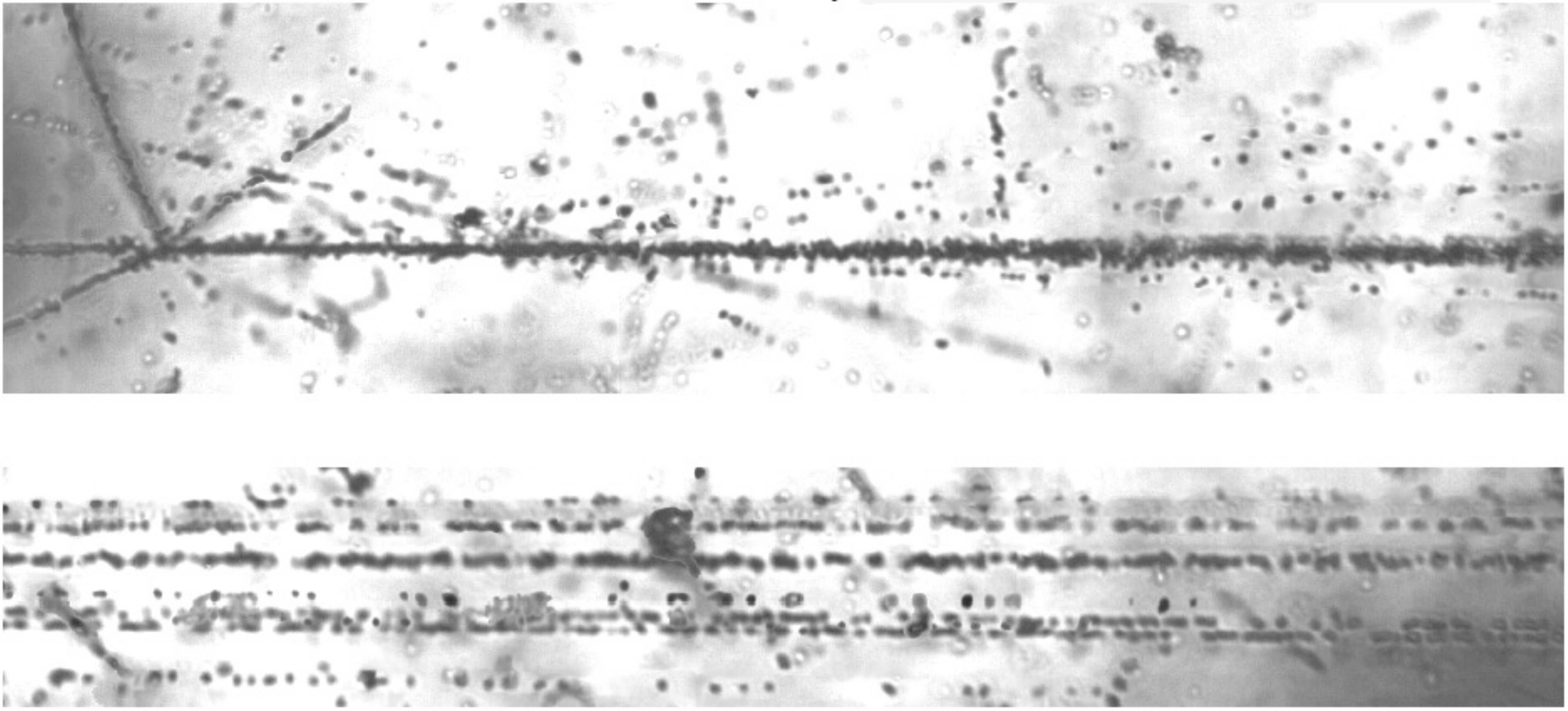}
    \caption{\label{Fig:2} Fragmentation of a 3.65$~A~$GeV $^{28}$Si nucleus in nuclear track emulsion.}
    \end{figure}

\indent The cluster ensembles produced in fragmentation of relativistic nuclei are best observed in nuclear track emulsion (NTE). As an example, Fig.~\ref{Fig:2} shows the macro photography of interaction in NTE of a 3.65$~A~$GeV $^{28}$Si nucleus. The granularity of the image is about 0.5~${\mu}$m. Of particular interest is a group of relativistic H and He fragments with the total charge $\sum$Z$_{fr}$~=~13. In the top photo one can see the fragment jet in a narrow cone accompanied by four singly charged relativistic particles in a wide cone and three fragments of the target nucleus. Moving in the direction of the jet fragments (bottom photo) allows three H and five He fragments to be distinguished. An intense \lq\lq track\rq\rq~on the bottom photo (third from top) splits into a pair of tracks with Z$_{fr}$~=~2 and the opening angle of about 2$\times10^{-3}$~rad, which corresponds to the $^8$Be nucleus decay. Such narrow decays are frequently observed in the fragmentation of relativistic nuclei. They testify to the completeness of observations across the spectrum of cluster excitations.\par

\indent According to NTE observations, the degree of dissociation of light nuclei as well as of the heaviest ones can reach a total destruction into the lightest nuclei and nucleons. Until now, information about this phenomenon has been fragmentary, and its interpretation has not been offered. Light nuclei are sources for the generation of the simplest configurations of the lightest clusters and nucleons. Being interesting by itself, their study provides a basis for understanding the dynamics of multiple fragmentations of heavy nuclei. The nuclear track emulsion exposed to relativistic radioactive nuclei makes it possible to diversify qualitatively the  \lq\lq tomography\rq\rq~of nuclear structure.\par

\indent The study of cluster structure by relativistic dissociation has both fundamental and practical importance. First of all, the probabilities with which the cluster states are shown in dissociation are related to the fundamental parameters of the ground and excited states of light nuclei. The knowledge of probabilities allows one to determine possible initial configurations of nuclear clusters, which is important for the analysis of the whole variety of nuclear reactions. Clustering is the basis of the underlying processes accompanying the phenomenon of the physics of nuclear isobars, hypernuclei and quark degrees of freedom. The ideas about nuclear clustering obtained in high-energy physics are important for applications in nuclear astrophysics, cosmic ray physics, nuclear medicine, and perhaps even nuclear geology. In particular, the probability distributions of the final cluster states may suggest new ways of multiple particle nuclear fusion, as inverse processes to their dissociation.\par

\indent At the JINR Nuclotron in 2002, the newly formed BECQUEREL collaboration launched  a program of irradiation of NTE stacks in the beams of relativistic isotopes of beryllium, boron, carbon and nitrogen, including radioactive ones (Fig.~\ref{Fig:1}). Coinciding with the name of the famous scientist, the project acronym indicates its key tasks$~-~\mathbf{Be}$ryllium ($\mathbf{B}$oron) $\mathbf{C}$lustering $\mathbf{Que}$st in $\mathbf{Rel}$ativistic Multifragmentation \cite{web1}. The physical design of the program consisted in a systematic verification of the assumption that in the dissociation of light relativistic nuclei it is possible to study the characteristics of their cluster structure. This idea is not obvious, and its implementation by means others than NTE face objective difficulties. Analysis of NTE exposures can best explore the structure and kinematical characteristics of a variety of ensembles of relativistic clusters. The ultimate goal of NTE application is the most complete identification and metrology of unusual configurations of clusters. Detailed information about the structure of dissociation will be very useful for the feasibility studies of electronic experiments with high statistics of events.\par
	
\indent Earlier observations among those discussed below were made in NTE exposures with the nuclei $^{12}$C \cite{Belaga}, $^{16}$O \cite{Andreeva}, $^{22}$Ne \cite{Naghy}, $^6$Li \cite{Adamovich1} and $^7$Li \cite{Adamovich2} and were carried out at the JINR Synchrotron in the 70-90s. Within the BECQUEREL project the peripheral interactions were analyzed in NTE (Fig. ~\ref{Fig:2}) exposed to the following set of nuclei: $^6$He \cite{Adamovich3}, $^{10}$B \cite{Adamovich4}, $^7$Be \cite{Peresadko}, $^{14}$N \cite{Shchedrina}, $^9$Be \cite{Artemenkov1,Artemenkov2}, $^{11}$B \cite{Karabova}, $^8$B \cite{Stanoeva}, $^9$C \cite{Krivenkov}, $^{10}$C, and $^{12}$N \cite{Artemenkov3,Artemenkov4,Kattabekov1,Kattabekov2,Mamatkulov}. These experimental results allow us to present a comprehensive picture of clustering for a family of nuclei at the beginning of the isotope table.\par

\indent The references to works cited in these lecture notes cover mainly the experimental results on the fragmentation of relativistic nuclei obtained with the NTE technique. It is recognized that this list cannot claim to be complete. Our goal is limited by the desire to give the initial presentation and generate interest in self-immersion in an exciting and promising topic of fragmentation of relativistic nuclei. Some of the unique materials on the subject were not  published sufficiently in the 70-90s due to circumstances beyond the authors' control, which makes their formal quoting difficult. Their preprints in Russian are stored on the BECQUEREL site \cite{web1}. We referred to them as to physical \lq\lq folklore\rq\rq~when writing these notes.\par

\section*{PHYSICS OF RELATIVISTIC NUCLEI}

\indent The BECQUEREL program owes its existence to a glorious era of research that deserves at least a brief reminder. The discovery of radioactivity by A.~H.~BECQUEREL at the same time made him the founder of the photographic method of its detection. Since then the searches for new phenomena in microphysics have been raising more and more new waves of interest in the use of nuclear photographs. Despite the known limitations in the statistics of the analyzed events, the classical method gives an objective topology of tracks in the full geometry, which allows one to see the prospects for technically advanced experiments. Events of multiple fragmentation of relativistic nuclei were observed as early as the 40s in NTE exposed to cosmic rays in the stratosphere \cite{Bradt}. Their photographs presented in the classic book by C.~H.~Powell, P.~H.~Fowler and D.~H.~Perkins \cite{Pfp}, among other fundamental observations can serve as a model of clarity in our time.   Our research is implemented in keeping with this tradition.\par

\indent Beams of light nuclei  of several $~A~$GeV were produced at the JINR Synchrophasotron in Dubna and at the BEVALAC of the Lawrence Berkeley Laboratory in the early 70s. Thus, prerequisites appeared for the application of the concepts and methods of high-energy physics for the development of the relativistic theory of atomic nuclei. At the same time experimental studies with the use of the NTE technique began at spectrometers and bubble chambers. Their main thrust was the search for the universal laws that describe the collisions of relativistic composite systems. Transition of spectra of nuclear fragments in the regime of limiting fragmentation and scale-invariant behavior was established. In the case of an uncorrelated formation of groups of relativistic fragments the description of their spectra could be reduced to the superposition of universal functions. However, meeting the generalizing principles the physics of relativistic fragmentation appears to be richer and deeper.\par

\indent A. M Baldin proposed to classify multiple particle production in nuclear collisions based on the relativistic-invariant description \cite{Baldin}. The particles are considered in the four-velocity space
\begin{equation}
\begin{split}
u_i~=~P_i/m_i,
\end{split}
\end{equation}
where $P_i$ are 4-momenta of particles participating in the reaction, and $m_i$ are their masses. Experimental data are presented in dimensionless invariant variables 
\begin{equation}
\begin{split}
b_{ik}~=~-~(P_i~/~mi~-~P_k~/~m_k)^2~=~-~(u_i~-~u_k)^2~=~2[(u_iu_k)~-~1]
\end{split}
\end{equation}
The variables $b_{ik}$ are directly related to the Lorentz factor of the relative motion of particles $\gamma_{ik}~=~(u_iu_k)$. In the range of relative velocities $b_{ik}~>>~1$, the hadrons involved in the process lose the role of quasiparticles, since the interaction of their constituents is so weakened that they can be considered within the framework of perturbative QCD. In the transition region $0.1~<~b_{ik}~<~1$, subnucleon degrees of freedom become important in the reconstruction of the structure and interactions of hadrons. The region $b_{ik}~<~10^{-2}$, corresponding to the interaction of weakly bound nucleon systems and nuclear clusters near the binding energy, is the domain of classical nuclear physics. It is a characteristic region for the physics of nuclear clustering. Invariant representation of the cluster kinematics can establish a connection with the findings of low-energy physics.\par

\indent The discovery of exotic nuclei at the BEVALAC accelerator brought the nuclear beams to the forefront of nuclear physics and led to the production of beams of radioactive nuclei in many accelerators. Entirely new phenomena were established in the structure of light radioactive nuclei and in nuclear reactions with their participation. Anomalously large radii of light nuclei, explained on the basis of nuclear structures, which consisted of spatially separated nucleons and nuclear cores, were observed.\par

\indent The Nuclotron, which replaced the JINR Synchrophasotron in the early 2000s, provides an opportunity to explore nuclear matter in the region $b_{ik}~<~10^{-2}$ for the optimal choice of the initial energy and the kinematics of detection. With the development of research in relativistic nuclear physics magneto-optical channels of particle transportation were built at this machine allowing secondary beams of 2$~A~$GeV/$c$ nuclei \cite{Rukoyatkin} to be formed. The channel used in our exposures has a length of about 50~m and consists of four bending magnets; its acceptance is about $2~-~3\%$.\par

\indent The nuclear track emulsion technique at the JINR Synchrophasotron began to be used in the 50s with  irradiations by 10 GeV protons \cite{Barashenkov}. Analysis of inelastic interactions of protons with nuclei of NTE composition pointed to the significant role of peripheral interactions. Often protons produced groups of mesons on Ag and Br nuclei which were visibly not destroyed. Later, these processes called coherent dissociation were studied in NTE irradiated by 70~GeV protons \cite{Antonova}. Similar reactions are possible in nucleus-nucleus interactions when the nucleus acts as a projectile, and the end result of coherent interaction is not the production of new particles, but the dissociation of the projectile nucleus. For the coherent dissociation of a projectile nucleus of the mass M$_0$ into a system of fragments with masses m$_i$ the threshold momentum of the nucleus is estimated as 
\begin{equation}
\begin{split}
p_{0min}\approx M_0B^{1/3}\Delta/\mu
\end{split}
\end{equation}
where $\mu$ is the mass of the $\pi$ meson, $B$ is the mass number of the target nucleus, and $\Delta~=~\sum m_i-M_0$ is mass defect with respect to the dissociation channel \cite{Belaga}. In particular, for the coherent dissociation of $^{12}$C$~\rightarrow~3\alpha$ in the Pb nucleus the estimate $p_{0min}$ is equal to approximately 300 MeV/$c$, and in the case of $^{16}$O$~\rightarrow~4\alpha$ $p_{0min}$ is roughly twice as much. Thus, the events of coherent dissociation of nuclei characterized by high thresholds should be investigated by experimental methods of high-energy physics.\par

\indent The establishment in the early 70s of relativistic nuclear physics was supported by the community which had rich experience in NTE applications. The particle accelerators opened a possibility of exploring the interactions of different nuclei of certain values of energy that allowed the spectra of relativistic fragments to be studied by the NTE technique. NTE was irradiated by nuclei that were first accelerated at the JINR Synchrophasotron, at the BEVALAC and later at the accelerators AGS (BNL) and SPS (CERN). The developed stacks of NTE pellicles were transferred for analysis to research centers worldwide in the spirit of traditions of the emulsion collaborations that arose as far back as in the pioneer period of cosmic ray research.\par

\begin{figure}
    \includegraphics[width=5in]{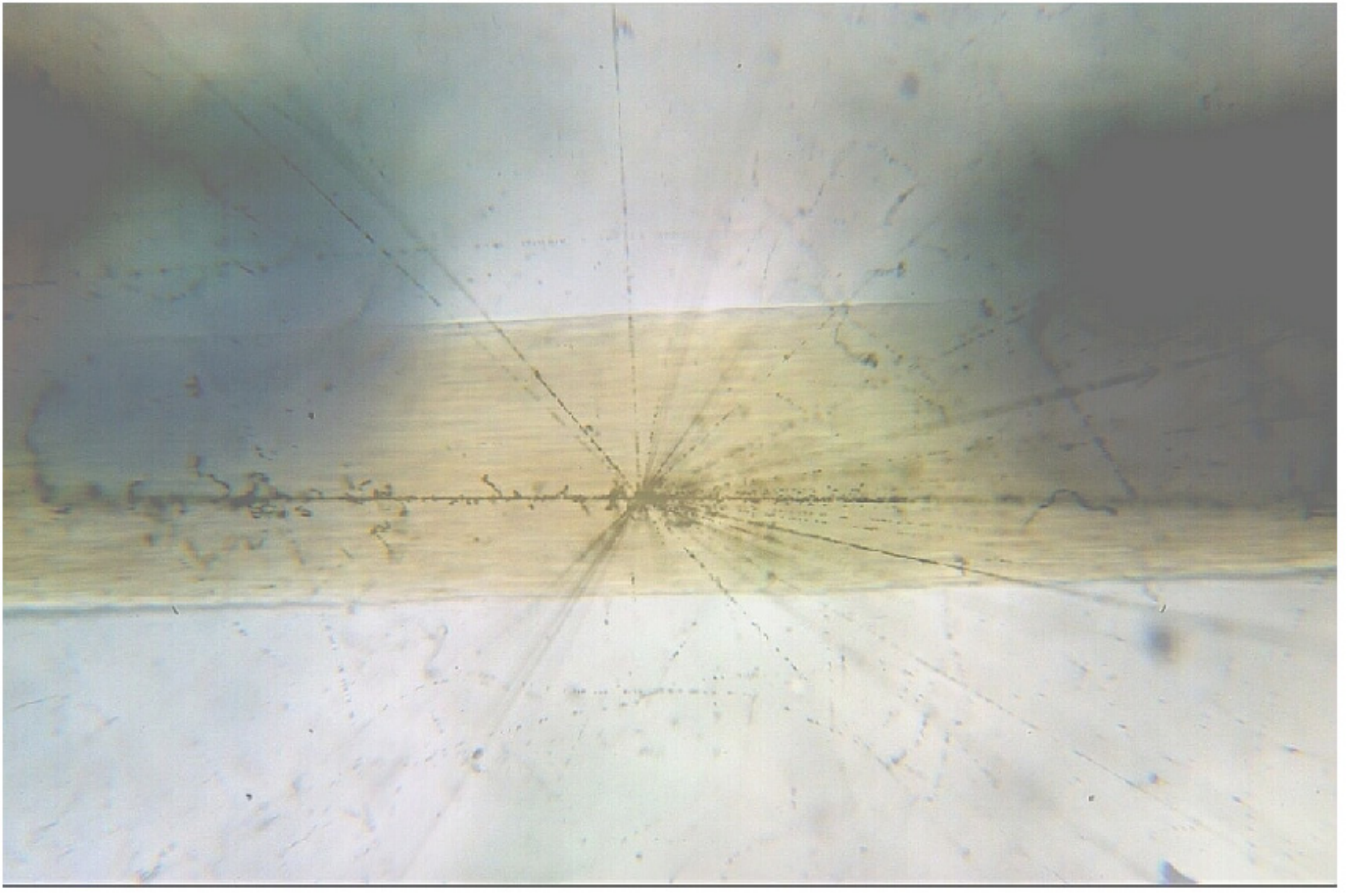}
    \caption{\label{Fig:3}  (Color online) Superposed photographs of a collision of a relativistic sulphur nucleus and a human hair obtained in the same scale by means of a microscope and a digital camera.}
\end{figure}

\indent The method received a motivation for further use because of its record-breaking resolution \cite{Pfp,Barkas}. It still retains uniqueness in the cone of relativistic fragmentation. The spatial resolution of the nuclear emulsion BR-2 (Russia) is 0.5~$\mu$m, and its sensitivity ranges from the most highly charged relativistic ions to singly charged relativistic particles. These features can be estimated in the photograph combining the pictures of the interaction of a relativistic sulfur nucleus and a human hair with a thickness of 60~$\mu$m~(Fig.~\ref{Fig:3}). Both images were obtained under identical conditions using a microscope and a digital camera. It can be argued that the nuclear emulsion gives the best projection of the events that occurred on the microcosm scale.\par

\indent Over time, the observation of such beautiful images was considered to be taken for granted. Demonstration of nucleus-nucleus interactions was replaced by the classification of tracks, not obvious to specialists in other techniques. The value of such a classification began to be forgotten with the weakening of interest in NTE caused by complexity of measurements. To make the results available to the perception, conservation of the patterns of peripheral interactions of relativistic nuclei was resumed in our video collection \cite{web1}.\par

\indent The emulsion method contributed to the establishment of the fundamental properties that characterize the collision of relativistic composite systems. As a rule, the event search was conducted for the primary tracks without selection providing systematized observations. However, this approach limits the statistics of rare events. Particular attention was given to central collisions as candidates for exotic events. The labor consuming analysis of its many tracks was motivated by searches for nuclear matter at the highest concentration of density and energy$~-~$the intranuclear cascade and shock waves in nuclear matter and, to the greatest extent, the quark-gluon plasma. The modern development of this area is widely known.\par

\begin{figure}
    \includegraphics[width=5in]{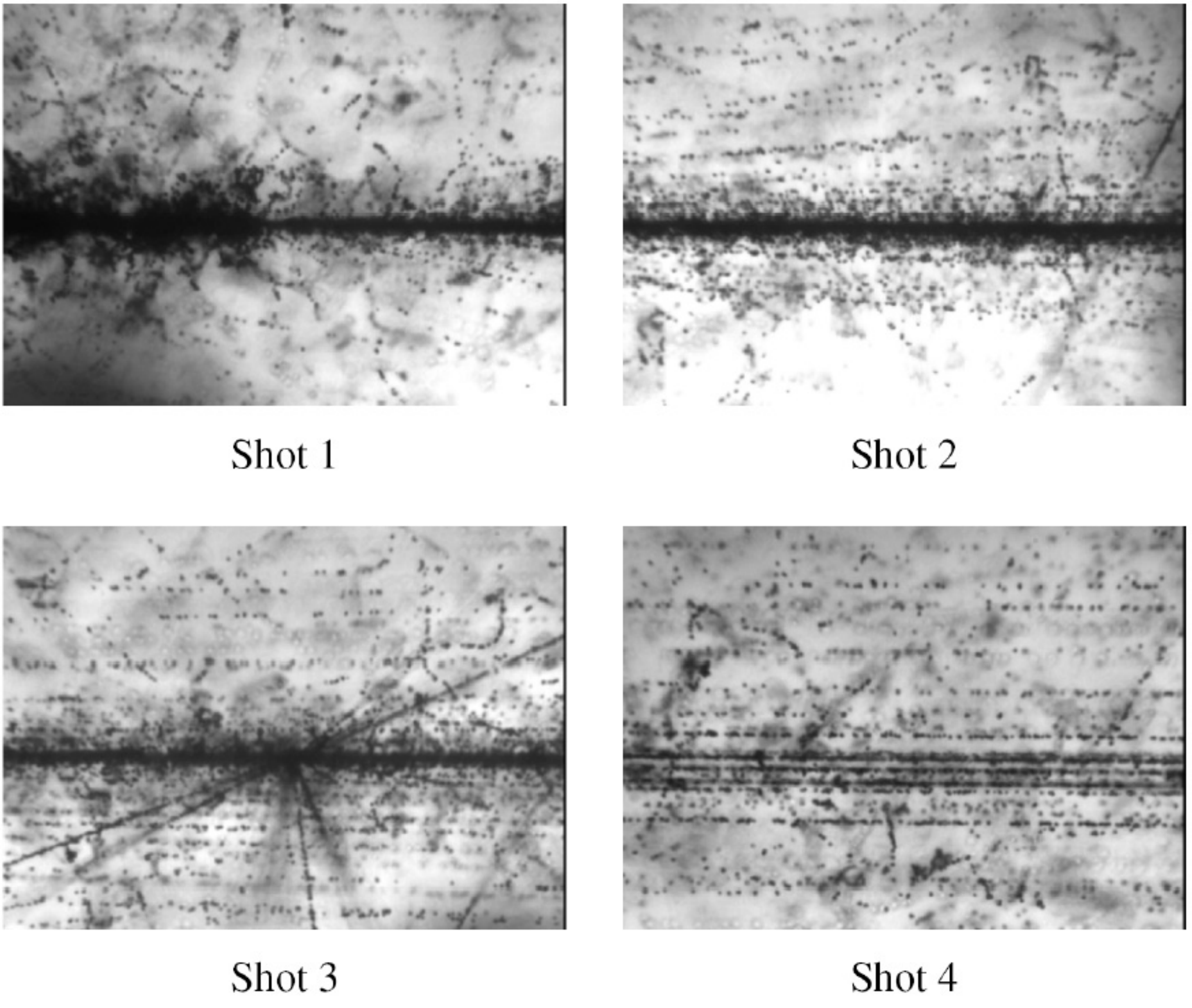}
    \caption{\label{Fig:4} Consecutively photographed event of the peripheral interaction of a 158$~A~$GeV $^{207}$Pb nucleus in nuclear track emulsion: primary nucleus track and interaction vertex followed by projectile fragment jet (Shot~1); jet core with apparent tracks of singly and doubly charged particles (Shot~2); jet core with a secondary interaction star (Shot~3); completely resolved jet core (Shot~4, 3~cm distance from the vertex).}
\end{figure}

\indent The results of the 70-90s retain the value in the aspect of relativistic fragmentation. Among the observed interactions of a few percent of events were the peripheral fragmentation of nuclei into the narrow jets of light nuclei, nucleon clusters and nucleons with a total charge close to the initial charge of the nucleus \cite{Belaga,Andreeva,Naghy,Adamovich1,Heckman,Friedlander,Jain,Singh1,Singh2,Baroni1,Baroni2,Adamovich5,Adamovich6,Cherry,Adamovich7}. Often, the peripheral events were not accompanied by the formation of fragments of the target nuclei, in the case of which there appeared an analogy with the coherent dissociation of protons proceeding at multiple smaller mass differences between the final and initial states. One of the most striking examples is given in Fig.~\ref{Fig:4}, which clearly shows the breakdown of ionization as a result of multiple fragmentation of the incident nucleus Au.\par

\indent Speaking about the nature of this phenomenon, it is possible to associate the probability of dissociation channels with the spectroscopic factors of the various cluster components of its ground state. These events indicate the disappearance of the Coulomb barrier of the nucleus and the exit of virtual clusters on the mass shell, followed by a rescattering. It is possible that the generation of fragment ensembles occurs not only in the states of the continuous spectrum. In the most \lq\lq delicate\rq\rq~collisions, population of the excited states above the decay thresholds is possible. In addition, nucleon clusters formed in the peripheral dissociation of relativistic nuclei may have a diffractive scattering. Thus, the peripheral collisions contain unique information about the quantum-mechanical aspects of the formation of the cluster ensembles. This assumption requires verifications with clearly interpretable examples. Positive findings will provide a basis for the development of ideas about the physics of multiple cluster systems.\par

\indent Despite their hidden aesthetics, peripheral interactions attracted a limited interest. Their study turned out to be in a shadow of \lq\lq romantic\rq\rq~physics of central collisions. No less important is the fact that, although the possibility of a relativistic approach to the study of nuclear structure was recognized, its application without a complete registration of relativistic fragments appeared to be limited. The apparent simplicity of the fragmentation cone study is deceptive. With respect to such peripheral interactions NTE remains the only means of observation that provide not only unique observation, but also a reasonable statistics. Of course, NTE does not provide momentum analysis. However, due to the development of relativistic physics of few-nucleon systems based on magnetic spectrometers and bubble chambers a variety of data about the fragmentation of relativistic nuclei may be attracted.\par
	
\indent Our study is aimed at exploring the coherent dissociation of neutron deficient nuclei, adjacent to the beginning of the table of isotopes (Fig.~\ref{Fig:1}), since the NTE technique offers special advantages for this. The following issues were raised:\par 
\begin{enumerate}
\item How does relativistic dissociation reflect the $\alpha$-cluster structure of light nuclei? 
\item How does $^{2,3}$H and $^3$He clustering manifest itself in relativistic dissociation? 
\item Is the population of cluster ensembles requiring nucleon rearrangement beyond $\alpha$-clustering is possible in relativistic dissociation? 
\item What is the proportion of nuclear diffractive and electromagnetic mechanisms of dissociation on heavy nuclei of NTE composition?
\end{enumerate}
	
\indent The stages of this study were closely related to the opportunities that arose at the JINR Nuclotron in the 2000s. In the final period of the operation of the JINR Synchrophasotron (1999), first experience of analysis  was obtained when NTE was exposed to a mixed secondary beam of $^6$He and $^3$H nuclei. Construction of the system of slow extraction of accelerated nuclei from the Nuclotron (2002) made it possible to perform irradiation by $^{10}$B nuclei. The $2\alpha~+~d$ clustering was established for the $^{10}$B dissociation which motivated the irradiation by $^{14}$N nuclei to study the $3\alpha~+~d$ clustering and later by $^{11}$B nucleus exposure to study the $2\alpha~+~t$ clustering. The interest in the $^{11}$B nucleus quickened analysis of the $\alpha~+~t$ clustering in the early $^7$Li irradiation. To develop ideas about $^3$He-based clustering, irradiation was carried out in the secondary beam of $^7$Be nuclei formed in charge-exchange reactions of primary $^7$Li nuclei (2004-2005). The acceleration of the $^{10}$B nuclei allowed secondary beams of $^9$Be and $^8$B isotopes to be created. The results of these exposures gave grounds for exposures in the beams of $^{9,10}$C and $^{12}$N isotopes formed in the fragmentation of primary $^{12}$C nuclei (2005-2006). The resumption of the use of nuclear emulsion has led to the survival of the NTE technology, to the preservation of the experience in data analysis, and to the involvement of young researchers.\par

\indent The next section presents the approaches taken to analyze the interactions of relativistic nuclei in emulsion and the key facts on the peripheral dissociation of light stable nuclei. Their combined use became the basis for the proposal of the BECQUEREL experiment for the study of radioactive nuclei.\par

\section*{DISSOCIATION OF RELATIVISTIC NUCLEI}

\subsection*{Advantages of the NTE technique}

\begin{figure}
    \includegraphics[width=4in]{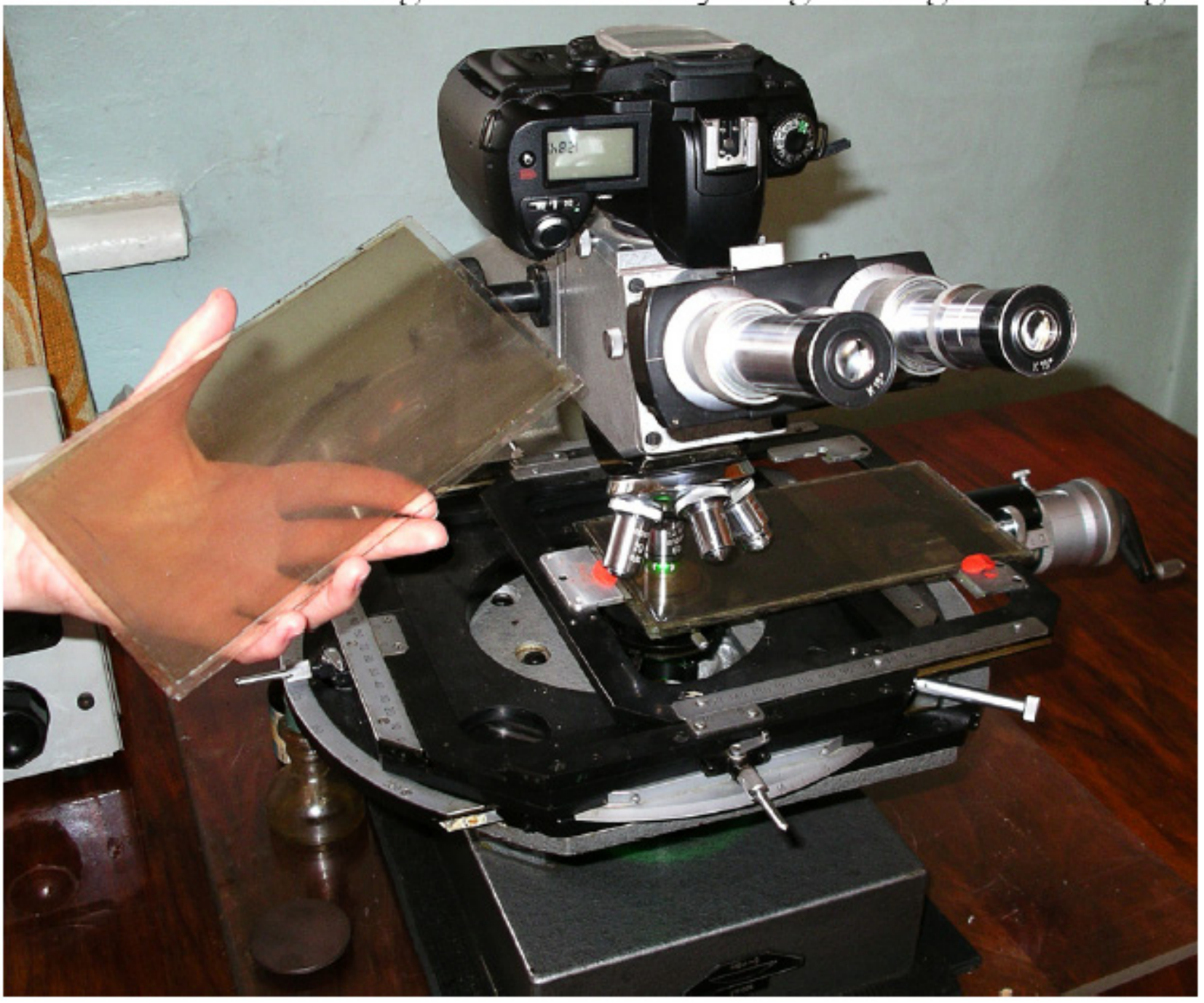}
    \caption{\label{Fig:5} Photograph of an NTE pellicle on a glass substrate and of a microscope with an installed photo camera.}
\end{figure}

\indent An emulsion chamber is assembled as a stack of pellicles 550 m$\mu$ thick and  10$\times$20~cm$^2$ in size (Fig.~\ref{Fig:5}). The factors in obtaining large event statistics are thickness reaching 80~g/cm$^2$ along the long side and complete  efficiency of charged particle detection. NTE contain Ag and Br nuclei as well as H nuclei in similar concentrations. By the density of hydrogen NTE is close to a liquid hydrogen target. This feature allows one to compare in the same conditions the disintegrations of projectile nuclei in nuclear diffractive and electromagnetic dissociation on heavy target nuclei as well as in collision with protons.\par

\indent The fragments of the relativistic nuclei are concentrated in a cone limited by the angle 
\begin{equation}
\begin{split}
\theta_{fr}~\approx~p_{fr}/p_0
\end{split}
\end{equation}
where $p_{fr}~=~0.2~$GeV/$c$ is a quantity characterizing the Fermi momentum of nucleons, and $p_0$ is the momentum per nucleon of projectile nucleus. If the beam is directed parallel to the pellicles, the tracks of all relativistic fragments can stay long enough in a single pellicle for 3-dimensional reconstruction. The distribution of events over the interaction channels with different composition of charged fragments (or the charged topology) is a direct feature of the fragmentation of relativistic nuclei. The results on charge topology of coherent dissociation for the relativistic nuclei $^{16}$O, $^{22}$Ne, $^{24}$Mg, $^{28}$Si and $^{32}$S are summarized in \cite{Andreeva1}.\par

\indent In NTE the angular resolution for the tracks of relativistic fragments is of the order of 10$^{-5}~$rad. Measurements of the polar angles $\theta$ of fragment emission are not sufficient for comparison of data for different values of the initial energy of nuclei. More generic is a comparison by the values of the transverse momentum $P_T$ of fragments with the mass number  $A_{fr}$ according to the approximation of 
\begin{equation}
\begin{split}
P_T~\approx~A_{fr}P_0\sin{\theta}
\end{split}
\end{equation}
which corresponds to conservation by the fragments of the velocity of the primary nucleus (or momentum $P_0$ per nucleon). Obviously, the most important is the $\theta$ angle resolution, since the $\theta$ distributions are \lq\lq pressed against\rq\rq~zero. For $\alpha$-cluster nuclei the assumption about the correspondence of a relativistic fragment with the charge Z$_{fr}~=~2$ to the $^4$He isotope is well justified. Separation of the isotopes $^3$He and $^4$He is required for neutron-deficient nuclei.\par

\indent In the fragmentation of the NTE nuclei, strongly ionizing target fragments (Fig.~\ref{Fig:2}) can be observed including $\alpha$ particles, protons with energy below 26~MeV energy and light nuclei$~-~$n$_b$~(b-particles), as well as non-relativistic protons above 26~MeV$~-~$n$_g$ (g-particles). In addition, the reactions are characterized by a multiplicity of mesons produced outside the cone of fragmentation$~-~$n$_s$ (s-particles). Using these parameters, conclusions can be drawn about the nature of the interaction.\par

\indent In  coherent dissociation events there are no fragments of the target nuclei (n$_b~=~0$, n$_g~=~0$) and charged mesons (n$_s~=~0$). Events of this type were informally named as \lq\lq white\rq\rq~stars due to the absence of tracks of strongly ionizing particles n$_h$ (n$_h~=~$n$_b~+~$n$_g$). \lq\lq White\rq\rq~stars are produced by nuclear diffraction and electromagnetic interactions on heavy target nuclei. Their share in the total number of inelastic events is a few percent. The name \lq\lq white\rq\rq~stars aptly reflects the \lq\lq breakdown\rq\rq~of ionization in the transition from the primary nucleus track to a narrow cone of the secondary tracks down to Z$_{pr}$ times. This feature constitutes the main difficulty for electronic techniques, since the greater the degree of dissociation at the event, the harder it is to register. In nuclear track emulsions the situation is quite opposite.\par

\begin{figure}
    \includegraphics[width=4in]{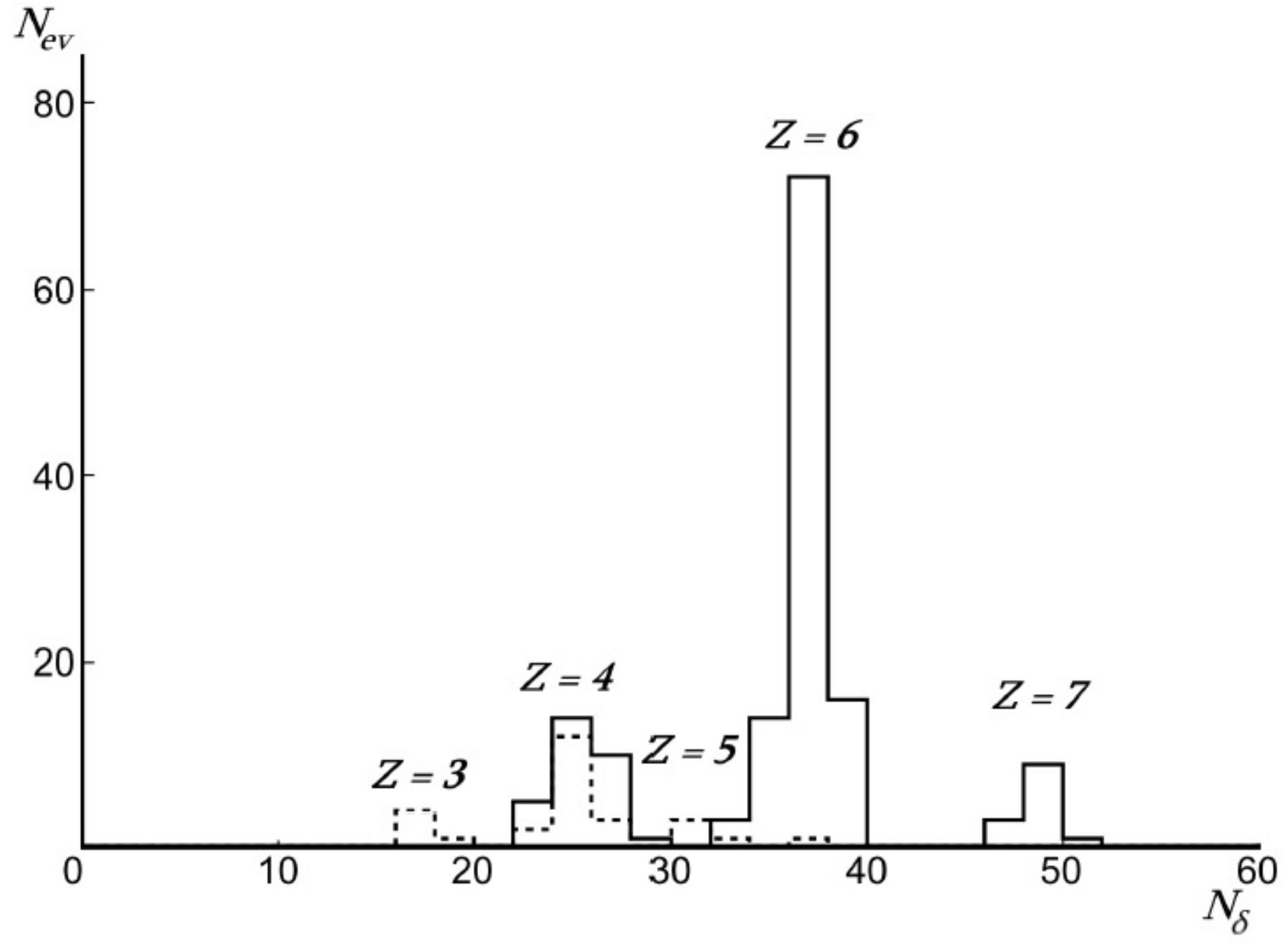}
    \caption{\label{Fig:6} Distribution of the beam particle tracks N$_{ev}$ (solid line) and secondary fragments (dashed line) with respect to the mean number of  $\delta$-electrons N$_{\delta}$, over 1~mm of the track length in nuclear track emulsion exposed to a mixed beam of $^7$Be, $^{10}$C and $^{12}$N nuclei.}
\end{figure}

\indent The practical task of determining charge topology is identification of fragment charges Z$_{fr}$. Due to 4-fold difference in ionization, the charges of relativistic fragments Z$_{fr}~=~1$ and 2 are determined already by visual search. The values of fragment charges Z$_{fr}~\ge~3$ are determined by the density of gaps on tracks or by the density of $\delta$-electrons N$_{\delta}$ depending on charges as Z$_{fr}^2$. A valuable condition is the conservation by relativistic fragments of the charge of the beam nuclei Z$_{pr}$, i.e. Z$_{pr}~=~\sum$Z$_{fr}$ for the interpretation of \lq\lq white\rq\rq~stars in NTE exposed to mixed secondary beams. It allows one to separate  in the beam the contribution of lighter nuclei with a similar charge to mass ratio. This criterion is fundamentally important for NTE exposures in beams with complex composition. An example of charge separation of the beam nuclei and secondary fragments in the mixed-beam exposure to  $^7$Be, $^{10}$C and $^{12}$N for events $\sum$Z$_{fr}~=~6$ and 7 is shown in Fig.~\ref{Fig:6} \cite{Kattabekov1,Kattabekov2,Mamatkulov}. In cases of light neutron-deficient nuclei the determination of charges allows one to determine their mass numbers.\par

\indent Relativistic H and He isotopes are identified by their values  $p\beta c$, where $p$is fragment momentum and $\beta c$ is its velocity. Due to \lq\lq quantization\rq\rq~of fragment momenta their mass numbers $A_{fr}$ are defined as $p_{fr}\beta_{fr}c/(p_0\beta_0c)$. The $p\beta c$ value is determined by the average angle of multiple Coulomb scattering estimated via the track offsets $\mid$D$\mid$ on 2-5 cm track sections. It is necessary to measure $\mid$D$\mid$ not less than in 100 points in order to achieve 20-30\% accuracy of $p\beta c$ determination comparable to the difference A$_{fr}$ for $^3$He and $^4$He. This labor-intensive method is not a routine procedure, and its use is justified in fundamentally important cases for limited number of fragment tracks.\par

\indent In particular, this method was applied in the analysis of NTE exposure \cite{Krivenkov} to 2$~A~$GeV/$c$ $^9$C nuclei in a situation when the $^3$He nuclei having the same magnetic rigidity as the $^9$C were predominant in the beam. The average value $<p\beta c>_{3He}$ for the beam $^3$He nuclei was (5.1$~\pm~$0.1)~GeV with RMS of 0.8~GeV, which is close to the expected value of 5.4~GeV (for $^4$He$~-~7.2~$GeV) and is acceptable for separation of the isotopes $^3$He and $^4$He. The \lq\lq white\rq\rq~stars with fragments of Z$_{fr}~=~5$ and 4 and with the beam particle charges Z$_{pr}~=~6$ found in this exposure were interpreted as $^9$C$~\rightarrow~^8$B$~+~p$ and $^7$Be$~+~2p$. Indeed, the distribution of particles Z$_{fr}~=~1$ has $<p\beta c>_H~=~(1.5~\pm~0.1)~$GeV and RMS of 0.4~GeV, which corresponds to protons.\par

\begin{figure}
    \includegraphics[width=4in]{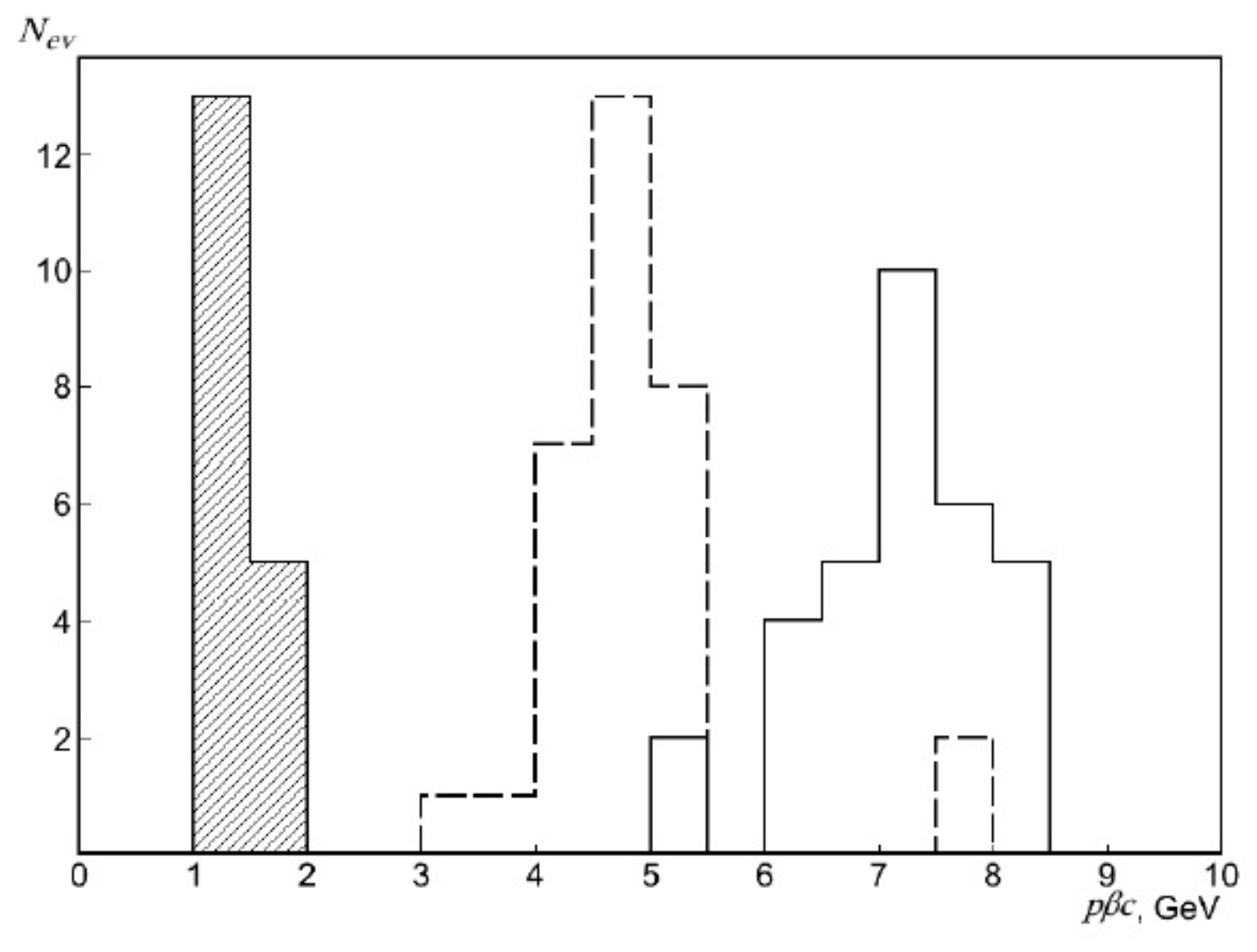}
    \caption{\label{Fig:7} Distribution of the measured values $p\beta c$ for fragments from \lq\lq white\rq\rq~stars $^{10}$C$~\rightarrow~$2He$~+~$2H (solid line$~-~$He, shaded histogram$~-~$H) and $^9$C$~\rightarrow~$3He (dashed line).}
\end{figure}

\indent The states of 3$^3$He became a central subject of study of the  coherent dissociation of $^9$C. Only for 22 He tracks in 16 found \lq\lq white\rq\rq~stars C$~\rightarrow~$3He it was possible to perform $p\beta c_{^3He}$ value measurements (Fig.~\ref{Fig:7}). The average value is $<p\beta c_{^3He}>~=~(4.9~\pm~0.3)$~GeV for RMS of 0.9~GeV and corresponds to the calibration of the $^3$He beam nuclei. Only for three $^3$He \lq\lq white\rq\rq~stars the determination of $p\beta c$ was possible for all of the fragments allowing these events to be identified as 3$^3$He most reliably.\par

\indent The values $p\beta c_{^4He}$ were measured for H and He tracks of 16 \lq\lq white\rq\rq~stars $^{10}$C$~\rightarrow~2$He$~+~$2H in the NTE exposed to a mixed beam of isotopes $^7$Be, $^{10}$C and $^{12}$N \cite{Kattabekov1,Kattabekov2,Mamatkulov} with the same momentum per nucleon as in the case of $^9$C. The dominance of $^1$H and $^4$He isotopes confirms the separation of $^{10}$C (Fig.~\ref{Fig:7}). In the case of He nuclei, 23 tracks were taken from the $^8$Be$_{g.s.}$ decays. For all He tracks measured in the exposure to $^9$C (including $^3$He calibration) the average value is $<p\beta c_{^3He}>~=~(5.0~\pm~0.1)$~GeV at RMS of 0.8~GeV, and in the of $^{10}$C it is $<p\beta c_{^4He}>~=~(7.9~\pm~0.2)~$GeV at RMS of 0.8~GeV. Thus, two groups of measurements clearly correspond to different He isotopes.  Fig.~7 shows the distribution of the measured values of $p\beta c$ for He fragments of the events $^9$C$~\rightarrow~3^3$He \cite{Krivenkov}. $^3$He and $^4$He fragments are clearly separated by $p\beta c$.\par

\begin{figure}
    \includegraphics[width=5in]{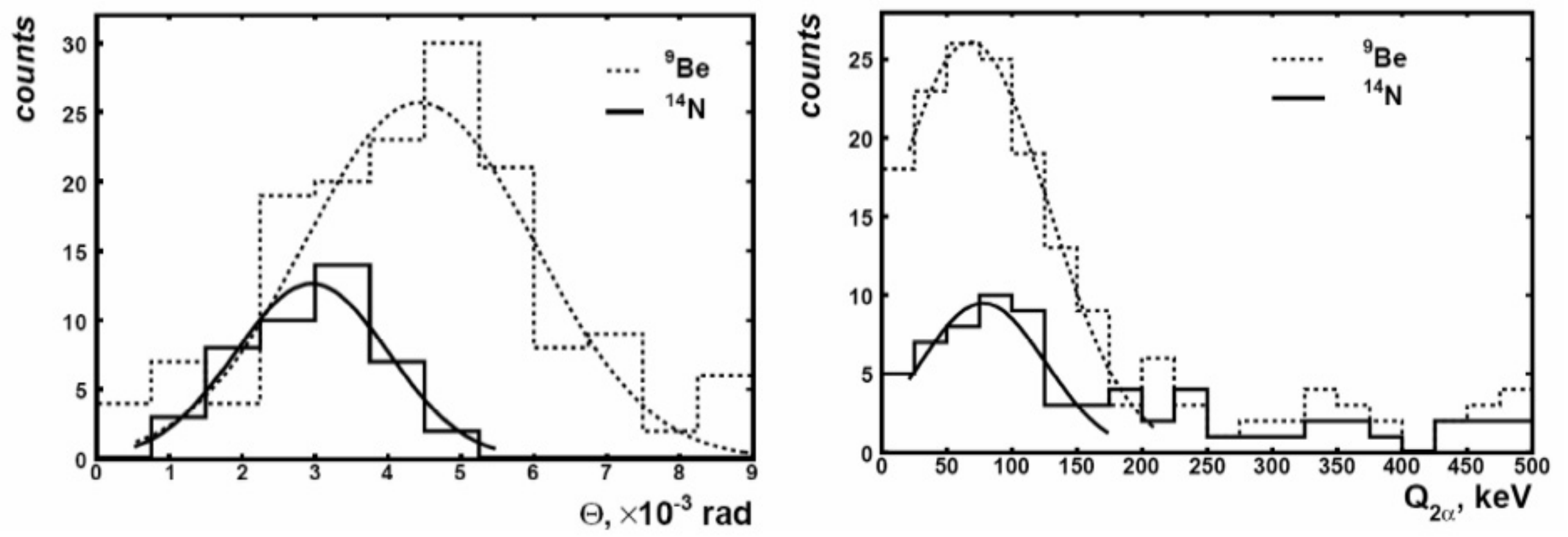}
    \caption{\label{Fig:8} Distribution of the opening angles $\Theta$ in $\alpha$-particle pairs (left) and energy Q$_{2\alpha}$ (right) for fragmentation events $^9$Be$~\rightarrow~^8$Be$_{g.s.}$(0$^+$) at 2$~A~$GeV/$c$ and $^{14}$N$~\rightarrow~^8$Be$_{g.s.}(0^+)$ at 2.9$~A~$GeV/$c$.}
\end{figure}

\indent The excitation energy of a fragment system Q is defined as the difference between the invariant mass of the fragmenting system M$^*$ and the mass of the primary nucleus M, i.e. Q~=~M$^*~-~$M. M$^*$ is the sum of all products of the fragment 4-momenta $P_{i,k}~$M$^{*2}~=~\sum(P_i~\cdot~P_k)$. 4-momenta $P_{i,k}$ are determined in the approximation of conservation of the initial momentum per nucleon by fragments. The opening angle distributions of $\alpha$-particle pairs $\Theta$ are superposed in Fig.~\ref{Fig:8} for the dissociation $^9$Be$~\rightarrow~^8$Be$_{g.s.}$ at 2$~A~$GeV/$c$ \cite{Artemenkov1,Artemenkov2} and for $^{14}$N$~\rightarrow~^8$Be$_{g.s.}$ at 2.9$~A~$GeV/$c$ \cite{Shchedrina}. Their average values differ significantly: (4.4$~\pm~$0.2)$\times10^{-3}$~rad and $(3.0~\pm~0.2)\times10^{-3}$~rad, which points to the sensitivity of the measurements to the reduction of the decay cone with increasing momentum. Overlaying when transformed to the Q$_{2\alpha}$ (Fig.~\ref{Fig:8}) points to on the identity of the source of narrow $\alpha$ pairs in both cases to $^8$Be$_{g.s.}(0^+$) decays with the average energy  $<Q_{2\alpha}>~=~(68~\pm~14)~$keV for $^9$Be and $(78~\pm~14)~$keV for $^{14}$N. Thus, the observation of the ground state decay of the $^8$Be nucleus shows a fine resolution of angle measurements as well as convenience of invariant representation.\par

\subsection*{Coherent dissociation of relativistic $^{12}$C and $^{16}$O nuclei}

\begin{figure}
    \includegraphics[width=6in]{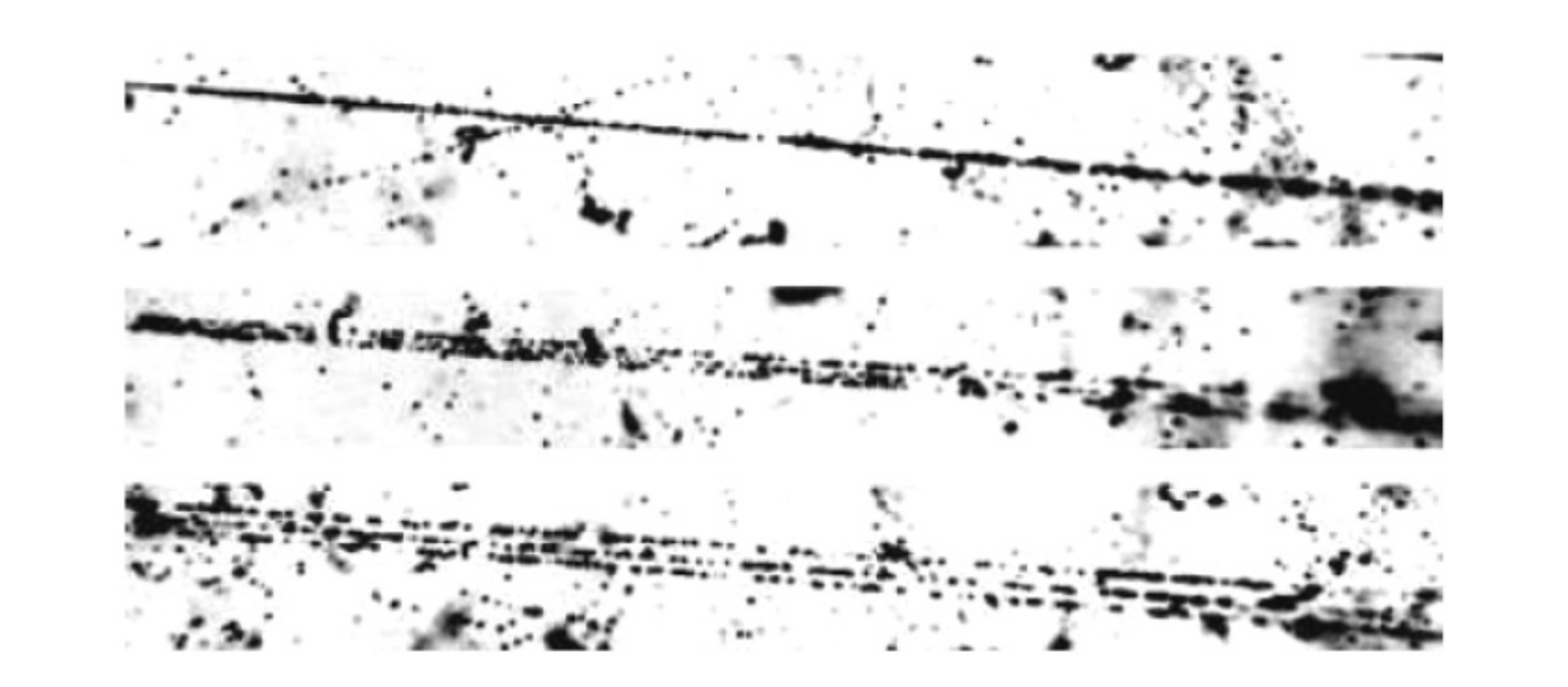}
    \caption{\label{Fig:9} Coherent dissociation $^{12}$C$~\rightarrow~$3He at 4.5$~A~$GeV/$c$ ; upper photo: an interaction vertex and a fragment jet; middle and lower photo: shifting from the vertex along the fragment jet allows three tracks of doubly charged fragments to be distinguished.}
\end{figure}

\indent At the JINR Synchrophasotron in the early 70s, NTE was exposed to 4.5$~A~$GeV/$c$ $^{12}$C nuclei (energy of 3.65$~A~$GeV). The statistics of 2468 interactions found along a 338 m scanned path of primary tracks included 28 \lq\lq white\rq\rq~stars. The only option for these stars was the cluster breakup $^{12}$C$~\rightarrow~3\alpha$ (threshold E$_{th}~=~7.3~$MeV) limited in the cone $\theta~<~3^{\circ}$ (example in Fig.~\ref{Fig:9}). Later the NTE was enriched with lead to enhance the electromagnetic dissociation \cite{Belaga}. The search for events was carried out in an accelerated manner over the NTE pellicle area. As a result, the statistics had already 72 \lq\lq white\rq\rq~stars $^{12}$C$~\rightarrow~3\alpha$. A key observation became relativistic $^8$Be decays that constituted at least 20\%. \par

\begin{figure}[t]
    \includegraphics[width=5in]{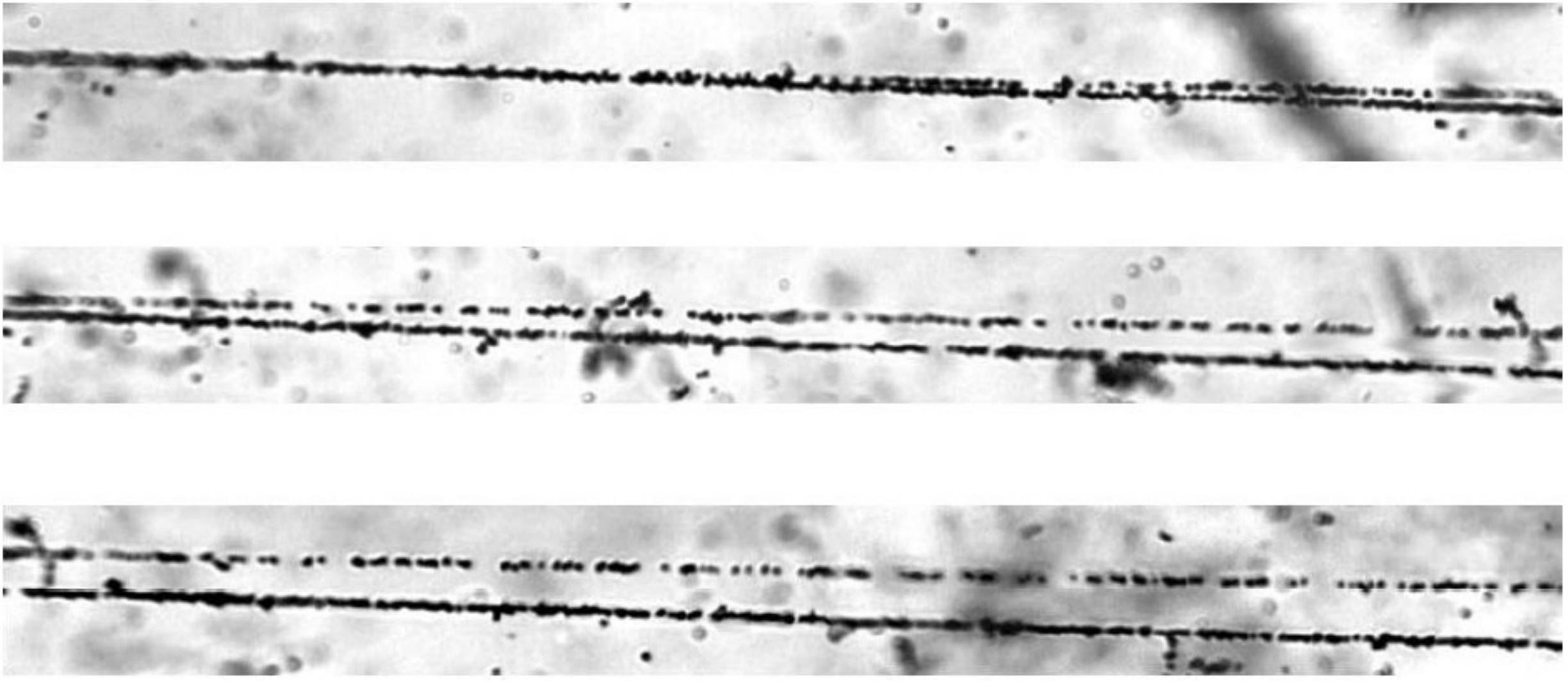}
    \caption{\label{Fig:10} Coherent dissociation $^{16}$O$~\rightarrow~$C$~+~$He at 4.5$~A~$GeV/$c$.}
\end{figure}

\begin{figure}[t]
    \includegraphics[width=5in]{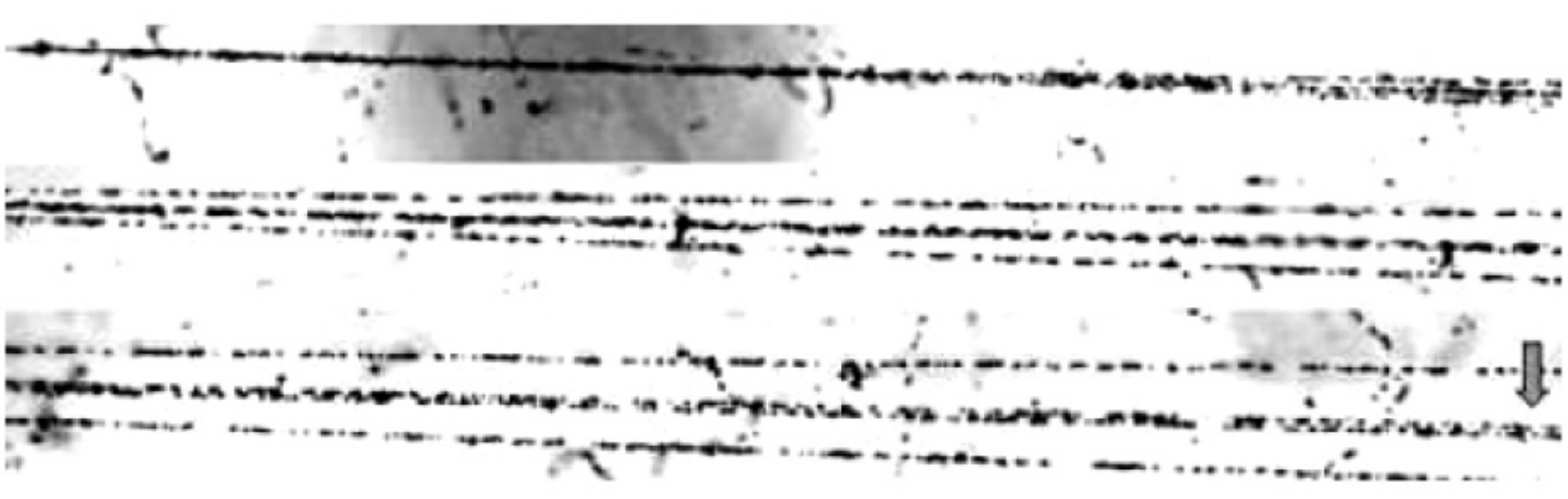}
    \caption{\label{Fig:11} Coherent dissociation $^{16}$O$~\rightarrow$2He$~+~^8$Be$_{g.s.}$ at 4.5$~A~$GeV/$c$; arrow points to tracks of the decay $^8$Be$_{g.s.}~\rightarrow~2\alpha$.}
\end{figure}

\indent The same approach was extended to the $^{16}$O nucleus. Table~1 shows an increased variety of channels. This distribution leads to a probability distribution. The channels C$~+~$He (E$_{th}~=~7.2~$MeV, example in Fig.~\ref{Fig:10}) and N$~+~$H (E$_{th}~=~12.1~$MeV) are leading. The statistics of \lq\lq white\rq\rq~stars $^{16}$O$~\rightarrow~4\alpha$ (example in Fig.~\ref{Fig:11}) that were found in an accelerated search reached 641 events \cite{Andreeva}, demonstrating in NTE  the possibility of studying processes with the cross-section $10^{-2}~-~10^{-3}$ of the inelastic cross-section. The probabilities of cascading channels defined by simulation were $\approx$25\% for the $^8$Be$~+~2\alpha$ and $\approx$20\% for 2$^8$Be. Thus, the relativistic 4$\alpha$-system proved to be strongly correlated.\par

\begin{table}
\caption{\label{Tabel:1} Charge topology of the fragments of the coherent dissociation of 4.5$~A~$GeV/$c$ $^{16}$O nuclei}
\begin{tabular}{c|c|c|c|c|c|c|c|c}
\hline\noalign{\smallskip}
~N+H~&~C+He~&~C+2H~&~B+3H~&~B+He+H~&~Be+He~&~Be+He+H~&~4He~&~3He+2H~\\
\noalign{\smallskip}\hline\noalign{\smallskip}
18 & 21 & 7 & 2 & 10 & 1 & 1 & 9 & 3 \\
\noalign{\smallskip}\hline
\end{tabular}
\end{table}

\subsection*{Features of the dissociation of heavier nuclei}

\begin{figure}
    \includegraphics[width=5in]{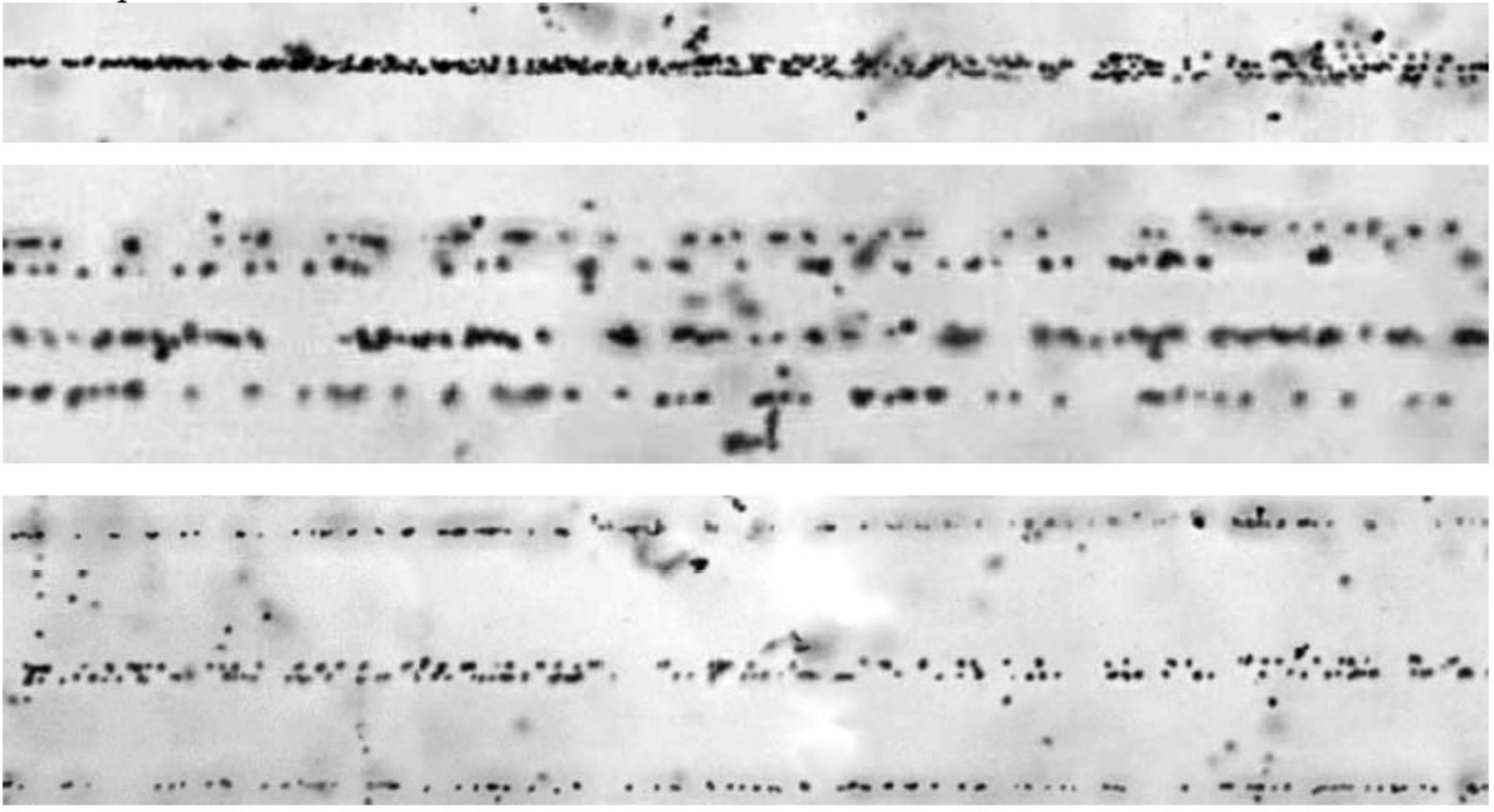}
    \caption{\label{Fig:12} Coherent dissociation $^{20}$Ne$~\rightarrow~$3He$~+~^8$Be$_{g.s.}(0^+)$ at 4.5$~A~$GeV/$c$.}
\end{figure}

\indent The progress in the development of the JINR Synchrophasotron as a source of relativistic nuclei achieved in the 80s has made it possible to perform exposures with the $^{22}$Ne, $^{24}$Mg, $^{28}$Si and $^{32}$S nuclei. The information received at that time about the peripheral fragmentation of nuclei retains its uniqueness and provides motivation for future experiments. We illustrate this statement, basing on the measurements of interactions of 3.22$~A~$GeV $^{22}$Ne nuclei. The statistics of events is traced in Table~2 for the channels $\sum$Z$_{fr}~=~10$ with multiplicities of the target fragments n$_b$ and n$_g$. There are channels present, starting from the separation of single fragments Z$_{fr}~=~1$ and 2 down to the destruction into the lightest nuclei (example in Fig.~\ref{Fig:12}). An obvious feature is the dominance of \lq\lq white\rq\rq~stars. Such distributions for relativistic Mg, Si and S nuclei have similar pattern.\par

\begin{table}
\caption{\label{Tabel:2} The distribution of the peripheral interactions of 3.22$~A~$GeV $^{22}$Ne nuclei over multiplicity of target fragments n$_b$ and n$_g$ (n$_s~=~0$); in parenthesis is share in $\%$}
\begin{tabular}{c|c|c|c|c|c|c}
\hline\noalign{\smallskip}
$n_b$&0&0&1&2&3&$>3$\\
$n_g$&0&1&0&0&0&0\\
\noalign{\smallskip}\hline\noalign{\smallskip}
F+H &~26(19.5)~& 9(15.0) &~13(44.8)~&~2~& $-$ & 1 \\
O+He & 54(40.6) &~19(31.7)~& 2(6.9) & $-$ &~1~&~1~\\
O+2H & 12(9.0) & 7(11.7) & $-$ & $-$ & $-$ & $-$ \\
N+He+H & 12(9.0) & 7(11.7) & 4(13.8) & 1 & $-$ & $-$ \\
N+3H & 3(2.3) & 3(5.0) & $-$ & $-$ & $-$ & $-$ \\
C+2He & 5(3.8) & 3(5.0) & 3(10.3) & 1 & $-$ & $-$ \\
C+2He+2H & 5(3.8) & 3(5.0) & 3(10.3) & $-$ & $-$ & $-$ \\
C+4H & 2(1.0) & $-$ & $-$ & $-$ & $-$ & $-$ \\
B+Li+H & 1(0.8) & $-$ & $-$ & $-$ & $-$ & $-$ \\
B+2He+H & 2(1.5) & 1(1.7) & $-$ & $-$ & $-$ & $-$ \\
B+He+H & 2(1.5) & 1(1.7) & $-$ & $-$ & $-$ & $-$ \\
B+5H & 1(0.8) & $-$ & 1(3.4) & $-$ & $-$ & $-$ \\
2Be+2H & $-$ & 1(1.7) & $-$ & $-$ & 1 & $-$ \\
~Be+Li+3H~& 1(0.8) & $-$ & $-$ & $-$ & $-$ & $-$ \\
Be+3H & 2(1.5) & $-$ & $-$ & $-$ & $-$ & $-$ \\
Be+He+4H & 1(0.8) & $-$ & $-$ & $-$ & $-$ & $-$ \\
Li+3He+H & $-$ & 1(1.7) & $-$ & $-$ & $-$ & $-$ \\
5He & 3(2.3) & $-$ & 1(3.4) & 2 & 1 & $-$ \\
4He+2H & 1(0.8) & 5(8.3) & 2(6.9) & $-$ & $-$ & $-$ \\
\noalign{\smallskip}\hline
\end{tabular}
\end{table}

\begin{figure}
    \includegraphics[width=4in]{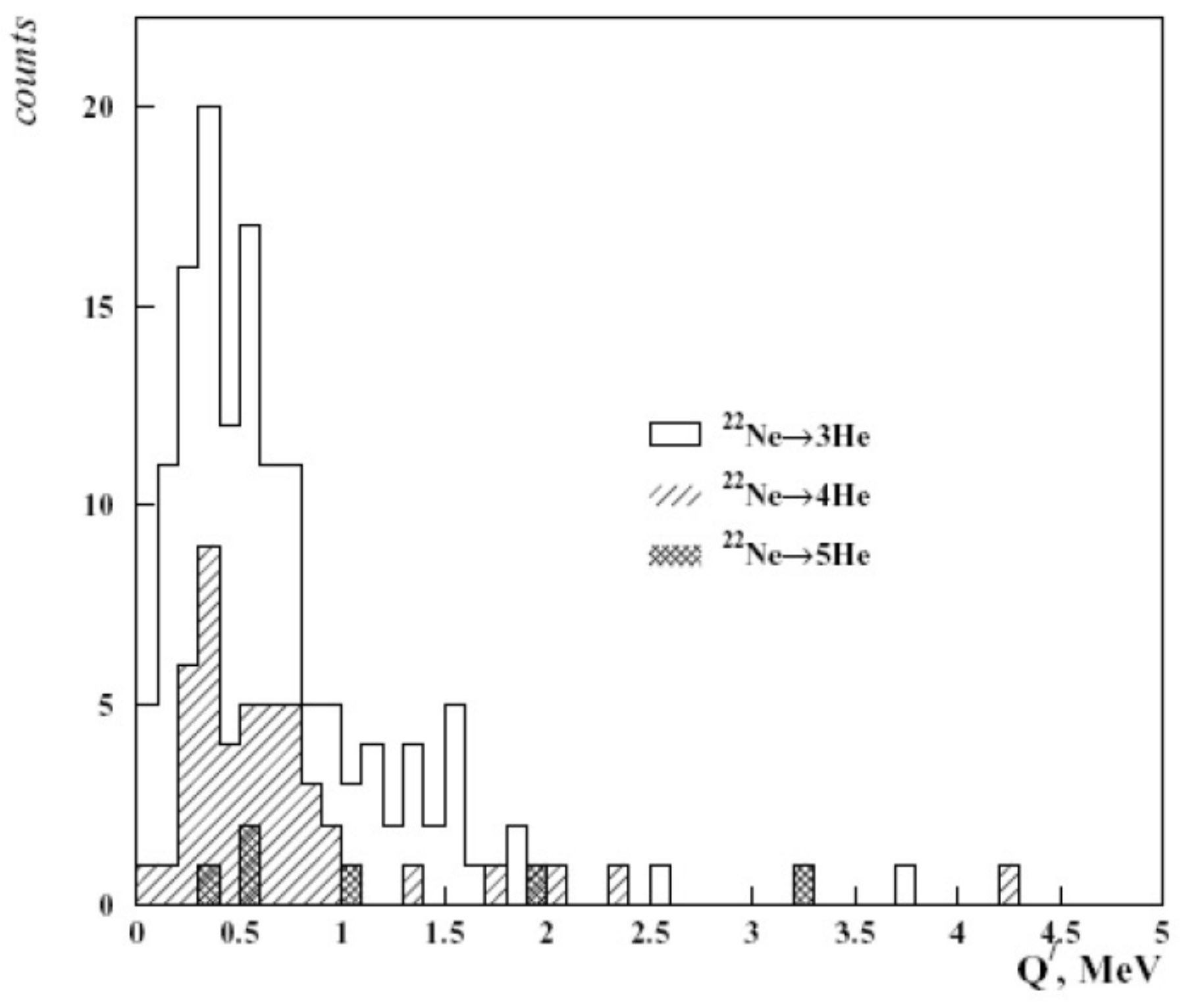}
    \caption{\label{Fig:13} Distribution of $\alpha$-particle pairs produced in the fragmentation $^{22}$Ne$~\rightarrow~$N$\alpha$ over energy  Q'~ (per nucleon of a fragment).}
\end{figure}

\indent A nuclear state analogous to the dilute Bose-Einstein condensate (BEC) can manifest itself in the formation of N$\alpha$-particle ensembles with a narrow velocity distribution in the center of mass. However, the c.m.s. definition is difficult enough, while the analysis of jets in the 4-velocity space $b_{ik}$ can represent N$\alpha$-systems in a universal way. Events $^{22}$Ne$~\rightarrow~$N$\alpha$ were selected, satisfying the criterion of $b_{ik}~<~10^{-2}$ for each $\alpha$-pair for N$_{\alpha}~=~3$, 4 and 5. Fig.~\ref{Fig:13} shows the distribution of normalized excitation energy Q'~=~Q/(4N$\alpha$). Despite the increase in multiplicity, N$\alpha$-jets remain similar. Three \lq\lq white\rq\rq~stars were found among the events of $^{22}$Ne$~\rightarrow~5\alpha$. Of these there were two \lq\lq golden\rq\rq~events containing all $\alpha$-particles within the 1$^{\circ}$ cone. For these two events, the values Q' are very small$~-~$400 keV and 600 keV per nucleon. The detection of these 5$\alpha$-states is an argument in favor of searching for $\alpha$-particle Bose-Einstein condensate in relativistic fragmentation.\par

\subsection*{Cluster structure of $^6$Li and $^7$Li nuclei}

\indent The data on the interactions of 4.5$~A~$GeV/$c$ $^6$Li nuclei \cite{Adamovich1} attracted our attention to the NTE technique for addressing the issues of cluster structure. The $^6$Li nucleus is the only among stable nuclei except the deuteron that are attributable to nuclei with exotic structure. Due to increased sizes and weak nucleon coupling the exotic nuclei are characterized by enhanced interaction cross-sections and narrowed momentum distributions of their cores in fragmentation. These properties of the $^6$Li nucleus are manifested in the relativistic fragmentation in NTE.\par

\begin{figure}
    \includegraphics[width=5in]{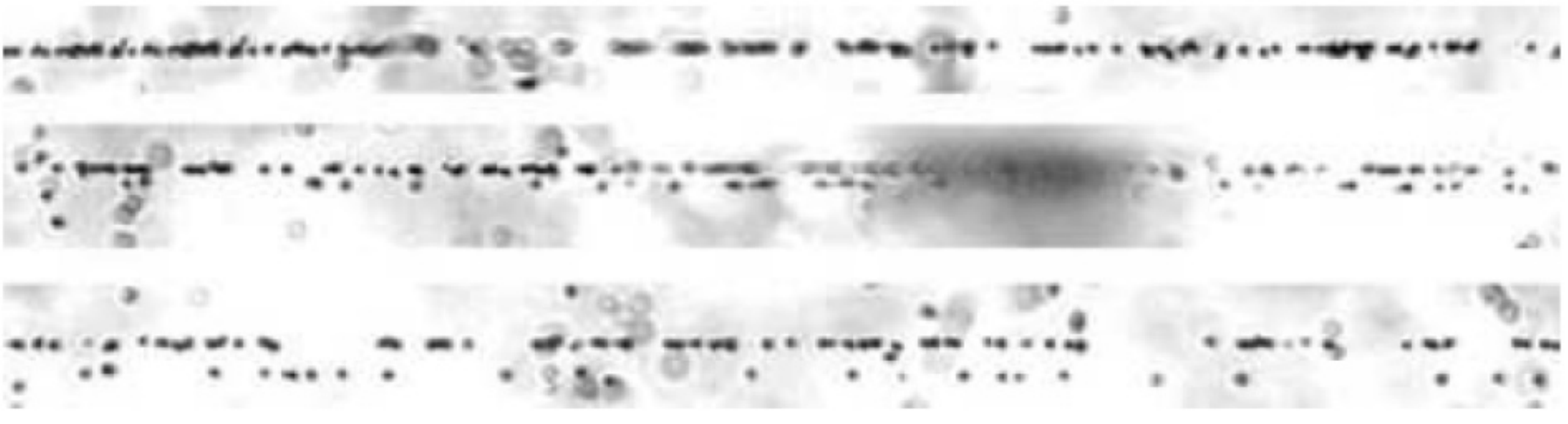}
    \caption{\label{Fig:14} Coherent dissociation $^6$Li$~\rightarrow~$He$~+~$H at 4.5$~A~$GeV/$c$.}
\end{figure}

\indent The free path with respect to inelastic interactions, which happened to be about 3~cm shorter than the one calculated by the Bradt-Peters formula ($\approx17~$cm) \cite{Bradt}, suggests an anomalously large radius of the $^6$Li nucleus. In the model of the geometric overlap of nuclear densities its value is equal to ($2.7~\pm~0.1$) Fermi which is consistent with the data on the  radius of the $^6$Li nucleus. A feature of the isotopic composition of $^6$Li fragments was an unusually high yield of deuterons nearly equal to the yield of protons, which was not observed in the fragmentation of the $^4$He, $^{12}$C, $^{22}$Ne, and $^{28}$Si nuclei. For the fragmentation $^6$Li$~\rightarrow~\alpha$ the value of the mean transverse momentum of $\alpha$ particles turned out to be reduced$~-~<P_{T_{\alpha}}>~=~(0.13~\pm~0.01)~$GeV/$c$, while for the interactions of $^{12}$C nuclei this value was $<P_{T_{\alpha}}>~=~(0.01~\pm~0.24$)~GeV/$c$. 31 \lq\lq white\rq\rq~stars in which fragments were completely identified can be regarded as \lq\lq golden\rq\rq~events (example in Fig.~\ref{Fig:14}). Among them there are 23 events corresponding to the dissociation channel $\alpha~+~d~$(E$_{th}~=~1.47~$MeV), and there are 4 events in the channels  $^3$He~+~$t$~(E$_{th}~=~15.8~$MeV) and $t~+~d~+~p~($E$_{th}~=~21.2~$MeV). Thus, the $\alpha~+~d$ cluster structure is clearly manifested.\par

\begin{figure}
    \includegraphics[width=4in]{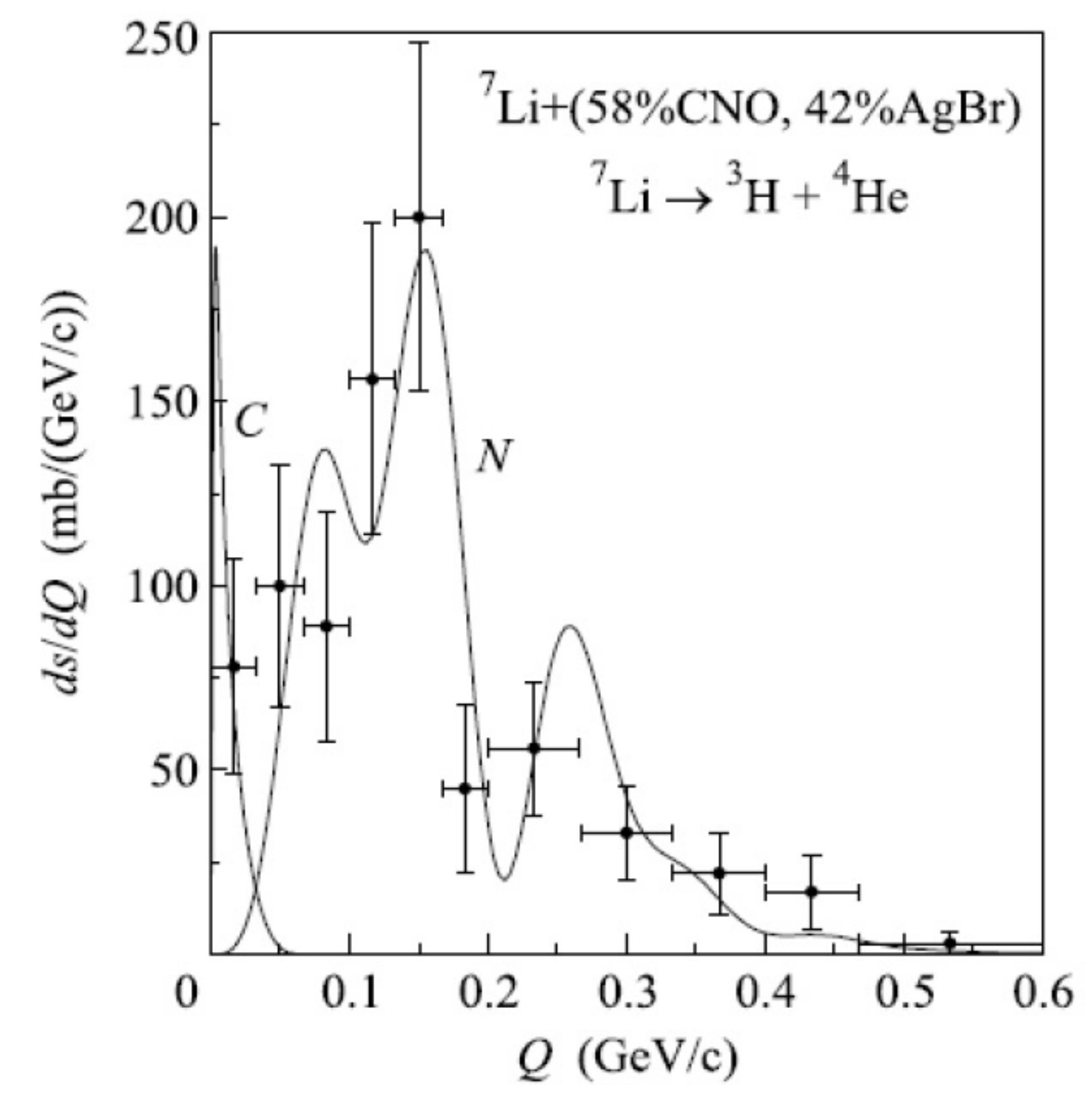}
    \caption{\label{Fig:15} Differential cross-section of the coherent dissociation $^7$Li$~\rightarrow~^4$He$~+~^3$H at 4.5$~A~$GeV/$c$ over the total transverse momentum Q \cite{Peresadko1}; experimental data and theoretical dependences of Coulomb (C) and nuclear diffractive (N) interactions.}
\end{figure}

\indent The question of the triton as a cluster was resolved based on an analysis of the \lq\lq white\rq\rq~stars $^7$Li$~\rightarrow~$He$~+~$H \cite{Adamovich1}. Determination of the masses of the relativistic fragments showed that 50\% of these events represent the channel $\alpha~+~t~$(E$_{th}~=~2.5~$MeV), while the channel $\alpha~+~d~+~n$ constitued 30\% (E$_{th}~=~$6.1~MeV) and $\alpha~+~p~+~2n~$(E$_{th}~=~7~$MeV)$~-~20\%$. These findings stimulated the analysis of the relationship of nuclear and electromagnetic diffraction mechanisms of cluster dissociation on a  mixture of NTE nuclei \cite{Peresadko1}. The first type of interaction for the $\alpha~+~t$ channel covers the total momentum range $50~<~P_T~<~500~$MeV/$c$, and the second$~-~$considerably narrower$~-~P_T~<~50~$MeV/$c$ (Fig.~\ref{Fig:15}).\par

\subsection*{Exposure in a mixed beam of $^6$He and $^3$H isotopes}

\indent Before the beginning of exposures under the BECQUEREL project experience was gained in the analysis of nuclear emulsion exposed to the beam \lq\lq cocktail\rq\rq~of a mixture of $^6$He and $^3$H nuclei \cite{Adamovich3,Pfp}. An extracted beam of 2.67$~A~$GeV/$c$~$^6$Li nuclei was directed to a plexiglass target located at the focal point of the beam transport channel. The $^6$He nucleus beam was formed by using the selection of products of the charge-exchange process $^6$Li$~\rightarrow~^6$He. The secondary particles produced almost at a zero angle were seized by the channel tuned to the selection of particles with charge to mass number Z$_{pr}$/A$_{pr}~=~1/3$. The percentage of $^6$He nuclei was about 1\%, and $^3$H nuclei were dominant.\par

\indent A few \lq\lq white\rq\rq~stars with a noticeable change in the direction of doubly charged tracks in which the $^6$He nucleus lost a neutron pair and emitted $\alpha$ particles were found in this exposure. The average transverse momentum of these $\alpha$ particles is $<P_{T_{\alpha}}>\approx35~$MeV/$c$. Thus, an indication for a drastically narrower distribution $P_{T_{\alpha}}$ for the  coherent dissociation of $^6$He was obtained, in which the neutron halo  is exhibited as a structural feature of this nucleus. However, the value of the $^6$He mean free path, including the registered coherent interactions, was 16.3$~\pm~$3.1~cm, being significantly greater than for $^6$Li. It can be assumed that excessive  mean range for $^6$He is due to lack of efficiency of observations of the  coherent dissociation $^6$He$~\rightarrow~^4$He$~+~2n$ (no more than 50\%). This assumption means that the contribution of coherent interactions is not less than 20\%. This experiment indicated the importance of selecting \lq\lq white\rq\rq~stars together with neutrons to determine the characteristics of the cluster structure. It should be recognized that in the case of neutron-rich nuclei per nucleon electronic experiments in the energy range of a few tens of $~A~$GeV  with the detection of neutrons by hadron calorimeters have the best prospects.\par

\section*{FIRST EXPOSURES AT THE JINR NUCLOTRON}

\subsection*{Dissociation of the $^{10}$B nucleus}

\begin{figure}
    \includegraphics[width=5in]{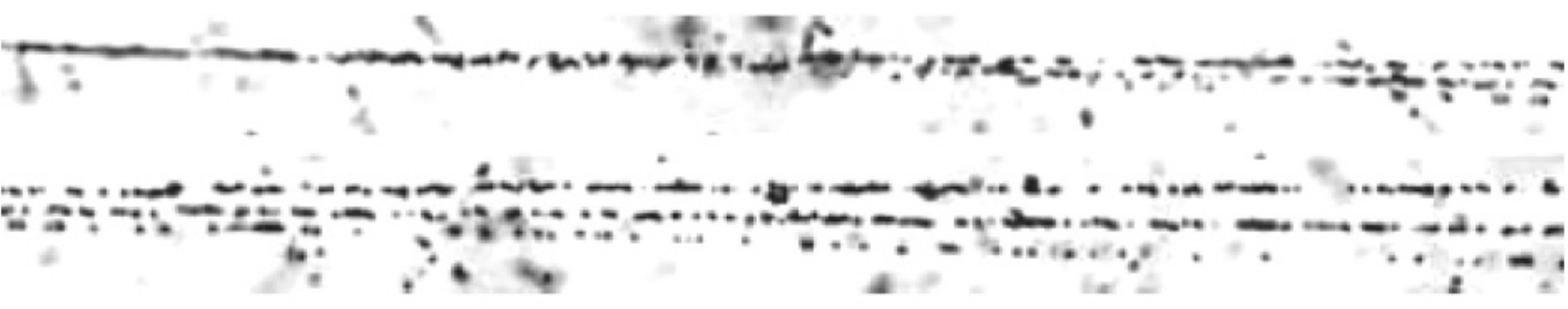}
    \caption{\label{Fig:16} Coherent dissociation $^{10}$B$~\rightarrow~$2He$~+~$H at 1.8$~A~$GeV/$c$.}
\end{figure}

\indent The $\alpha~+~d$ clustering   of the $^6$Li nucleus, which was demonstrated with remarkable detail \cite{Adamovich1}, led to an idea to identify a more complicated clustering $~-~$ $2\alpha~+~d$ $~-~$ in the next odd-odd nucleus$~-~^{10}$B \cite{Adamovich4}. The thresholds of separation of nucleons and lightest nuclei are close for this nucleus$~-~$E$_{th}(^6$Li$~+~\alpha)~=~4.5~$MeV, E$_{th}(^8$Be$~+~d)~=~6.0~$MeV, E$_{th}(^9$Be$~+~p)~=~6.6~$MeV. It was found that in approximately 65\% of peripheral interactions ($\sum$Z$_{fr}~=~5$, n$_s~=~0)$ of 1$~A~$GeV $^{10}$B nuclei occur via the 2He$~+~$H channel  (example in Fig.~\ref{Fig:16}). A singly charged particle in$~\approx$40\% of these events is the deuteron. The abundant deuteron yield is comparable with the $^6$Li case and points to the deuteron clustering in the $^{10}$B nucleus. Events in the He$~+~$3H channel  constitute 15\%. 10\% of the events contain both  Li and He fragments. The presence (or absence) of fragments of the target nucleus has practically no effect on the charge topology of the projectile nucleus fragmentation.\par

\indent Just 2\% of the events contain fragments with charges Z$_{fr}~=~4$ and 1, i.e. the $^9$Be nucleus and the proton. This \lq\lq negative\rq\rq~observation merits attention because it serves as a test of the relation of the shell and cluster description of the $^{10}$B nucleus. Indeed, the spin of this nucleus is equal to 3, which explains the $p$-shell filling order. Removal of a proton from the p-shell leads to the formation of the $^9$Be nucleus with spin 3/2. Thus, the separation of the proton does not require the transfer of the angular momentum. However, this channel is suppressed, which indirectly favors the leading role of the $2\alpha~+~d$ structure in the $^{10}$B ground state.\par

\indent A valuable finding of the exposure is an event of the coherent dissociation $^{10}$B$\rightarrow$3He. Associated with the rearrangement of  nucleons in $\alpha$ clusters, the process $^{10}$B$\rightarrow2^3$He$+^4$He could proceed via the charge-exchange reaction $^{10}$B$\rightarrow^6$Li$+^4$He$\rightarrow^3$H$+^3$He$+^4$He$\rightarrow$2$^3$He$+^4$He (E$_{th}~=~20~$MeV). By the charge composition this event is almost certainly identified as $^{10}$B$\rightarrow2^3$He$+^4$He, since the threshold of breakup of the second $\alpha$ cluster $^{10}$B$\rightarrow3^3$He$+n$ is even 16~MeV higher. The measurements of multiple scattering of the He tracks have confirmed this interpretation.\par

\subsection*{Dissociation of the $^{11}$B nucleus}

\indent The determining role of the $^3$H cluster in the  fragmentation of $^7$Li motivated a study of the triton cluster in the breakups of 2.75$~A~$GeV $^{11}$B nuclei \cite{Karabova}. The experiment was aimed at the channels with low thresholds of cluster separation$~-~$E$_{th}(^7$Li$~+~\alpha)~=~8.7~$MeV, E$_{th}(2\alpha~+~t)~=~11.2~$MeV and E$_{th}(^{10}$Be$~+~p)~=~11.2~$MeV. A leading channel, 2He + H, was also established for the $^{11}$B nucleus. Similarly to the case of $^{10}$B, a large proportion of tritons in the $^{11}$B \lq\lq white\rq\rq~stars favor its existence as a cluster. However, the increasing excess of neutrons that require (as in the case of $^7$Li) an increasing volume of measurements of multiple scattering leads to a decrease in the effectiveness of our approach.\par

\indent Eight \lq\lq white\rq\rq~stars of the charge-exchange reaction $^{11}$B$\rightarrow^{11}$C$^*\rightarrow^7$Be$~+~^4$He have been found. Charge exchange events through other channels were not observed. This fact demonstrates that while a three-body channel leads in $^{10}$B and $^{11}$B breakups, the  two-body leads in the $^{11}$C case. These observations motivate a direct study of $^{11}$C dissociation through the channels $^7$Be$~+~\alpha$~(E$_{th}~=~7.6~$MeV), $^{10}$B$~+~p~($E$_{th}~=~8.7~$MeV) and $^3$He$~+~2\alpha~($E$_{th}~=~9.2~$MeV).\par

\indent One should note the practical value of information about the $^{11}$C structure for nuclear medicine. In contrast to the $^{12}$C nucleus there should also be a significant contribution of the $^7$Be nucleus in the final states of $^{11}$C fragmentation. This circumstance leads to less \lq\lq spreading\rq\rq~of ionization from $^{11}$C fragmentation products.\par

\subsection*{Dissociation of the $^7$Be nucleus}

\begin{figure}
    \includegraphics[width=5in]{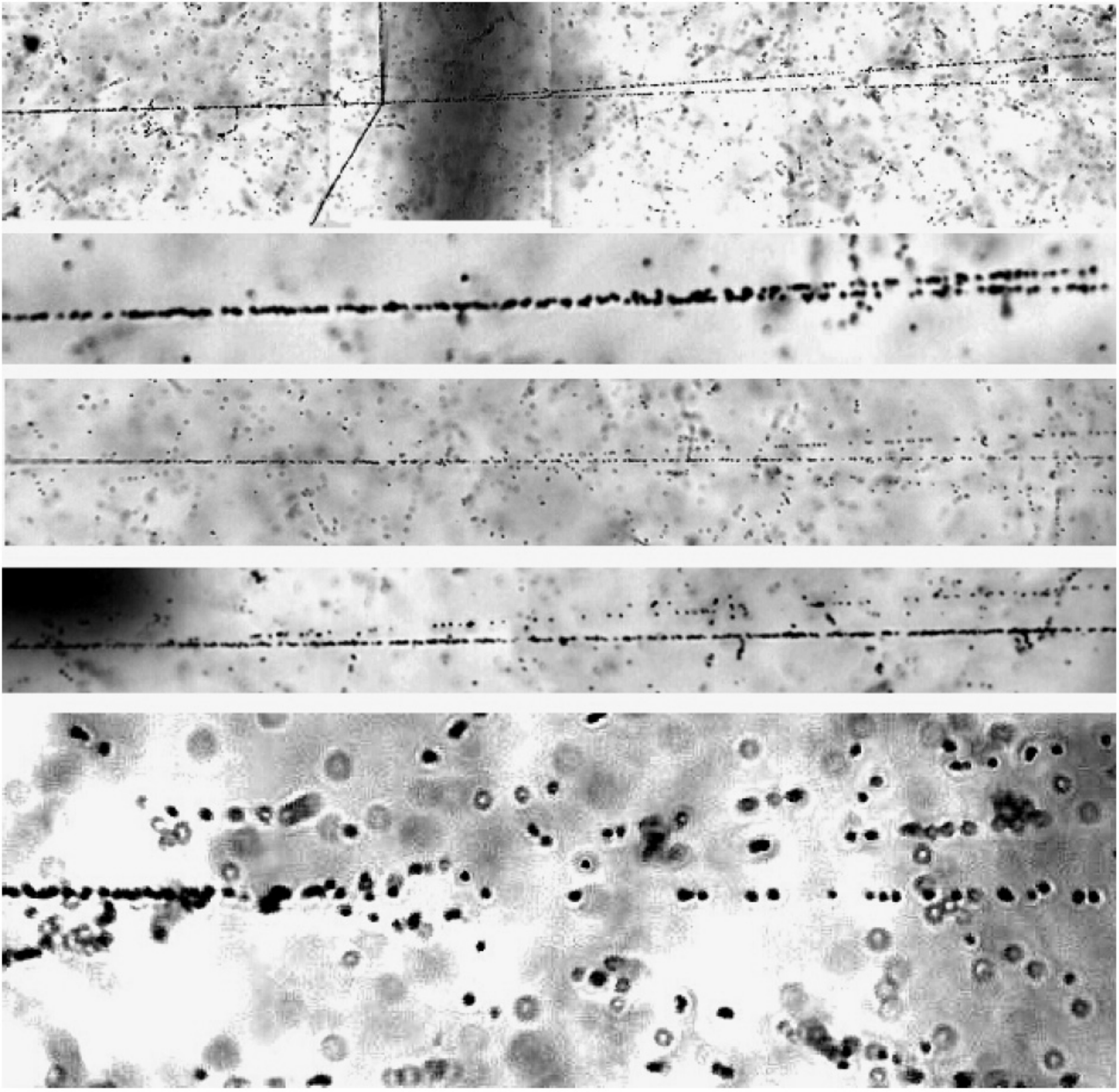}
    \caption{\label{Fig:17} Examples of events of the peripheral dissociation of $^7$Be nuclei at 2$~A~$GeV/$c$; top photo: splitting on two He fragments with production of a pair of target nucleus fragments; below: \lq\lq white\rq\rq~stars 2He, He$~+~$2H, Li$~+~$H and 4H.}
\end{figure}

\indent The next stage was peripheral interactions of the $^7$Be nuclei obtained in charge-exchange reactions of 1.2$~A~$GeV $^7$Li nuclei \cite{Peresadko,Pfp}. The numbers of events in various channels of $^7$Be fragmentation with the charge topology $\sum$Z$_{fr}~=~4$ are presented in Table~3 (examples in Fig.~\ref{Fig:17}). Statistics of 94 coherent N$_{ws}$ (n$_h~=~0$) and 55 non-coherent events N$_{tf}$ (n$_h~>~0$) is presented. Dissociation thresholds for the given channels E$_{th}$ are indicated (MeV).\par

\begin{table}
\caption{\label{Tabel:3} Distribution of $^7$Be interactions over identified fragmentation channels $\sum$Z$_{fr}~=~4$.}
\begin{tabular}{c|c|c|c|c|c|c|c|c|c|c}
\hline\noalign{\smallskip}
Channel&$^4$He+$^3$He&$^3$He+$^3$He&$^4$He+2$p$&$^4$He+$d$+$p$&$^3$He+$2p$&$^3$He+$d$+$p$&$^3$He+$2d$&$^3$He+$t$+$p$&$3p$+$d$&$^6$Li+$p$\\
E$_{th}$,MeV&(1.6)&(22.2)&(6.9)&(12.9)&(29.9)&(29.5)&(25.3)&(21.2)&(35.4)&(5.6)\\
\noalign{\smallskip}\hline\noalign{\smallskip}
N$_{ws}$ & 30 & 11 & 13 & 10 & 9 & 8 & 1 & 1 & 2 & 9\\
(\%)& (31) & (12) & (14) & (11) & (10) & (9) & (1) & (1) & (2) & (10)\\
\noalign{\smallskip}\hline\noalign{\smallskip}
N$_{tf}$ & 11 & 7 & 9 & 5 & 9 & 10 &   &   & 1 & 3\\
(\%)& (20) & (12) & (16) & (9) & (16) & (19) &   &   & (2) & (6)\\
\noalign{\smallskip}\hline
\end{tabular}
\end{table}

\indent  Approximately 50\% of the dissociation events occur without neutron emission, i.e., when $\sum$A$_{fr}~=~7$. In general, the coherent dissociation $\sum$Z$_{fr}~=~4$ and $\sum$A$_{fr}~=~7$ is determined by the configuration of $^4$He$~+~^3$He in the $^7$Be structure. The channels with a high threshold, in which there is no $^4$He cluster play a noticeable role. The statistics of the channels with He clusters shows a weak dependence on the values of dissociation thresholds. Apparently, the role of the $^3$He cluster in the $^7$Be nucleus goes beyond the $^4$He$~+~^3$He bond. Table~3 gives suggestions for the probabilities of possible configurations in the $^7$Be ground state including unobserved neutrons.\par

\section*{FRAGMENTATION OF THE $^9$Be NUCLEUS}

\indent The $^9$Be nucleus having the properties of the loosely bound system $2\alpha~+~n$ is the \lq\lq cornerstone\rq\rq~of cluster physics. Due to its low neutron separation threshold, dissociation of $^9$Be can be a source of unstable $^8$Be nuclei. The $^8$Be isotope is known as the only nucleus whose ground state is characterized as the $\alpha$-particle Bose condensate. Investigation of the $^9$Be nucleus fragmentation in $\alpha$-particle pairs seems to be an obvious starting point towards more complicated N$\alpha$-systems. However, there is a practical obstacle on the way of studying this stable nucleus. Beryllium is a toxic element which makes  immediate acceleration of $^9$Be nuclei impossible. Therefore, a secondary beam of relativistic $^9$Be nuclei was obtained in the fragmentation reaction $^{10}$B$~\rightarrow~^9$Be \cite{Artemenkov1,Artemenkov2,Pfp}. The share of $^9$Be nuclei was approximately 2/3, while 1/3 fell on He and Li isotopes.\par

\indent In the two-body model used for the calculation of the magnetic moment \cite{Parfenova1,Parfenova2} of the $^9$Be nucleus, the latter is represented as a bound state of the neutron and $^8$Be core in the 0$^+$ (g.s.) and 2$^+$ states with neutron separation thresholds being E$_{th}~=~1.67$ and 4.71~MeV. The weights of these states are 0.535 and 0.465. Therefore, in the $^9$Be dissociation it is possible to observe the $^8$Be 0$^+$ and 2$^+$ states with a similar intensity and in the simplest terms. In the $^8$Be nucleus there is a clear separation in energy E$_{ex}$ and width $\Gamma$ of the ground 0$^+$ (E$_{ex}~=~92~$keV, $\Gamma~=~5.6~$eV), the first 2$^+$ (E$_{ex}~=~3.1~$MeV, $\Gamma~=~1.5~$MeV) and second excited 4$^+$ (E$_{ex}~=~11.4~$MeV, $\Gamma~=~3.5~$MeV) states. Observation of these states can serve as a test of NTE spectroscopic capabilities. The  excitation structure of $^9$Be itself is much more complicated$~-~$there are 10 levels from the threshold to 12~MeV. There is uncertainty about the contribution of the + $^5$He state.\par

\indent An accelerated search for $^9$Be$~\rightarrow~$2He events was carried out \lq\lq along the strips\rq\rq. Focusing on a simple topology allowed bypassing the complicated problem of the identification of the secondary beam nuclei. As a result of scanning, 500 $\alpha$-particle pairs were found in the projectile fragmentation cone. Measurements of immersing angles and angles in the emulsion plane were performed for all $\alpha$-pair tracks which made it possible to determine the pair opening angles $\Theta$. A peculiarity of the resulting $\Theta$ distribution is the formation of two peaks. About 81\% of the events formed two roughly equal groups$~-~$\lq\lq narrow\rq\rq~$\alpha$-pairs in the interval $0<\Theta_{n(arrow)}<10.5~$mrad and \lq\lq wide\rq\rq~ones $~-~15.0<\Theta_{w(ide)}<45.0~$mrad. The remaining 19\% of the events are classified as \lq\lq intermediate\rq\rq~pairs $10.5<\Theta_{m(ediuum)}<15.0~$mrad and \lq\lq wider\rq\rq~pairs$~-~45.0<\Theta_{v(ery)w(ide)}<114.0~$mrad.\par

\begin{figure}
    \includegraphics[width=4in]{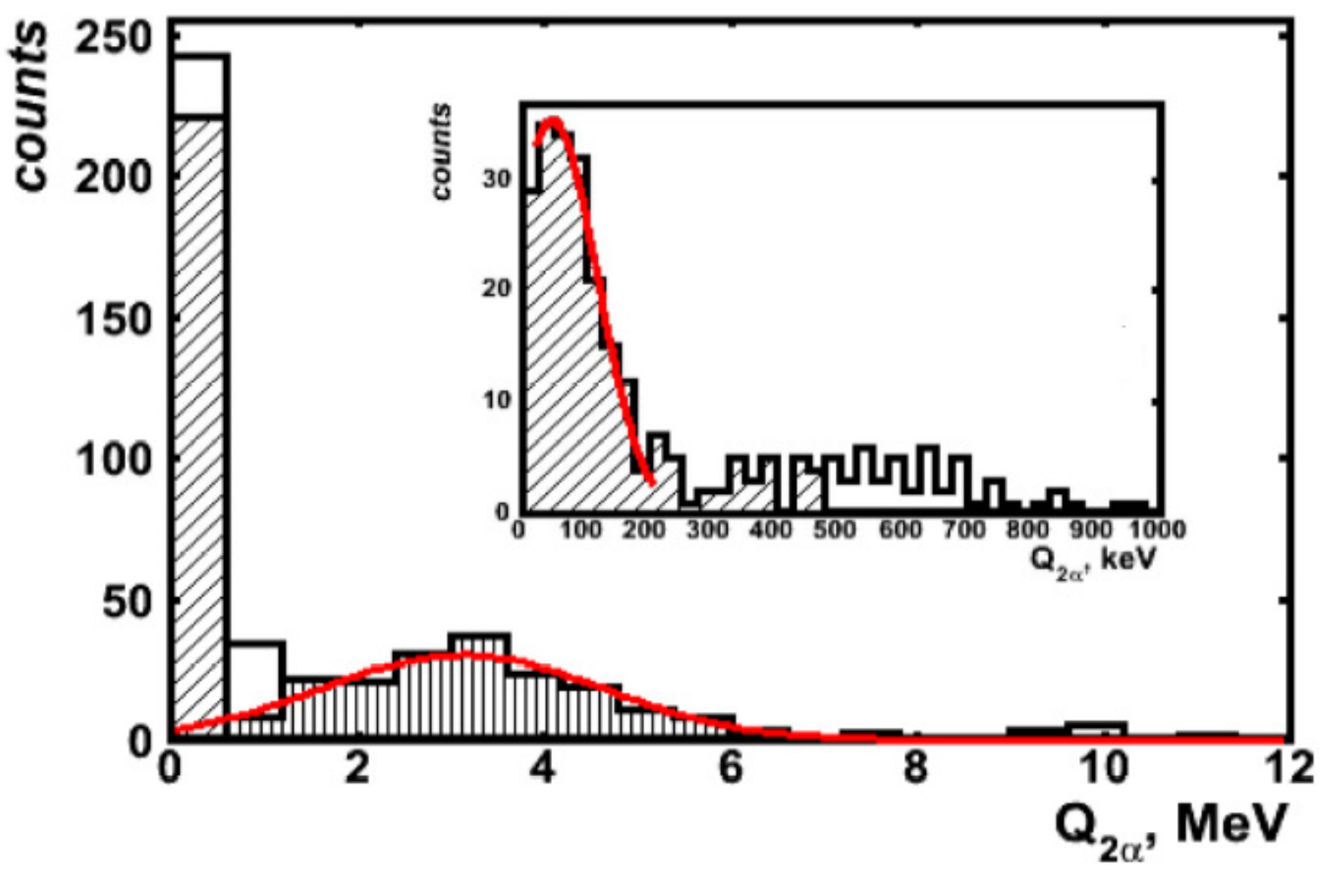}
    \caption{\label{Fig:18} Distribution of events of the peripheral fragmentation $^9$Be$~\rightarrow~2\alpha$ at 2$~A~$GeV/$c$ over energy Q$_{2\alpha}$; obliquely shaded histogram~-~events with opening angles $\Theta_n$; vertically shaded histogram~-~events with opening angles $\Theta_w$; solid line~-~total distribution of opening angles $\Theta$; on insertion~-~magnified distribution Q$_{2\alpha}$ for angular region $\Theta_n$.}
\end{figure}

\indent The physical meaning of this observation is explicitly manifested in the distribution of the $\alpha$-pair energy Q$_{2\alpha}$ $($Fig.~\ref{Fig:18}).~(75$\pm$10)\% of events with \lq\lq narrow\rq\rq~opening angles $\Theta_n$ are characterized by mean $<$Q$_{2\alpha}>=(86\pm4)~$keV with a standard deviation $\sigma($Q$_{2\alpha})~=~(48~\pm~2)~$keV. This value $<$Q$_{2\alpha}>$ corresponds to the $^8$Be$_{g.s.}$ 0$^+$ state decay. The value $\sigma($Q$_{2\alpha})$ can serve as an estimate of resolution. For events with \lq\lq wide\rq\rq~opening angles $\Theta_w$ the value $<$Q$_{2\alpha}>$ is equal to (3.1$~\pm~$0.11)~MeV with $\sigma($Q$_{2\alpha}$)=(1.30$\pm$0.08)~MeV. In this case $<$Q$_{2\alpha}>$ and $\sigma($Q$_{2\alpha}$) correspond to the $^8$Be 2$^+$ state. Events with \lq\lq intermediate\rq\rq~opening angles $\Theta_m$, may be associated with the formation of $^5$He, and $\Theta_{vw}~-~$with the decay of the $^8$Be 4$^+$ state. For events $\Theta_{vw}$ an important factor is the accuracy of the determination of energy and of identification of He isotopes. Thus, the formation of $\Theta_n$ pairs is matched to decays of the $^8$Be 0$^+$ ground state and $\Theta_w$ pairs$~-~$of the first excited 2$^+$ state. The shares of the events $\Theta_n$ and $\Theta_w$ constitute 0.56$\pm$0.04 and 0.44$\pm$0.04, respectively. These values demonstrate the compliance with the weights of the $^8$Be 0$^+$ and 2$^+$ states adopted in \cite{Parfenova1,Parfenova2} and point to the presence of these states as components of the $^9$Be ground state.\par

\indent In two important cases the events can be attributed to the target nucleus that participated in the interaction. First, these are \lq\lq white\rq\rq~stars (n$_b$~+~n$_g$~=~0) due to interactions on the heavy target nuclei Ag and Br. Second, these are events with single g-particles accompanying interactions with the H nuclei.
Approximately 80\% of the \lq\lq white\rq\rq~stars $^9$Be$~\rightarrow~2\alpha$ are characterized by the Rayleigh distribution parameter $\sigma_{AgBr}(P_{Tsum})~=~(77~\pm~7)~$MeV/$c$. This value is explainable within the framework of the statistical model for the fragment with mass number A~=~8 and the outer neutron in  $^9$Be. When the  radius of the $^9$Be nucleus is 2.5~Fermi the corresponding value of the dispersion of the neutron momentum distribution should be equal to $\sigma_0~=~81.4~$MeV/$c$. The remaining 20\% of the Ag-Br events are associated with a large angle scattering of \lq\lq narrow\rq\rq~$\alpha$-particle pairs ($^8$Be$_{g.s.}$) with $\sigma_{AgBr}(P_{Tsum})~=~(267~\pm~45)~$MeV/$c$. The $P_{Tsum}$ distribution of 88\% events for the H group is characterized by $\sigma_H(P_{Tsum})~=~(126~\pm~23)~$MeV/$c$. This value indicates that the breakup $^9$Be$~\rightarrow~2\alpha$ on protons corresponds to a harder interaction (less peripheral) than in the case of Ag and Br nuclei.\par

\indent Significant statistics of \lq\lq white\rq\rq~stars allow checking whether there is a correlation between the $\alpha$-pair momentum transfer $P_{Tsum}$ and the emergence of the $^8$Be nucleus in the ground and excited states. Samples from the intervals $\Theta_n$ and $\Theta_m~+~\Theta_w~+~\Theta_{vw}$ are described by the Rayleigh distribution with parameters $\sigma_{AgBr}(P_{Tsum})~=~(75~\pm~9)~$MeV/$c$ and (80$~\pm~$10)~MeV/$c$. Thus, there is no significant difference of the $P_{Tsum}$ distributions for coherent dissociation events via the 0$^+$ and 2$^+$  states of the $^8$Be nucleus. In general, the data can be viewed as evidence that the nuclear structure of $^9$Be has with high probability a core in the form of the two states of the $^8$Be nucleus and an outer neutron. These results are consistent with the descriptions of the $^9$Be nucleus structure suggesting the presence of the 0$^+$ and 2$^+$ states  of the $^8$Be nucleus with approximately equal weights.\par

\section*{PERIPHERAL INTERACTIONS OF $^{14}$N NUCLEI}

\indent The $^{14}$N nucleus is of interest as intermediate between the cluster  nucleus $^{12}$C and the doubly magic  nucleus $^{16}$O. The study of $^{14}$N nuclei can expand understanding of the evolution of increasingly complex structures beyond the $\alpha$-clustering. The information about the structure of $^{14}$N has an applied value. As a major component of the Earth's atmosphere the $^{14}$N nucleus can be a source of the light rare earth elements Li, Be and B, as well as of deuterium. Generation of these elements occurs as a result of bombardment of the atmosphere during its lifetime by high-energy cosmic particles. Therefore, the cluster features of the $^{14}$N fragmentation can determine the abundances of lighter isotopes. Beams of $^{14}$N nuclei can be used in radiation therapy, which also gives a practical interest in obtaining detailed data about the characteristics of the $^{14}$N fragmentation.\par

\indent For the first time the fragmentation of relativistic $^{14}$N nuclei was studied in NTE exposed at the Bevatron in the 70s \cite{Heckman}. Limitations in measurement of angles and fragment identification \cite{Heckman} motivated a study of the dissociation of 2.9$~A~$GeV/$c$ $^{14}$N nuclei in NTE exposed at the JINR Nuclotron \cite{Shchedrina}. The starting task was to reveal the role of external nucleon clustering in the form of a deuteron. This type of clusterization is expected for odd-odd light stable nuclei, whose number is small.\par

\indent Events were selected in which the total charge of the fragments $\sum$Z$_{fr}$ was equal to the projectile nucleus charge Z$_{pr}~=~7$ and there were no produced mesons (see Table~4). The main contribution is provided by the channels C$~+~$H, 3He$~+~$H, and 2He$~+~$3H (77\%). The share of events C$~+~$H (E$_{th}~=~7.6~$MeV) is sufficiently significant$~-~$25\%. The share of B$~+~$He  events (E$_{th}~=~20.7~$MeV) turned out to be small$~-~$only 8\%. A significant reduction in the proportion of deuterons relative to protons in comparison with  $^6$Li and $^{10}$B nuclei was demonstrated. A leading role both for \lq\lq white\rq\rq~stars and events with the formation of target fragments is taken by the multiple channel $^{14}$N$~\rightarrow~$3He$~+~$H (E$_{th}~=~15~$MeV) having a probability of about 35\%. Thus, the $^{14}$N nucleus manifests itself as an effective source of 3$\alpha$-systems. It was found that 80\% of the 3$\alpha$ ensembles correspond to the excitations of the $^{12}$C nucleus from the breakup threshold to 14~MeV. $^{14}$N produces fragments in the channel 3He$~+~$H via the  formation of $^8$Be with approximatelly 20\% probability. Events $^{11}$C$~+~^3$H, $^6$He$~+~^4$He$~+~^3$He$~+~p$, $^4$He$~+~2^3$He$~+~d$ have been identified; for these partial rearrangement of the $\alpha$-structure is necessary.\par

\begin{table}
\caption{\label{Tabel:4} Distribution of the peripheral interactions of $^{14}$N nuclei over the configurations $\sum$Z$_{fr}~=~7$ including \lq\lq white \rq\rq~stars N$_{ws}$ and events N$_{tf}$ with target fragments}
\begin{tabular}{c|c|c|c|c|c|c|c|c|c}
\hline\noalign{\smallskip}
Channel&C+H&B+He&B+2H&Be+He+H&Li+4H&Li+He+2H&2He+3H&3He+H&He+5H\\
\noalign{\smallskip}\hline\noalign{\smallskip}
N$_{ws}$& 16 & 5 & 5 & 2 & 1 &   & 6 & 21 & 5 \\
N$_{tf}$& 24 & 4 & 3 & 5 & 2 & 3 & 21 & 35 & 3 \\
\noalign{\smallskip}\hline
\end{tabular}
\end{table}

\section*{COHERENT DISSOCIATION OF $^8$B NUCLEI}

\begin{figure}
    \includegraphics[width=5in]{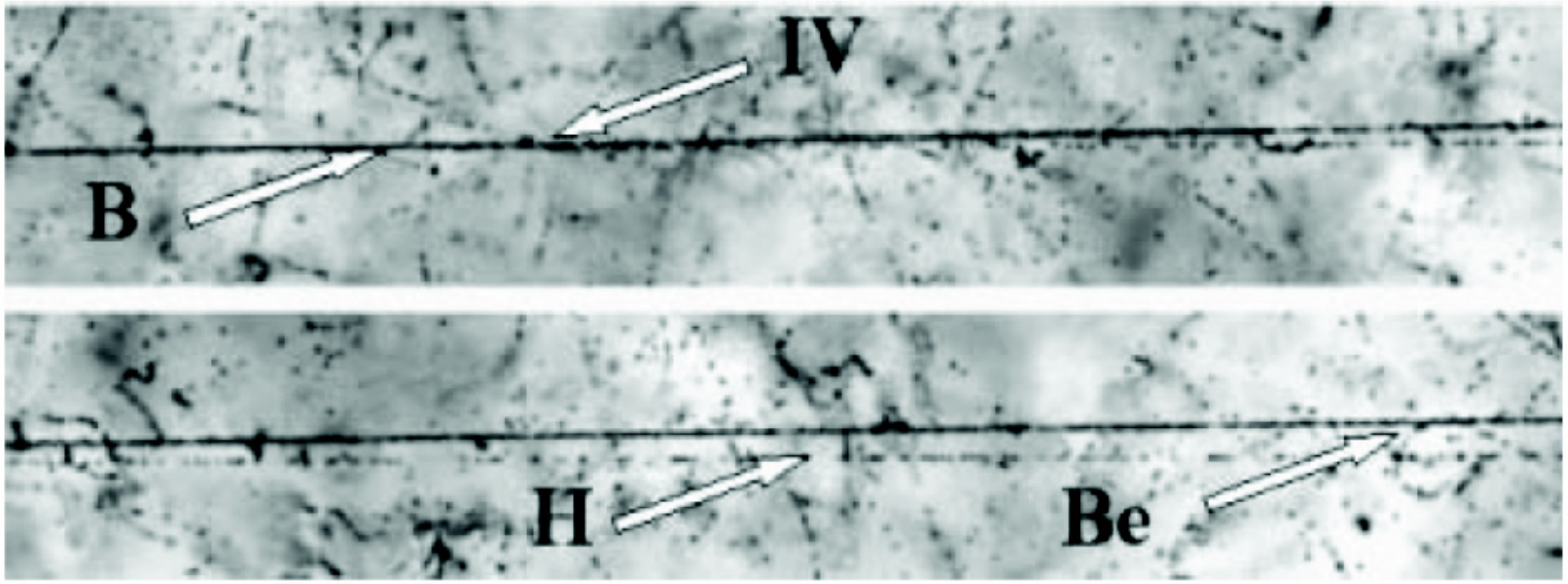}
    \caption{\label{Fig:19} Coherent dissociation $^8$B$~\rightarrow^7$Be$~+~p$ at 2$~A~$GeV/$c$ (IV is interaction vertex).}
\end{figure}

\begin{figure}
    \includegraphics[width=5in]{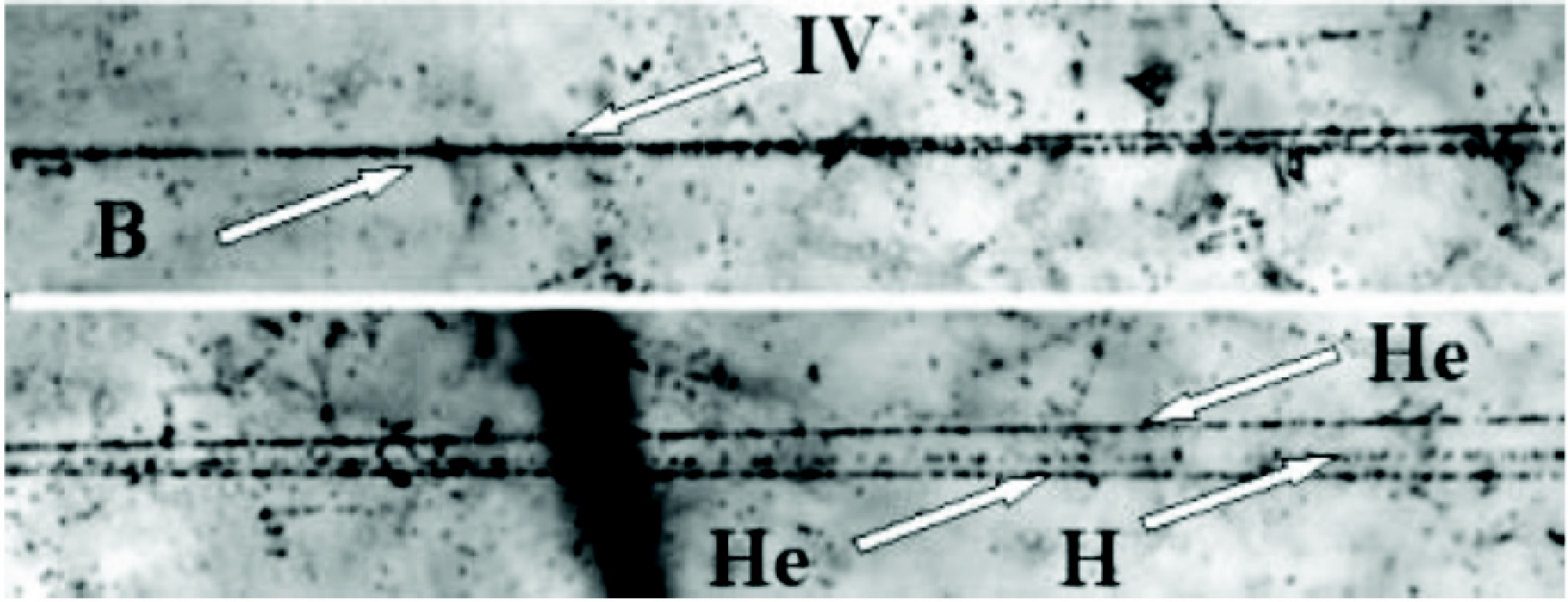}
    \caption{\label{Fig:20} Coherent dissociation $^8$B$~\rightarrow~$2He$~+~$H at 2$~A~$GeV/$c$.}
\end{figure}

\indent $^8$B fragments produced by 1.2$~A~$GeV $^{10}$B nuclei were selected for exposure of NTE \cite{Stanoeva,Pfp}. The charge composition of the relativistic fragments for the events $\sum$Z$_{fr}~=~5$ accompanied by target nucleus fragments and (or) produced mesons N$_{tf}$ and \lq\lq white\rq\rq~stars N$_{ws}$ (examples in Fig.~\ref{Fig:19} and~\ref{Fig:20}) show a qualitative difference (Table~5). The main conclusion is that the contribution of the dissociation $^8$B$~\rightarrow~^7$Be$~+~p$ is leading among \lq\lq white\rq\rq~stars. This situation is qualitatively different from the  dissociation of the $^{10}$B isotope. Data on N$_{ws}$ may be useful as estimates of the probabilities of few body configurations in the $^8$B ground state.\par

\begin{table}
\caption{\label{Tabel:5} Distribution of the peripheral interactions of $^8$B nuclei over the configurations $\sum$Z$_{fr}~=~5$ }
\begin{tabular}{c|c|c|c|c}
\hline\noalign{\smallskip}
~Channel~&~B~&~Be+H~&~2He+H~&~He+3H~\\
~E$_{th}$, MeV~&~~~&~(0.138)~&~(1.72)~&~(6.9)~\\
\noalign{\smallskip}\hline\noalign{\smallskip}
~N$_{ws}$ (\%)~& 1 (2) &~25 (48)~&~14 (27)~&~12 (23)~\\
~N$_{tf}$ (\%)~&~11 (19)~&~8 (14)~&~17 (29)~&~22 (38)~ \\
\noalign{\smallskip}\hline
\end{tabular}
\end{table}

\indent Due to the record low binding energy of the external proton (E$_{th}~=~138~$keV), the $^8$B nucleus  is the most sensitive probe of the electromagnetic interaction with the target nucleus. In the center of mass of the system $^7$Be$~+~p$ the average transverse momenta of the particles is $<P_T^*>~=~(62~\pm~11)~$MeV/$c$ at RMS of 54~MeV/$c$. This small value indicates a weak bond of the proton and the core. The distribution of the total transverse momenta of the pairs in the \lq\lq white\rq\rq~stars has an average value of $<P_T(^8$B$^*)>~=~(95~\pm~15)~$MeV/$c$ at RMS of 73~MeV/$c$, and a significantly greater one for events with target nucleus fragments or produced mesons $<P_T(^8$B$^*)>~=~(251~\pm~29)~$MeV/$c$ at RMS of 112~MeV/$c$.\par

\indent Analysis of angular correlations allowed establishing the criteria of the electromagnetic dissociation events by the total transverse momentum $P_T(^8$B$^*)~<~150~$MeV/$c$, energy Q$_{pBe}~<~5~$MeV and  by the azimuth angle $\varepsilon_{pBe}~>~\pi/2$ between the fragments. Because of  Z$^2$ dependence of the electromagnetic cross-section on a nucleus target charge species, the proportional contribution can be assumed from  Ag and Br nuclei. Then the obtained cross-sections comprise $\sigma_{Ag}~=~(81~\pm~21)~$mb and $\sigma_{Br}~=~(44~\pm~12)~$mb. Analysis of the ratio of the Coulomb and nuclear dissociation and stripping in the dissociation of $^8$B$~\rightarrow~^7$Be$~+~p$ for the Pb target up to the energy of $\approx$2$~A~$GeV was carried out in \cite{esbensen}. Extrapolation $\sigma_{Ag}$ to the Pb nucleus leads to the value $\sigma_{Pb}~=~(230~\pm~60)~$mb, which is close to the theoretical value of $\approx$210~mb \cite{esbensen}.\par

\section*{COHERENT DISSOCIATION OF $^9$C NUCLEI}

\indent The $^9$C nucleus became the next studied object on the proton border of nuclear stability. The coherent dissociation of $^9$C can proceed through the channels $^8$B$~+~p$ (E$_{th}~=~1.3~$MeV) and $^7$Be$~+~2p$ (E$_{th}~=~1.4~$MeV) as well as the $^7$Be core breakups (E$_{th}~>~3~$MeV). Besides,  the population of the 3$^3$He system, which has a relatively low formation threshold (about 16~MeV), is possible by means of neutron rearrangement from the $^4$He cluster to  $^3$He cluster being formed. Probability of the transition $^9$C$~\rightarrow~$3$^3$He can point to the 3$^3$He component weight in the $^9$C ground state and may be important in calculating the characteristics of the $^9$C nucleus based on the cluster wave functions taking into account such a deeply bound state. Being a non-trivial cluster excitation, the 3$^3$He state may be important for the development of nuclear astrophysics scenario with one more initial state of the fusion reaction similar to the 3$\alpha$-process. An intriguing problem is to find a resonant 2$^3$He state in the $^9$C$~\rightarrow~3^3$He dissociation similar to the dissociation $^{12}$C$~\rightarrow~^4$He$^8$Be.\par

\indent In the study of $^9$C interactions there is a need to overcome two practical problems. First, the $^3$He nuclei, having the same ratio of the charge Z$_{pr}$ to the mass number A$_{pr}$, are dominant in the generated beam. Thus, it was important to avoid NTE overexposure to $^3$He nuclei. Second, it was necessary to ensure the $^9$C dominance over the contributions of the $^{10,11}$C isotopes. A comparative analysis of the coherent dissociation of the studied neighboring isotopes helped this problem to be solved.\par

\indent $^{12}$C$^{6+}$ ions, created by a laser source, were accelerated to 1.2$~A~$GeV and extracted to the production target. Further, the secondary beam tuned for selection of $^9$C nuclei was guided on the emulsion stack \cite{Krivenkov,Pfp}. With dominance of C nuclei, the beam contained an insignificant admixture of $^6$Li, $^7$Be and $^8$B.\par

\indent The main branch of the coherent dissociation is represented by events $\sum$Z$_{fr}~=~6$, which is to be expected due to the dominance of C nuclei in the beam. The most valuable is the analysis of the channels corresponding to the $^9$C nucleus dissociation with the lowest thresholds $^8$B$~+~p$ and $^7$Be$~+~2p$, as well as the 3He channel. The events in the last channel could be eligible for the coherent dissociation $^9$C$~\rightarrow~3^3$He. The events Z$_{pr}~=~6$ and Z$_{fr}~=~5$ and 4 are interpreted as $^9$C$~\rightarrow~^8$B$~+~p$ and $^7$Be$~+~2p$. The events 2He$~+~$2H and He$~+~$4H are dominant (Table~6). In the case of $^9$C, events in these channels occur with approximately equal probability as expected due to the dissociation of the $^7$Be core \cite{Peresadko}. This ratio does not correspond to the isotope $^{10}$C, for which the probability of the 2He$~+~$2H channel is approximately by an order of magnitude higher than for the He$~+~$4H channel \cite{Kattabekov1,Mamatkulov}. Besides, \lq\lq white\rq\rq~stars $^6$Li$~+~3p$ and 6H  produced as a result the dissociation of the $^7$Be core were observed .\par

\begin{table}
\caption{\label{Tabel:6} Distribution of \lq\lq white\rq\rq~stars N$_{ws}$ of $^9$C nuclei over the configurations $\sum$Z$_{fr}~=~6$}
\begin{tabular}{c|c|c|c|c|c|c }
\hline\noalign{\smallskip}
 Channel&~B+H~&~Be+2H~&~3He~&~2He+2H~&~He+4H~&~6H~\\
\noalign{\smallskip}\hline\noalign{\smallskip}
N$_{ws}$& 15 & 16 & 16 & 24 & 28 & 6 \\
\noalign{\smallskip}\hline
\end{tabular}
\end{table}

\begin{figure}
    \includegraphics[width=4in]{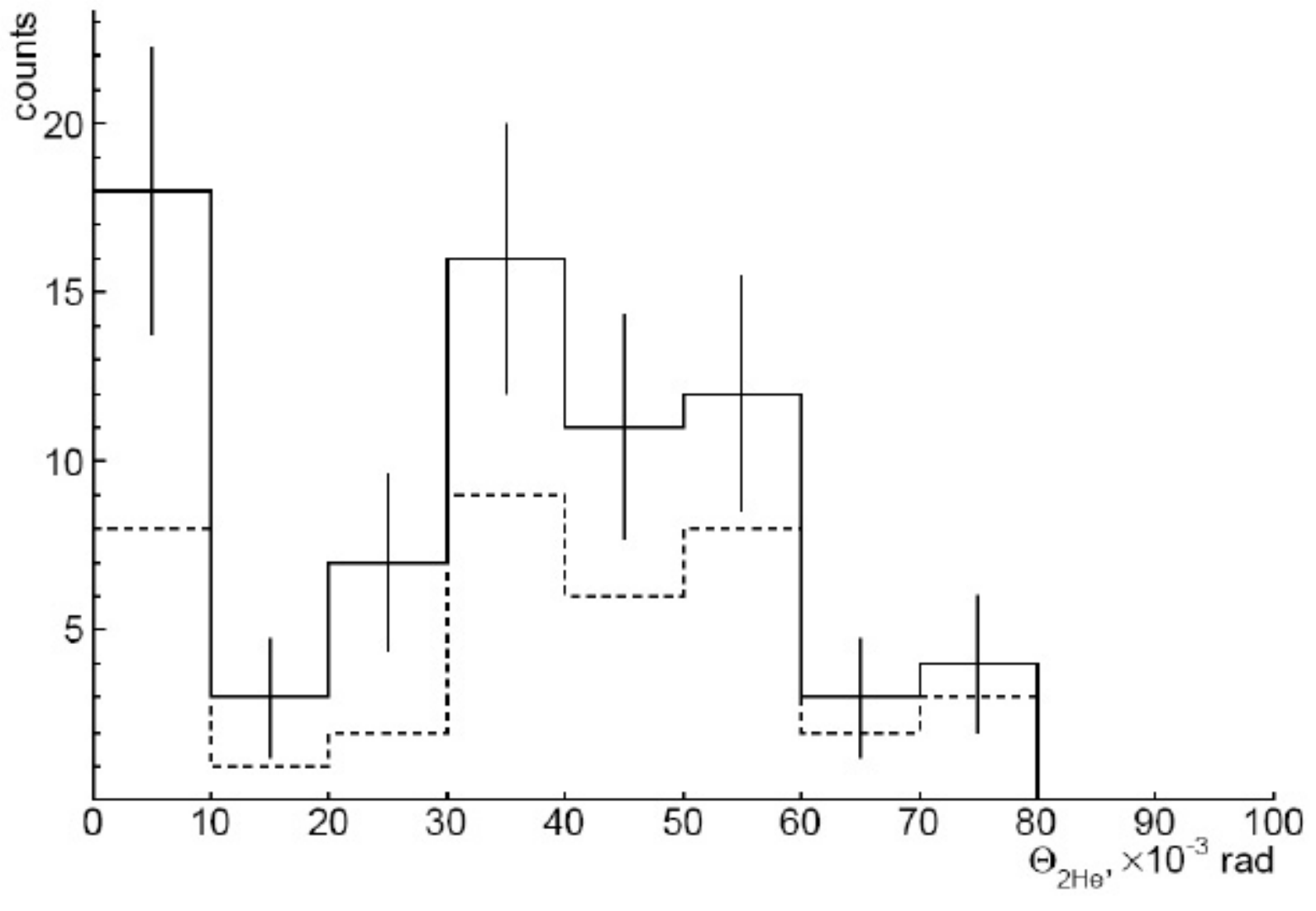}
    \caption{\label{Fig:21} Total distribution of opening angles $\Theta_{2He}$ between the relativistic He fragments in the \lq\lq white\rq\rq~stars $^9$C$~\rightarrow~3^3$He and in the events $^8$B$~\rightarrow~2$He$~+~$H with the formation of target nucleus fragments or mesons; dotted line indicates the contribution of \lq\lq white\rq\rq~stars $^9$C$~\rightarrow~3^3$He.}
\end{figure}

\indent The 3$^3$He states are the central subject of the current study. The dissociation probability via this channel ($\approx$14\%) is comparable to the nucleon separation channels. The significant probability of the coherent dissociation channel $^9$C$~\rightarrow~3^3$He makes it an effective source for the search for a resonant 2$^3$He state near the threshold analogous to the $^8$Be ground  state.  The opening angle distribution $\Theta_{2He}$ of the fragment pairs in the \lq\lq white\rq\rq~stars $^9$C$~\rightarrow~3^3$He is shown in Fig.~\ref{Fig:21}. The main part corresponding to 30 pairs of 2He is described by a Gaussian distribution with parameters $<\Theta_{2He}>~=~(46~\pm~3)\times10^{-3}~$rad at RMS of 16$\times$10$^{-3}$ rad. The corresponding energy distribution is limited to the region Q(2$^3$He)$~<~20~$MeV.\par

\indent Eight narrow 2He pairs with opening angles limited to $\Theta_{2He}~<~10^{-2}$ rad are reliably observed thanks to the NTE resolution. They are allocated in a special group with an average of $<\Theta(2^3$He$)>~=~(6~\pm~1)\times10^{-3}~$rad at RMS of 3$\times10^{-3}~$rad. The energy distribution has a mean value $<$Q$(2^3$He$)>~=~(142~\pm~35)~$keV at RMS of 100~keV. Thus, despite the low statistics, this distribution points to an intriguing possibility of the existence of a resonant 2$^3$He state slightly above the mass  threshold of 2$^3$He \cite{Artemenkov5}.\par

\indent To test a possible 2$^3$He resonance (conventionally called \lq\lq dihelion\rq\rq), an analysis of data on the $^8$B nucleus \cite{Stanoeva} was carried out. Events $^8$B$~\rightarrow~2$He$~+~$H accompanied by target nucleus fragments or mesons were selected in order to enhance the effect. This condition provides an effective selection of interactions with neutron knocking out from the $^4$He cluster in the $^8$B nucleus. Thus, the distribution $\Theta(2^3$He) takes the same view as in Fig.~\ref{Fig:4} and also includes a separate group of narrow pairs with $<\Theta(2^3$He$)>=(4.5\pm0.5)\times10^{-3}~$rad (RMS~1.5$\times10^{-3}~$rad), corresponding to the case of the \lq\lq white\rq\rq~stars $^9$C$~\rightarrow~3^3$He. The total distribution of the opening angles $\Theta_{2He}$ between the relativistic He fragments in the \lq\lq white\rq\rq~stars $^9$C$~\rightarrow~3^3$He and in the  events $^8$B$~\rightarrow~2$He$~+~$H with the formation of target nucleus fragments or mesons shown in Fig.~\ref{Fig:21} enhances evidence for the existence of a near-threshold 2$^3$He resonance. Moreover, the question arises about the nature of a broad peak with maximum near $\Theta(2^3$He$)~\approx(40~-~50)\times10^{-3}~$rad. Possibly in this $\Theta$ region  the decays 2$^3$He are similar to the decay of the $^8$Be 2$^+$ state  \cite{Artemenkov3,Artemenkov4}.\par

\indent Of course, this finding is worth studying and testing with much higher statistics. One of its more technically simple options may be the dissociation of $^7$Be$~\rightarrow~2^3$He with a neutron knock out and the formation of fragments of target nuclei or mesons. However, it is possible that the \lq\lq dihelion\rq\rq~formation is associated with the presence of a 2$^3$He component in the $^9$C and $^8$B structures. In the $^7$Be nucleus such a component can be suppressed, which means the suppression of \lq\lq dihelion\rq\rq~formation in the fragmentation of this nucleus. Therefore it is important to implement a search for the 2$^3$He resonance with larger statistics using fragmentation of low-energy $^9$C and $^8$B nuclei. Pointing to the existence of\lq\lq dihelion\rq\rq~, this observation motivates the search for a mirror state of the $^3$H pair$~-~$\lq\lq ditriton\rq\rq.\par

\section*{COHERENT DISSOCIATION OF $^{10}$C AND $^{12}$N NUCLEI}

\subsection*{Exposure to a mixed beam of $^{12}$N, $^{10}$C and $^7$Be nuclei}

\indent A secondary beam containing $^{12}$N, $^{10}$C and $^7$Be nuclei can be formed by selection of products of charge-exchange and fragmentation reactions of relativistic $^{12}$C nuclei. Such a composition is not so much desirable but unavoidable since the Z$_{pr}$/A$_{pr}$ ratios of these nuclei differ by only 3\%. Separation of these nuclei is not possible in a channel with the momentum acceptance of 2-3\% , and they are simultaneously present in the beam, forming the so-called \lq\lq beam cocktail\rq\rq. The contribution of $^{12}$N nuclei is small relative to $^{10}$C and $^7$Be nuclei in accordance with the charge-exchange and fragmentation cross-sections. Because of the momentum spread, $^3$He nuclei can penetrate into the channel. For the neighboring nuclei $^8$B, $^9$C and $^{11}$C the difference of Z$_{pr}$/A$_{pr}$ from $^{12}$N is about 10\%, which causes their suppression in the secondary beam. An event-by-event identification of $^{12}$N in the exposed NTE is possible for \lq\lq white\rq\rq~stars by fragment topologies and beam nucleus charges determined by $\delta$-electron counting on the beam tracks. In the case of dominant $^{10}$C nuclei it is sufficient to make sure that the contribution of the neighboring C isotopes by the overall pattern of the  composition of \lq\lq white\rq\rq~stars is small.\par

\indent Based on these considerations it was suggested to expose NTE to a mixed beam of 2$~A~$GeV/$c$ $^{12}$N, $^{10}$C and $^7$Be nuclei \cite{Kattabekov1,Pfp}. The amplitude spectrum from a scintillation counter installed in the location of NTE irradiation pointed to the dominance of He, Be, C isotopes and to a small admixture of N nuclei in the substantial absence of $^8$B nuclei. A stack of 15 NTE layers was exposed to a secondary beam with such a composition. The initial stage of analysis was to search for beam tracks with charges Z$_{pr}~=~1,~2$ and Z$_{pr}~>~2$. The ratio of beam tracks Z$_{pr}~=~1,~2$ and Z$_{pr}~>~2$ was$~\approx$1:3:18. Thus, the contribution of $^3$He nuclei  decreased dramatically  in this exposure as compared with the $^9$C case.\par

\indent The analysis presented below is based on the search for events along the tracks of primary particles with charges visually valued as Z$_{pr}~>~2$ over a length of about 1088 m. As a result, 7241 inelastic interactions were found, including 608 \lq\lq white\rq\rq~stars containing only relativistic particle tracks in the  angular cone $\theta_{fr}~<~11^{\circ}$. In the \lq\lq white\rq\rq~stars, which might be created by $^{12}$N nuclei, the average densities of $\delta$-electrons N$_{\delta}$ were measured on the tracks of the beam nuclei and secondary fragments with charges Z$_{fr}~>~2$. As was shown in the study of the nuclei $^8$B \cite{Stanoeva} and $^9$C \cite{Krivenkov}, the application of this method allows one to eliminate the contribution from the charge-exchange reactions with production of mesons of accompanying lighter nuclei. The dominance of C nuclei in this irradiation has made such selection particularly relevant and has justified the use of a cumbersome procedure of $\delta$-electron counting.\par

\subsection*{Dissociation of $^{10}$C nuclei}

\indent The $^{10}$C nucleus is the only example of a stable 4-body structure in which the removal of any of the constituent clusters or nucleons leads to an unbound state condition. The breakup threshold of the $^{10}$C$~\rightarrow~2\alpha~+~2p$ process is E$_{th}~=~3.73~$MeV. The next threshold via $^8$Be$_{g.s.}~+~2p$ is slightly higher$~-~$E$_{th}~=~3.82~$MeV. Knocking out one of the protons (E$_{th}~=~4.01~$MeV) leads to the formation of an unstable $^9$B nucleus, which decays into a proton and a $^8$Be nucleus. By way of $\alpha$-cluster separation (E$_{th}~=~5.10~$MeV) a $^6$Be resonance can be formed, its decay energy being 1.37~MeV. The decay of $^6$Be via the $^5$Li resonance is impossible, because the threshold for the formation of $^5$Li$_{g.s.}~+~p$ is 0.35~MeV higher than the $^6$Be ground state. In addition, the channel $^5$Li$_{g.s.}~+~\alpha$ is closed since this threshold is 1.5~MeV higher than the $^9$B ground state. Therefore, in the $^{10}$C dissociation the resonances $^6$Be$_{g.s.}$ and $^5$Li$_{g.s.}$ can only be produced directly and not in  cascade decays of $^9$B.\par

\begin{figure}
    \includegraphics[width=5in]{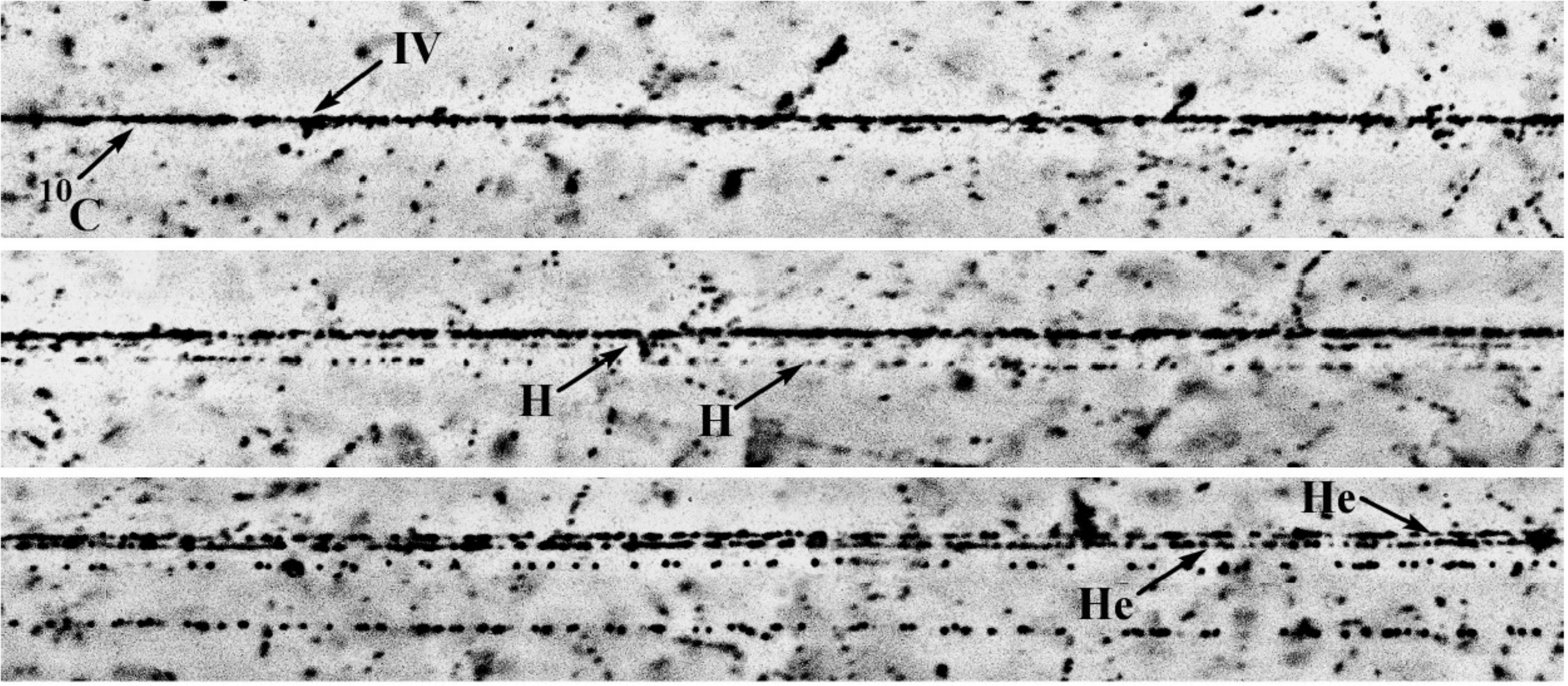}
    \caption{\label{Fig:22} Coherent dissociation $^{10}$C$~\rightarrow~p~+~^9$B$_{g.s.}$ at 2$~A~$GeV/$c$.}
\end{figure}

\indent Events $\sum$Z$_{fr}~=~6$ were selected among the found peripheral interactions \cite{Kattabekov1,Mamatkulov}. Their distribution on the charge topology is presented in Table~7. The subject of the analysis was a sample consisting of 227 \lq\lq white\rq\rq~stars N$_{ws}$. A peculiarity of this class of events is the dominance of the channel 2He$~+~$2H, which is indeed the most expected one for the $^{10}$C isotope. The channels N$_{ws}$ requiring destruction of $\alpha$-clustering in $^{10}$C nuclei and having substantially higher thresholds are manifested with much lower probabilities. The macro photography of a typical event is shown in Fig.~\ref{Fig:22}. The interaction vertex in which a group of fragments formed is marked in the top photo. Further, one can distinguish two H (middle photo) and two He fragments (bottom photo). The most remote track originated in the dissociation $^{10}$C$~\rightarrow~^9$B$_{g.s.}~+~p$. The other tracks correspond to the decay of the unbound $^9$B nucleus. The pair of the He tracks corresponds to the following decay of another unbound $^8$Be nucleus.\par

\indent Comparison of the N$_{ws}$ topology distribution with the version for the 627 $^{10}$C N$_{tf}$ events accompanied by the production of mesons,  fragments of target nuclei or recoil protons, points to the \lq\lq turning on\rq\rq~of the He$~+~$4H channel in the latter case (Table~7). First of all, a much smaller perturbation of the $^{10}$C cluster structure in the \lq\lq white\rq\rq~stars with the respect to the N$_{tf}$ case is confirmed. In addition, the comparison shows that the probabilities of the fragmentation channels beyond the \lq\lq pure\rq\rq~clustering $2\alpha-2p$ do not differ too much in the cases N$_{ws}$ and N$_{tf}$ (Table~7). This fact indicates the existence in the $^{10}$C structure of a small admixture of virtual states with participation of deeply bound cluster-nucleon configurations.\par

\begin{table}
\caption{\label{Tabel:7} Distribution over the charge configurations of relativistic fragments $\sum$Z$_{fr}~=~6$ of $^{10}$C fragmentation events for \lq\lq white\rq\rq~stars N$_{ws}$ and collisions with produced mesons, target fragments or recoil protons N$_{tf}$}
\begin{tabular}{c|c|c|c|c|c|c|c|c}
\hline\noalign{\smallskip}
 &~2He+2H~&~He+4H~&~3He~&~6H~&~Be+He~&~B+H~&~Li+3H~&~C+n~\\
\noalign{\smallskip}\hline\noalign{\smallskip}
N$_{ws}$ & 186 & 12 & 12 & 9 & 6 & 1 & 1 &  \\
(\%) & (81.9) & (5.3) & (5.3) & (4.0) & (2.6) & (0.4) & (0.4) &  \\
\noalign{\smallskip}\hline\noalign{\smallskip}
N$_{tf}$ & 361 & 160 & 15 & 30 & 17 & 12 & 2 & 30 \\
(\%) & (57.6) & (25.5) & (2.4) & (4.8) & (2.7) & (1.9) & (0.3) & (4.8) \\
\noalign{\smallskip}\hline
\end{tabular}
\end{table}

\indent Angular measurements were carried out for the tracks of the \lq\lq white\rq\rq~stars 2He$~+~$2H. The Rayleigh distribution parameters which describe the statistics of thel angles of fragment emission are equal to $\sigma_{\theta_H}~=~(51~\pm~3)\times10^{-3}~$rad and $\sigma_{\theta_{He}}~=~(17~\pm~1)\times10^{-3}~$rad. These values are consistent with those of the statistical model \cite{Feshbach,Goldhaber} $\sigma_{\theta_p}~\approx47\times10^{-3}~$rad and $\sigma_{\theta_{\alpha}}~\approx19\times10^{-3}$ rad for $^1$H and $^4$He fragments. Measurements of the angles allow  the transverse momenta of the fragments and their ensembles to be estimated. The distribution of the \lq\lq white\rq\rq~stars 2He$~+~$2H for the full transverse momentum $P_T$ is described by the Rayleigh distribution with parameter $\sigma_{P_T}(2\alpha~+~2p)~=~(161~\pm~13)~$MeV/$c$. Such a value is expected for the diffraction dissociation \cite{Peresadko1}.\par

\begin{figure}
    \includegraphics[width=5in]{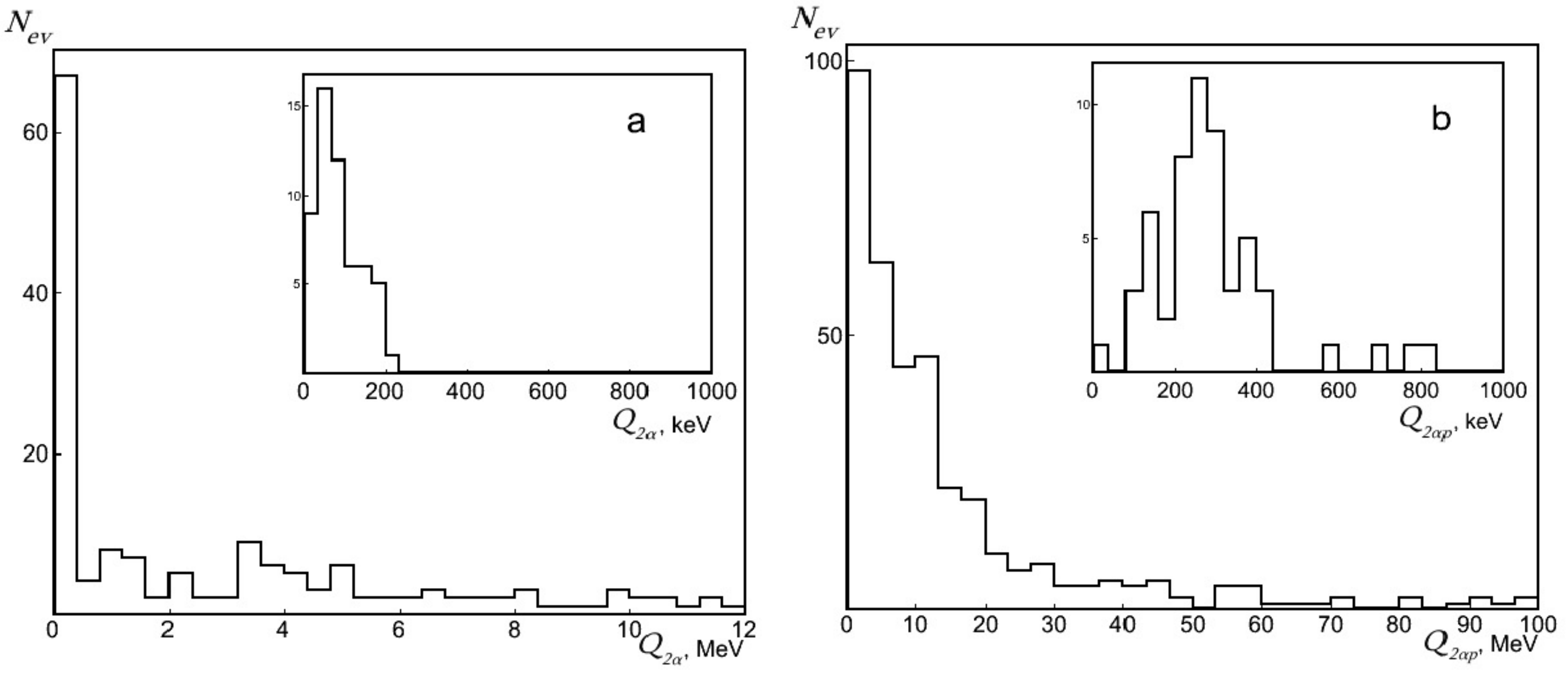}
    \caption{\label{Fig:23} Distributions of the \lq\lq white\rq\rq~stars $^{10}$C$~\rightarrow~2\alpha~+~2p$ over energy Q$_{2\alpha}$ of pairs 2$\alpha$ ($a$) and over Q$_{2\alpha p}$ of triples $2\alpha~+~p$ ($b$); on insertions$~-~$magnified distributions Q$_{2\alpha}$ and Q$_{2\alpha p}$.}
\end{figure}

\indent The distribution of these events over the energy Q$_{2\alpha}$ of the 2$\alpha$ pairs and Q$_{2\alpha p}$ of the 2$\alpha~+~p$ triples is shown in Fig.~\ref{Fig:23}. In 68 of them 2$\alpha$ pairs with emission angles not exceeding 10$^{-2}~$rad are observed. The distribution Q$_{2\alpha}$ of these 2$\alpha$ pairs with an average $<$Q$_{2\alpha}>~=~(63~\pm~30)~$keV at RMS of 83~keV allows concluding that the formation of $^8$Be$_{g.s.}$ is observed in these events. In turn, the distribution Q$_{2\alpha p}$ indicates that the dissociation $^{10}$C$~\rightarrow~2\alpha~+~2p$ is accompanied by the formation of unbound $^9$B nuclei. The average value $<Q_{2\alpha p}>~=~(254~\pm~18)~$keV at RMS of 96~keV corresponds to the energy and width of the decay $^9$B$_{g.s.}~\rightarrow~^8$Be$_{g.s.}~+~p$. A clear correlation between Q$_{2\alpha}$ and Q$_{2\alpha p}$ points to the cascade process $^{10}$C$~\rightarrow~^9$B$~\rightarrow~^8$Be. The contribution of these decays allows concluding that the $^9$B nucleus manifests itself with a probability of (30$~\pm~$4)\% in the $^{10}$C structure. Earlier, the $^9$B nuclei from the fragmentation of $^{12}$C were reconstructed in an experiment with transverse orientation of NTE pellicles \cite{Toshito}.\par

\indent For 40 events $^{10}$C$~\rightarrow~^9$B (73\%) the Rayleigh parameter $\sigma_{P_T}(^9$B) of the distribution over the total transverse momentum $P_{T2\alpha p}$ of the 2$\alpha~+~p$ triples is (92$~\pm~$15)~MeV/$c$. It corresponds to the value of 93~MeV/$c$ expected in the statistical model. Within this model the radius of the region emission of an outer proton by the $^{10}$C nucleus is R$_p~=~(2.3~\pm~0.4)~$Fermi which does not contradict to the value derived from the geometric overlap model \cite{Bradt} based on measurements of inelastic cross-sections. The $^9$B decays unaccounted herein belong to  $^9$B  scatterings at large angles as compared to the angular decay cone.\par

\indent The $\sigma_{P_T{}^9B}$ and R$_p$ values can be compared with the data on the fragmentation $^{10}$C$~\rightarrow~^9$C. These events are classified as interactions in which target nucleus fragments or mesons are generated, while a heavy relativistic fragment retains the primary nucleus charge (Table~7). In 21 interactions of this type no more than one b- or g-particle was observed, which allows them to be attributed to neutron knockouts The distribution of the transverse momentum $P_{T^9C}$ values of $^9$C nuclei is described by the Rayleigh parameter $\sigma_{P_{T^9C}}~=~(224~\pm~49)~$MeV/$c$. Thus, the $P_{T^9C}$ spectrum appears to be much harder than the  $P_{T2\alpha p}$ spectrum of $^9$B. This fact is associated with the knocking out of neutrons that are bound much more strongly than the outer protons. On the other hand, the knockout of a neutron by a proton is, generally speaking, not a peripheral process, but rather a \lq\lq probing\rq\rq~of the overall density of a projectile nucleus. The radius of a neutron knockout region is (1.0$~\pm~$0.2) Fermi by the statistical model. Of course, this is a naive estimate. Nevertheless, it points to a more compact \lq\lq package\rq\rq~of neutrons than protons in the $^{10}$C nucleus.\par

\begin{figure}
    \includegraphics[width=4in]{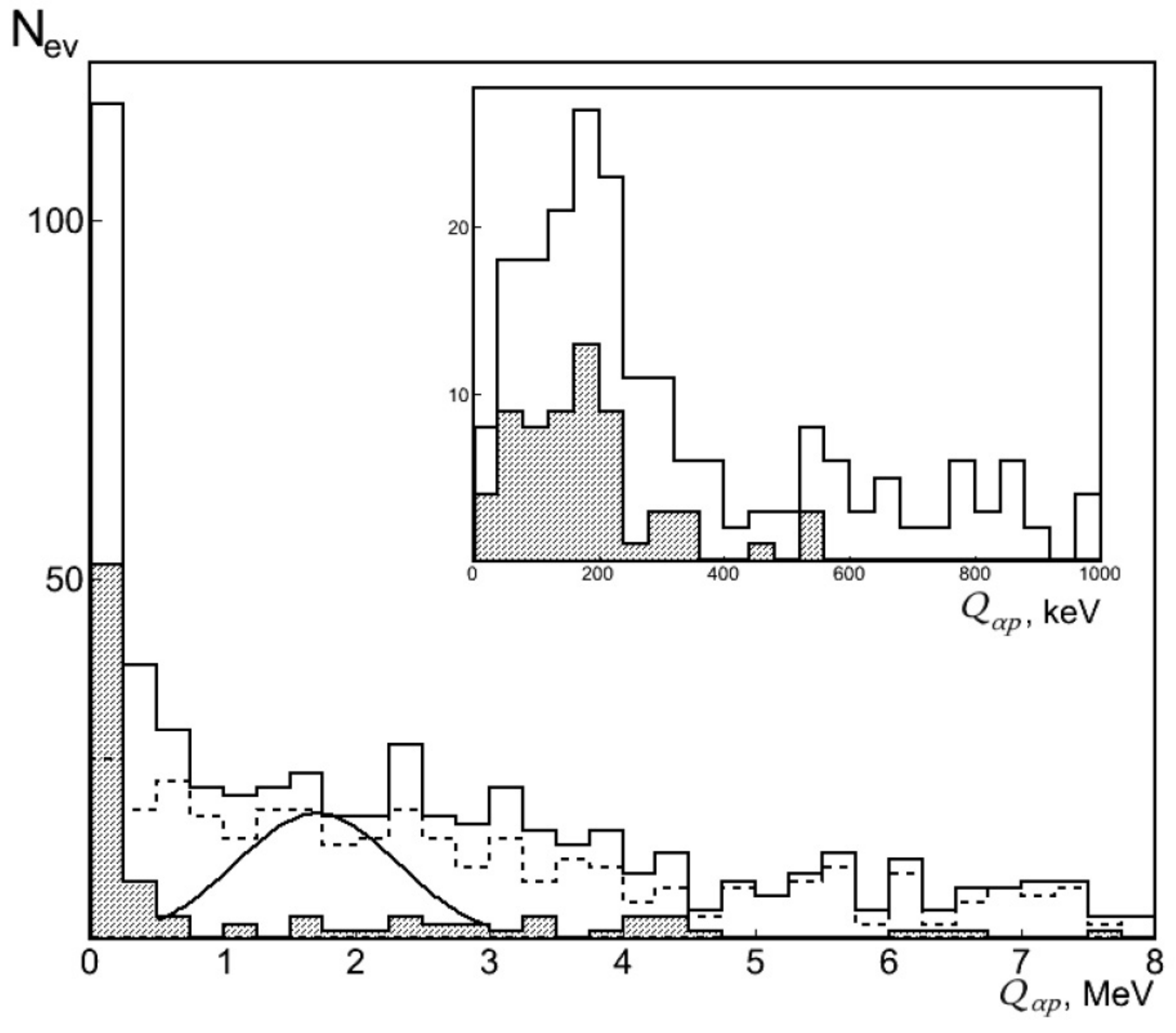}
    \caption{\label{Fig:24} Distribution over energy Q$_{\alpha p}$ of $\alpha p$ pairs in \lq\lq white\rq\rq~stars $^{10}$C$~\rightarrow~2\alpha~+~2p$; solid line$~-~$histogram of all combinations Q$_{\alpha p}$; shaded histogram$~-~$Q$_{\alpha p}$ with $^9$B and $^8$Be production; dashed histogram$~-~$Q$_{\alpha p}$ without $^9$B and $^8$Be production; the curve indicates the expected position of  the $^5$Li resonance; on insertion$~-~$magnified distribution Q$_{\alpha p}$.}
\end{figure}

\indent The distribution of opening angles $\Theta_{\alpha p}$ for 736 $\alpha p$ pairs allows  the resonance decay contribution $^5$Li$_{g.s.}~\rightarrow~\alpha p$ to be estimated (Fig.~\ref{Fig:24}.). The features of $\Theta_{\alpha p}$, which are a narrow peak and a broad maximum, are clarified in the Q$_{\alpha p}$ energy distribution of $\alpha p$ pairs. The peak, pinned to zero, reflects $^9$B decays. The $\alpha p$ pairs of the region 20$\times10^{-3}~<~\Theta_{\alpha p}~<~45\times10^{-3}~$rad are grouped in Q$_{\alpha p}$, corresponding to $^5$Li$_{g.s.}$ decays. Their distribution is described by a Gaussian with a mean value of (1.9$~\pm~$0.1)~MeV with $\sigma$ of 1.0~MeV, which is consistent with the decay energy (1.7~MeV) and the width (1.0~MeV) of the $^5$Li$_{g.s.}$ resonance. About 110 pairs of $\alpha p$ can be attributed to the $^5$Li$_{g.s.}$ decays. There is a small contribution from the $^6$Be resonance decays at the intermediate  values of Q$_{\alpha p}$ which are lower than those of the $^5$Li$_{g.s.}$ decay.\par

\indent Among the \lq\lq white\rq\rq~stars (Table~7) the events Be$~+~$He and $^3$He are observed having thresholds E$_{th}~=~15~$MeV and 17~MeV for the $^{10}$C nucleus. Identification of the He fragments by the $p\beta c$ parameter confirms their interpretation as $^7$Be$~+~^3$He and 2$^3$He$~+~^4$He and does not contradict the assumption that it was exactly the $^{10}$C nuclei that were dissociated. The population of these states requires a rearrangement of the neutrons from one of the $\alpha$-particle clusters to a $^3$He cluster to be produced. Another interpretation points to the presence in the $^{10}$C ground state of deeply bound cluster states $^7$Be$~+~^3$He and 2$^3$He$~+~^4$He with a weight of 8\%.\par

\indent An inverse \lq\lq packaging\rq\rq~2$^3$He$^4$He$~\rightarrow~^7$Be$^3$He$~\rightarrow~2p2^4$He$~\rightarrow~^{10}$C will result in a powerful release of energy. Replacing of one more $^3$He nucleus by $^4$He gives a state close to the $^{11}$C ground state. The  formation of $^{10,11}$C isotopes in astrophysical $^3$He$~-~^4$He mediums leads one to $^{10,11}$B isotopes. Their abundance in cosmic rays can be indicative of nucleosynthesis in $^3$He and $^4$He mixtures. Such an assertion is not commonly accepted. Boron isotopes are believed to be generated in the bombardment of carbon stars by high-energy particles or in the splitting of heavier nuclei of cosmic rays. Nevertheless, the studies of $^3$He states with various isotopic compositions can add new information to the already known scenarios of nucleosynthesis.\par

\subsection*{Coherent dissociation of $^{12}$N nuclei}

\indent Clustering of the insufficiently explored $^{12}$N nucleus is the next goal in the further development of the $^7$Be, $^8$B and $^{9,10,11}$C studies in the relativistic dissociation approach. In an astrophysical aspect its existence provides an alternative scenario for the synthesis of the $^{12}$C isotope via the fusion $^{11}$C$~+~p$. For $^{12}$N \lq\lq white\rq\rq~stars, the channels $^{11}$C$~+~p$ (E$_{th}~=~0.6~$MeV), $^8$B$~+~^4$He (E$_{th}~=~8~$MeV) and $p~+~^7$Be$~+~^4$He (E$_{th}~=~7.7~$MeV) and the channels associated with the dissociation of the $^7$Be core are expected to play a leading role. The threshold of the channel $^3$He$~+~^9$B$_{g.s.}$ is located at E$_{th}~=~10~$MeV. A small difference in the binding energy compared with the channels containing fragments Z$_{fr}~>~2$ suggests a possible duality of the $^{12}$N nucleus. On the one hand, its basis can be represented by the bound $^7$Be and $^8$B nuclei, on the other hand by the unbound $^8$Be and $^9$B nuclei. Therefore, a particular feature of the coherent $^{12}$N dissociation could be a competing contribution of $^8$Be and $^9$B decays.\par

\indent Measurements of the charges of the beam nuclei Z$_{pr}$ and relativistic fragments Z$_{fr}~>~2$ in the candidate events of the $^{12}$N dissociation made it possible to select 72 \lq\lq white\rq\rq~stars which satisfy the condition Z$_{pr}~=~7$ and $\sum$Z$_{fr}~=~7$ \cite{Kattabekov1,Kattabekov2}. The charge topology distribution of these stars is shown in Table~8. Accidentally, the mass numbers A$_{fr}$ become definite for isotopes Z$_{fr}~>~2$. According to the \lq\lq white\rq\rq~star statistics, the share of $^{12}$N nuclei in the beam is estimated to be 14\%, while those of $^{10}$C and $^7$Be nuclei are about 43\% each (excluding H and He nuclei). These values do not reflect the ratio of the cross-sections of the charge exchange and fragmentation reactions and have a technical importance. The significant contribution to the beam of charge-exchange products $^{12}$C$~\rightarrow~^{12}$N compared with $^{10}$C and $^7$Be fragments of $^{12}$C is explained by the fact that the beam was tuned to the ratio Z$_{pr}/$A$_{pr}~=~5/12$ of $^{12}$N, which is slightly different from the values for  $^{10}$C and $^7$Be.\par

\begin{table}
\caption{\label{Tabel:8} Distribution of the $^{12}$N \lq\lq white\rq\rq~stars; middle row$~-~$ selection with the condition $\theta_{fr}~<~11^{\circ}$, bottom row$~-~\theta_{fr}~<~6^{\circ}$}
\begin{tabular}{c|c|c|c|c|c|c|c}
\hline\noalign{\smallskip}
~He+5H~&~2He+3H~&~3He+H~&~$^7$Be+3H~&~$^7$Be+He+H~&~$^8$B+2H~&~$^8$B+He~&~C+H~\\
\noalign{\smallskip}\hline\noalign{\smallskip}
9 & 24 & 2 & 10 & 9 & 11 & 3 & 4  \\
\noalign{\smallskip}\hline\noalign{\smallskip}
2 & 12 & 2 & 5 & 8 & 9 & 3 & 4 \\
\noalign{\smallskip}\hline
\end{tabular}
\end{table}

\indent For a further selection of events containing specifically $^{12}$N fragments (not \lq\lq participants\rq\rq), the condition on the angular cone of coherent dissociation was enhanced to $\theta_{fr}~<~6^{\circ}$, which is determined by a \lq\lq soft\rq\rq~constraint on the nucleon Fermi momentum. In the distribution of 45 selected events (Table~8) the share of the channels with heavy fragments Z$_{fr}~>~2$ reaches approximately 2/3, and the contribution of the channels containing only He and H fragments is quite significant. A noticeable contribution of a very \lq\lq fragile\rq\rq~$^8$B points to a \lq\lq cold\rq\rq~fragmentation with minimal perturbation of the $^{12}$N structure. As judged by the facts of approximate equality of the probabilities of the channels 2He and He + 2H in the dissociation of the $^7$Be nucleus \cite{Peresadko}, $^7$Be core of $^8$B \cite{Stanoeva} and $^9$C \cite{Krivenkov}, one would expect that for the $^{12}$N nucleus the probabilities of the channels 2He$~+~$3H and 3He$~+~$H are nearly equal. In contrast, the statistics in the 2He$~+~$3H channel turned out to be unexpectedly large.\par

\indent Angular measurements were used to study the contribution of $^8$Be decays. Only two candidates for $^8$Be$_{g.s.}$ decays were found in the distribution on the opening angle $\Theta_{2He}$ for the \lq\lq white\rq\rq~stars 2He$~+~$3H and 3He$~+~$H. Thus, the  contribution of $^8$Be$_{g.s.}$ to the $^{12}$N structure is estimated to be only 4$~\pm~2$\%. For the neighboring nuclei $^{12}$C \cite{web1}, $^{10}$C \cite{Kattabekov1,Mamatkulov}, $^{10}$B \cite{Adamovich4} and $^{14}$N \cite{Shchedrina} it amounted to about 20\%. The data on $\Theta_{2He}$ for $^{12}$N do not exclude a possibility of dissociation via $^8$Be 2$^+$ state decays. The latter question requires statistics at a new level.\par

\indent When searching for an analogy between $^9$C and $^{12}$N nuclei by replacing one of the outer protons in the system $2p~+~^7$Be by an $\alpha$ cluster, there arises the following difficulty. The probability of channels, which require the splitting of the outer $\alpha$ cluster in the $^{12}$N nucleus, roughly coincides with the values for channels that can be associated with the separation of only $\alpha$ cluster. A \lq\lq simple\rq\rq~picture of the $^{12}$N nucleus as a $p~+~^7$Be$~+~^4$He structure  appears to be insufficient. It is most likely that the cluster structure of the $^{12}$N ground state constitute a complex mixture of the $^7$Be core states and all possible configurations of H and He nuclei.\par

\section*{STOPPED RADIOACTIVE NUCLEI}

\begin{figure}
    \includegraphics[width=4in]{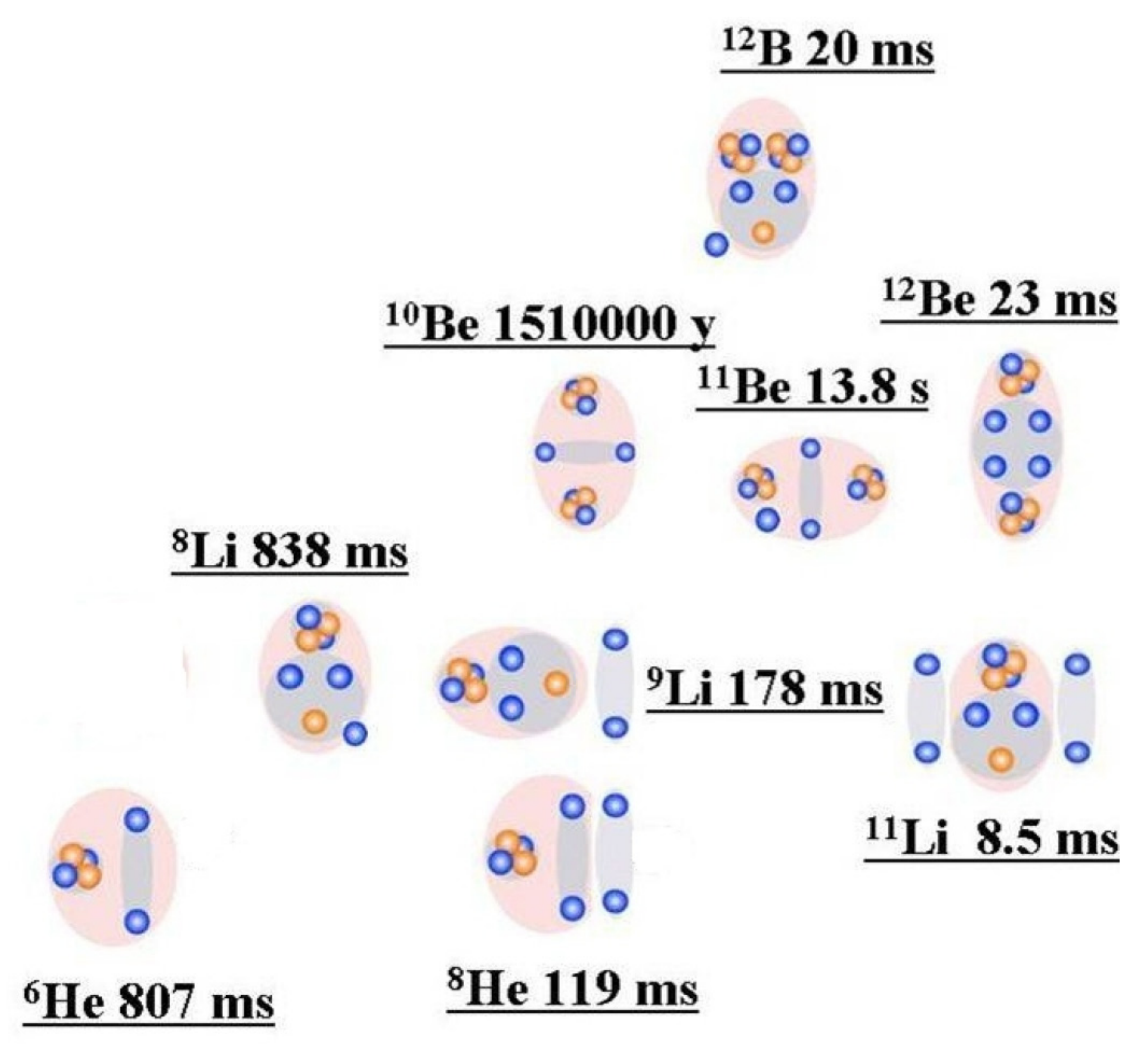}
    \caption{\label{Fig:25} (Color online) Diagram of cluster degrees of freedom in stable and neutron-deficient nuclei; lifetimes of isotopes are indicated.}
\end{figure}

\indent Studies of nuclei along the neutron stability border formed an area of research$~-~$the physics of nuclei with exotic structure (Fig.~\ref{Fig:25}). New phenomena in the structure of such nuclei and in nuclear reactions with their participation have been discovered. Great progress has been made in studying the structure of the nuclei $^6$He, $^8$He, $^{11}$Li and $^{14}$Be \cite{Aumann}. Small values of the binding energy allow  the structure of exotic nuclei to be determined as molecule-like. Evidences for their abnormally large radii which are interpreted as the formation of spatially separated clusters and nucleons have been received.\par

\indent The exotic nature of the structure has been established in the measurement of interaction cross-sections of relativistic nuclei with neutron excess that were found to be enhanced in comparison with the geometric type dependence. However, the relativistic energy range turned out to be inconvenient for deeper investigations of these nuclei. For an increasingly greater neutron excess in the study of relativistic nuclei it would be required to accelerate increasingly heavier nuclei with large intensities. Therefore, research with moving neutron-rich nuclei shifted to low-energy accelerators, where advantages  exist for magnetic analysis and neutron detection.\par

\indent In the energy range of nuclei several MeV per nucleon, there is a possibility of implantation of radioactive nuclei into detector material. Of course, in this approach daughter nuclei are investigated rather than the nuclei themselves. In this respect it is worth mentioning the known, although somewhat forgotten, possibilities of NTE for the detection of slow radioactive nuclei. More than half a century ago, \lq\lq hammer\rq\rq~tracks from the decay of $^8$Be nuclei through the first excited state 2$^+$ of about 2.0 MeV were observed in NTE. They occurred in the $\alpha$ decays of stopped $^8$Li and $^8$B fragments, which in turn were produced by high-energy particles \cite{Pfp}. Another example is the first observation of the $^9$C nucleus from the decay 2$\alpha~+~p$ \cite{Swami}. When used with sufficiently pure secondary beams, NTE appears to be an effective means for a systematic study of the decay of light nuclei with an excess of both neutrons and protons. In NTE the directions and ranges of the beam nuclei and slow products of their decay can be measured, which provides a basis for $\alpha$ spectrometry. A question of major importance is to supplement the 3$\alpha$  spectroscopy of $^{12}$N and $^{12}$B decays \cite{H1,H2,H3} with data on 3$\alpha$ angular correlations.\par

\begin{figure}
    \includegraphics[width=4in]{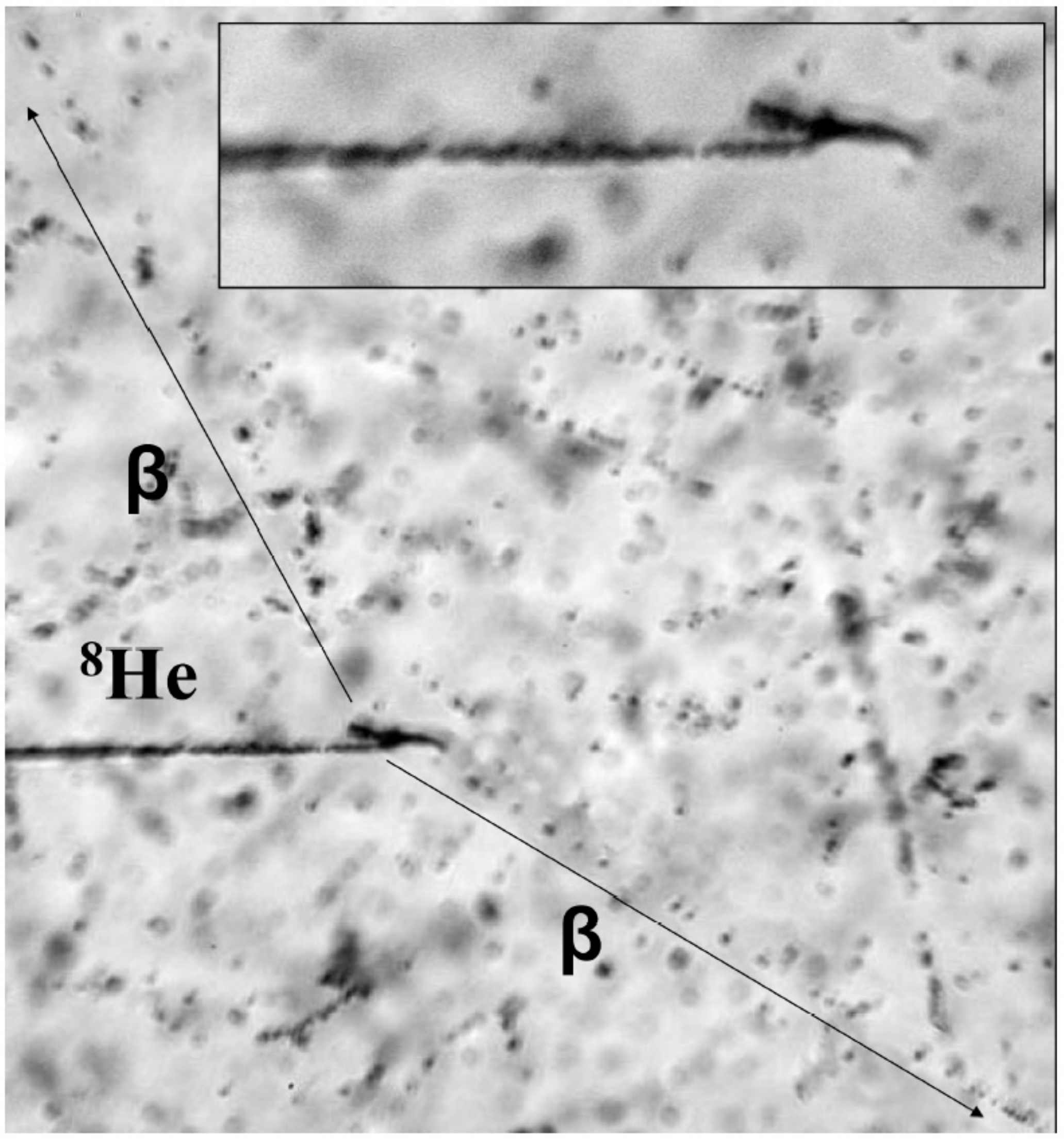}
    \caption{\label{Fig:26} Decay of a stopped $^8$He nucleus; arrows indicate directions of emission of relativistic electrons; on insertion$~-~$magnified decay vertex with a pair of $\alpha$-particle tracks (ranges of about 5~$\mu$m).}
\end{figure}

\indent In March 2012 NTE was exposed at the Flerov Laboratory of Nuclear Reactions (JINR) at the ACCULINNA spectrometer  \cite{Rodin,ACC}. The beam in use was enriched by 7$~A~$MeV $^8$He nuclei. A 107 $\mu$m thick NTE pellicle was oriented at a 10$^{\circ}$ angle during irradiation, which provided approximately a five-fold effective thickness increase. Fig.~\ref{Fig:26} shows a decay of the $^8$He nucleus  stopped in NTE. For ten minutes of irradiation, statistics of about two thousand of such decays was obtained. It is pleasant to note that the used NTE have been recently reproduced by the enterprises \lq\lq Slavich\rq\rq~(Pereslavl-Zalessky, Russia) \cite{Slavich}.\par

\begin{figure}
    \includegraphics[width=4in]{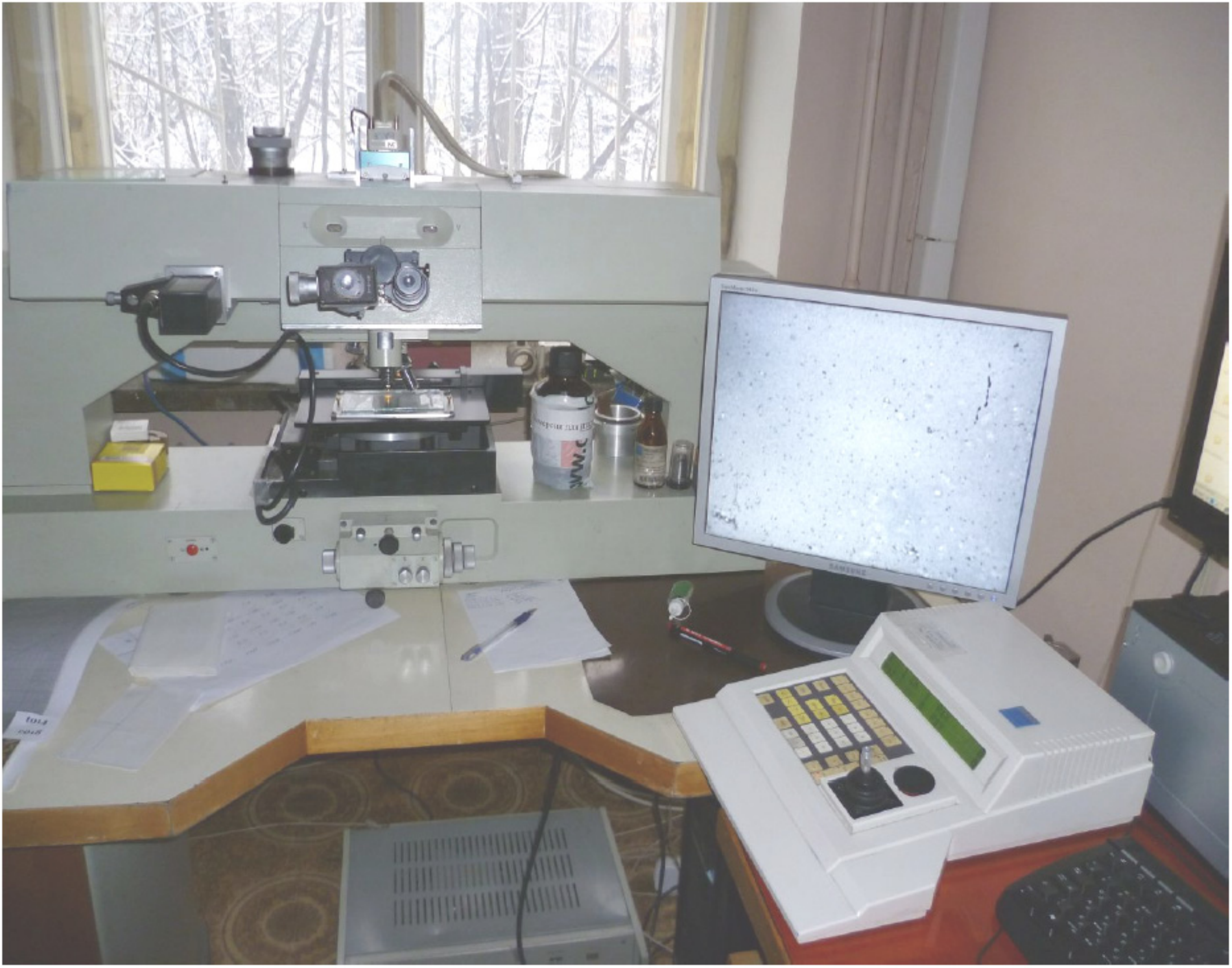}
    \caption{\label{Fig:27} (Color online) Automated microscope PAVICOM-2 (FIAN, Moscow).}
\end{figure}

\indent The use of automated microscopes in searching for and measuring such decays will open the possibility of an unprecedented level of detail and statistics. One of such microscopes is PAVICOM-2 (Fig.~\ref{Fig:27}) of FIAN (Moscow). The PAVICOM complex \cite{Aleksandrov} was originally designed for handling NTE exposed to Pb nuclei at the SPS accelerator (CeRN). Currently, almost all types of solid-state track detectors (emulsions, x-ray films, mylar, plastic, crystals) are handled at the PAVICOM. Automatic analysis of nuclear decays appears to be an exciting prospect  for application of  the PAVICOM team experience. In this way a synergy can be achieved from the classical technique culture combined with modern technology.\par

\section*{HIGH-ENERGY FRONTIER}

\begin{figure}
    \includegraphics[width=4in]{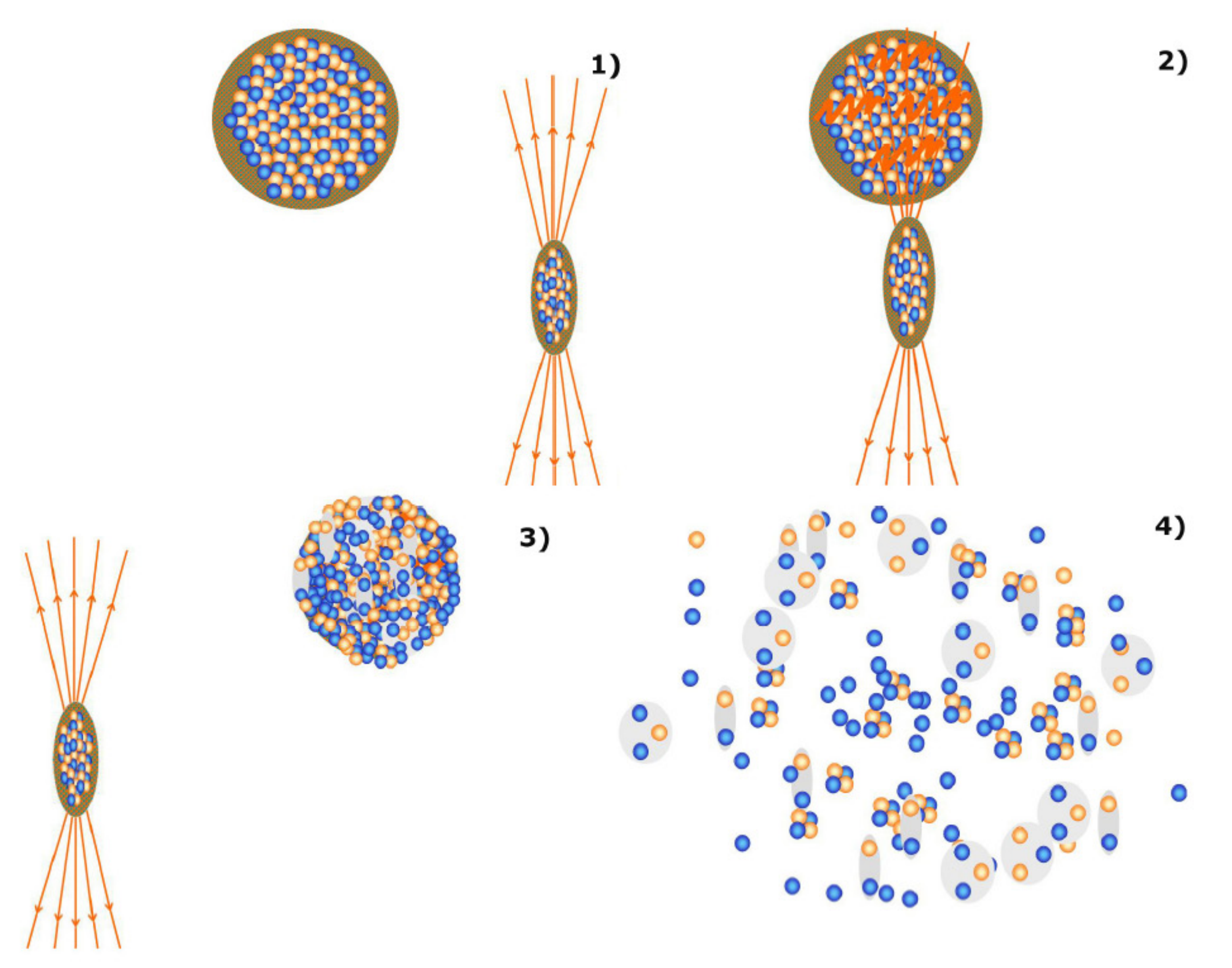}
    \caption{\label{Fig:28} (Color online) Scenario of coherent dissociation of a heavy nucleus in the electromagnetic field of a heavy target nucleus. The nuclei approach each other with an impact parameter larger than their radii (1). The intersection of electromagnetic field of the target nucleus leads to absorption of several virtual photons and to excitation of the projectile nucleus (2). The projectile nucleus turns into an ensemble of lightest fragments and nucleons (3). The ensemble breaks down (4).}
\end{figure}

\indent The presented studies of light nuclei are only the first step toward complex cluster-nucleon ensembles He$~-~$H$~-~n$ produced in the dissociation of heavy nuclei. The question that has to be answered is what kind of physics underlies the "catastrophic" destruction shown in Fig.~\ref{Fig:4}?  Events of dissociation of relativistic nuclei down to a complete destruction into the lightest nuclei and nucleons without visible excitation of target nuclei were reliably observed in NTE for Au and Pb and even U projectile nuclei \cite{Friedlander}. The existence of this phenomenon is certain. It is possible that it confirms the essential role of the long-range quantum electrodynamics interaction. The charges of relativistic heavy nuclei make possible multiphoton exchanges and transitions in many-particle states (Fig.~\ref{Fig:28}), which are almost impossible to observe in electron-nucleus interactions.\par

\indent The predicted dependence of these processes on the target nucleus charge has the form Z$^{2n}$, where $n$ is the number of virtual photons in the interaction \cite{Llope}. Experimentally, such a phenomenon can be established by an enhanced cross-section dependence on the target nucleus charge using the hadron calorimeter method. An alternative scenario of coherent dissociation consists in virtual meson exchanges. In any case the excitation of multiple nuclear giant resonances can give rise to unexpected and even exotic configurations of nucleons and clusters in the final states of decays of these resonances.\par

\indent The phenomenon of electromagnetic dissociation of relativistic nuclei was discovered in Berkeley in the 70s, when the fragmentation of $^{12}$C and $^{16}$O nuclei in a variety of isotopes was studied at 1.05A and 2.1$~A~$GeV \cite{Heckman2}. A sharp rise of the cross-sections was observed as compared with the overlap dependence of colliding nuclei. The observed effect was explained by the Z$^2$ dependence on the target charge and was described by the equivalent photon method using the data on the cross-sections of photon-nucleus interactions. For the $^{18}$O nucleus at 1.7$~A~$GeV, the fragmentation cross-sections with separation of one or two nucleons were obtained in interactions with nuclei from Be to U \cite{Olson}. Despite a relatively high threshold for nucleon separation (above 12~MeV), an increase  of Coulomb-type (Z$^2$) cross-section was also observed. Channels with lower thresholds remained unreachable (for instance, $^{18}$O($\gamma$, $\alpha$)$^{14}$C with E$_{th}~=~6~$MeV). However, the electromagnetic nature of the effect was revealed in an obvious way.\par

\indent Observations of coherent dissociation in nuclear emulsion and fragmentation in magnetic spectrometers stimulate ideas of experiments with neutron-rich nuclei at energies above 10$~A~$GeV, when an effective identification of relativistic nuclei and neutrons becomes possible in segmented hadron calorimeters. Identification of the dissociation channels $^6$Li$~\rightarrow~^3$He$~+~t$, $^9$C$~\rightarrow~3^3$He and $^{10}$C$~\rightarrow2^3$He$~+~\alpha$ raises the problem of search for mirror transitions with replacement of helions ($^3$He) with tritons. The probabilities of the coherent dissociation channels $^6$He$~\rightarrow~2t$, $^9$Li$~\rightarrow~3t$ and $^{10}$Be$~\rightarrow~2t~+~\alpha$ will allow establishing the role of deeply bound configurations with triton participation. On the other hand, the triton is a long-lived nucleus. The generation and subsequent fusion of tritons in astrophysical processes can lead to new branches of the synthesis of neutron-rich nuclei. For the study of cluster ensembles with participation of tritons the calorimetric method provides an alternative to low-energy nuclear physics approaches.\par

\indent The possibility of the existence of a cluster of four neutrons or a tetraneutron $^4n$ is under discussion \cite{Marques,Nesterov,Timofeyuk1,Timofeyuk2,Simenog,Bertulani,Orr}. Even being unstable, the $^4n$ state can be manifested as a resonance. A calorimeter-based experiment on the photodisintegration of $^8$He nuclei above 10$~A~$GeV produced in fragmentation of relativistic  $^{12}$C nuclei \cite{Anderson} will allow a search for the tetraneutron to be accomplished.\par

\section*{CONCLUSIONS}

\indent Thanks to its record spatial resolution and sensitivity, the method of nuclear track emulsions allowed carrying out a \lq\lq tomography\rq\rq~ for a whole family of light nuclei, including neutron deficient ones. In the case of peripheral interactions a relativistic scale of collisions of nuclei not only does not impede investigation of the cluster aspects of nuclear structure, but also offers advantages for studying few-particle ensembles. The facts collected in \lq\lq mosaic\rq\rq~in these notes can serve as experimental \lq\lq lighthouses\rq\rq~ for developing theoretical concepts of nuclear clustering as well as for planning new experimental studies with relativistic nuclei.\par

\indent In the $^{10}$B and $^{11}$B dissociation the three-body channels 2He$~+~$H are dominant (about~75\%). For the $^{10}$B nucleus an enhanced deuteron yield is observed, which is comparable with the $^6$Li nucleus case and points to the deuteron clustering in $^{10}$B. A large share of tritons in the dissociation of the $^7$Li and $^{11}$B nuclei points to the triton clustering in these nuclei. Observation of the $^{11}$B coherent charge exchange only for the two-body channel $^7$Be$~+~^4$He($^{11}$C$^*$) points to a sensitivity of the dissociation to the peculiarities of the mirror nuclei.\par

\indent In the  coherent dissociation of $^6$He nuclei, an average transverse momentum of $\alpha$ particles is about 35~MeV/$c$. Its value, which is noticeably smaller than in the inclusive $^6$He fragmentation, shows that it is desirable to use most peripheral interactions in studies of the neutron halo in nuclei.\par

\indent The share of $^3$He fragments in the $^7$Be dissociation, which is twice exceeds the content of $^4$He fragments, points to clustering based on the helion ($^3$He nucleus). It is most clearly manifested in the leading role of the coherent dissociation $^4$He$~+~^3$He. At the same time the role of the $^3$He cluster is beyond  partnership in the bond $^4$He$~+~^3$He, and the presence of more complex configurations involving $^3$He in the $^7$Be structure is possible.\par

\indent The fragmentation $^9$Be$~\rightarrow~2\alpha$ occurs mainly (80\%) via the 0$^+$ and 2$^+$states of the $^8$Be nucleus  with close probabilities. There is no difference between the total transverse momentum distributions of $\alpha$ pairs for the coherent dissociation via these states. These facts support the $^9$Be concept suggesting the presence of superposition of the $^8$Be 0$^+$ and 2$^+$ states with close probabilities in its ground state.\par

\indent In the peripheral fragmentation of $^{14}$N nuclei the channel $^{14}$N$~\rightarrow~$3He$~+~$H is  dominant (50\%) and manifests itself as an effective source of 3$\alpha$ ensembles. The formation of 80\% of 3$\alpha$ triples corresponds to $^{12}$C excitations from the threshold up to 14~MeV. With a probability of about 20\% the $^{14}$N nucleus forms fragments via the $^8$Be nucleus.\par

\indent The contribution of $^7$Be$~+~p$ in the coherent dissociation  of $^8$B is dominant. The contribution of few-body configurations consisting of He and H nuclei in the $^8$B structure is estimated to be 50\%. In electromagnetic dissociation of $^8$B nuclei a limiting value of the total transverse momentum of pairs $^7$Be$~+~p$ does not exceed 150~MeV/$c$.\par

\indent A particular feature for the $^9$C nucleus is events of coherent dissociation into three $^3$He nuclei, the probability of which is approximately equal to the values for the channels with the separation of one or a pair of protons (about 14\%). This observation points to a considerable contribution of 3$^3$He component to the $^9$C ground state. In the channel $^9$C$~\rightarrow~3^3$He. pairs of $^3$He nuclei with opening angles less than 10$^{-2}~$rad are observed, which indicates the possibility for the existence of a resonant state 2$^3$He (\lq\lq dihelion\rq\rq) with a decay energy of (142$~\pm~$35)~keV.\par

\indent For the $^{10}$C nucleus the share of the coherent dissociation events $2\alpha~+~2p$ is about 80\%. About 30\% of them belong to the channel $^9$B$_{g.s.}~+~p$ with a subsequent decay $^8$Be$~+~p$.\par

\indent There are no obviously leading channels in the coherent dissociation of $^{12}$N nuclei. At the same time there is an intensive formation of $^7$Be and $^8$B fragments. Most probably, the role of the $^{12}$N core can be attributed to the $^7$Be nucleus.\par

\indent Further advance to heavier neutron-deficient isotopes by means of the emulsion method remains promising, although it is getting more difficult. In this way, a further increase of the diversity of the ensembles $p-^3$He$-\alpha$ under study is possible.\par

\indent In general, the presented results confirm the hypothesis that the known features of light nuclei define the pattern of their relativistic dissociation. The probability distributions of the final configuration of fragments allow  their contributions to the structure of the investigated nuclei to be evaluated. These distributions have an individual character for each of the presented nuclei appearing as their original \lq\lq autograph\rq\rq. The nuclei themselves are presented as various superpositions of light nuclei-cores, the lightest nuclei-clusters and nucleons. Therefore, the selection of any single or even a pair of configurations would be a simplification determined by the intention to understand the major aspects of nuclear reactions and nuclear properties rather than the real situation. The  data presented are intended to help estimate the degree and effects of such simplifications.\par

\indent The approach based on the dissociation of relativistic nuclei, opens new horizons in the study of the cluster structure of nuclei and unbound cluster systems. At present only first steps which  nevertheless are quite necessary have been made. Dissociation of relativistic nuclei leads to the appearance of multiple particle combinations with kinematical characteristics that are of interest in nuclear astrophysics and that cannot be formed in other laboratory conditions. On the other hand, in multiple dissociations of neutron-rich nuclei into light fragments the presence of a significant neutron component becomes unavoidable which is caused by a symmetrical composition of light nuclei. Thus, there is a prospect of exploration of polyneutron states. Besides, an applied interest appears here too.\par

\indent Thus, producing new knowledge, nuclear photography awakens \lq\lq nuclear imagination\rq\rq. One cannot exclude that the completeness of the observations provided by the nuclear track emulsion may remain unattainable for the electronic detection methods. In this case, conclusions of emulsion studies will allow one to recognize their limitations and give confidence to \lq\lq rich\rq\rq~experiments with a great variety of detectors.\par

\section*{ACKNOWLEDGEMENTS}
\indent The author considers it his pleasant duty to extend appreciation to his colleagues for the BECQUEREL project. Drawing attention to the cited publications appears to be a suitable form for such an acknowledgement. Nevertheless, the support given by Prof. A. I. Malakhov (JINR) deserves special thanks. It was under his leadership that the beam extraction system from the superconducting Nuclotron was put into operation, making possible our exposures.\par
\indent The author is sincerely grateful to Prof. C. Beck (University of Strasbourg) for his invitation to write this review. My senior friend Prof. S. P. Kharlamov (Lebedev Physical Institute, Moscow) gave a critical analysis of the initial version. I. G. Zarubina (JINR) made the first proof-text version and prepared illustrations. D. O. Krivenkov (JINR) performed makeups of these notes and of the quoted publications. I. S Baldina and O. K. Kronshtadtov (JINR) edited the English version.\par
\indent The preparation of the review was supported by Grant 12-02-00067 from the Russian Foundation for Basic Research and by grants from the Plenipotentiaries of Bulgaria and Romania to JINR.\par

\end{document}